\documentclass[universe,review,accept,pdftex,moreauthors]{Definitions/mdpi} 
\firstpage{1} 
\pubvolume{1}
\issuenum{1}
\articlenumber{0}
\pubyear{2022}
\copyrightyear{2022}
\externaleditor{Academic Editor: Antonino Del Popolo} 
\datereceived{27 April 2022} 
\dateaccepted{20 August 2022} 
\datepublished{} 
\hreflink{https://doi.org/} 
\pdfoutput=1
\usepackage{amsfonts,amssymb,amscd,amsmath}
\usepackage{graphicx}
\usepackage{color}
\usepackage{epsfig}
\usepackage{soul}
\usepackage{bm} 
\DeclareGraphicsExtensions{.pdf,.jpg,.png,.gif}
\usepackage[utf8]{inputenc}

\newcommand{\beql}[1]{
  \begin{equation}\label{eq:#1}}
\newcommand{\eeq}{
  \end{equation}}
\newcommand{\eq}[1]{
  \mbox{Equation~\eqref{eq:#1}}}
\renewcommand{\sec}[1]{
  \mbox{Section~\ref{sec:#1}}}

\definecolor{brightpink}{rgb}{1.0, 0.0, 0.5}

 \theoremstyle{mdpi}
 \newcounter{thm}
 \newcounter{ex}
 \newcounter{re}
\Title{Cosmology from Strong Interactions}
\TitleCitation{Cosmology from Strong Interactions}
\Author{Andrea Addazi$^{1,2}$, Torbj\"orn Lundberg $^{3}$, Antonino Marcian\`o $^{2,4}$, Roman Pasechnik $^{3,}$* and Michal~\v{S}umbera $^{5}$}
\AuthorCitation{Addazi, A.; Lundberg, T.; Marcian\`o, A.; Pasechnik, R.; \v{S}umbera, M.}
\address{$^{1}$ \quad
Center for Theoretical Physics, College of Physics Science and Technology, Sichuan University, Chengdu~610065, China; addazi@scu.edu.cn 
\\
$^{2}$ \quad
Laboratori Nazionali di Frascati INFN, Via Enrico Fermi 54, 00044 Frascati RM, Italy; marciano@fudan.edu.cn
\\
$^{3}$ \quad
Department of Astronomy and Theoretical Physics, Lund University, SE-223 62 Lund, Sweden; torbjorn.lundberg@thep.lu.se
\\
$^{4}$ \quad
Center for Field Theory and Particle Physics,
Department of Physics, Fudan University, \mbox{Shanghai 200433, China} 
\\
$^{5}$ \quad
Nuclear Physics Institute CAS, 25068 \v{R}e\v{z}, Czech Republic; sumbera@ujf.cas.cz}

\corres{Correspondence: roman.pasechnik@thep.lu.se}

\abstract{
The wealth of theoretical and phenomenological information about Quantum Chromodynamics at short and long distances collected so far in major collider measurements has profound implications in cosmology. We provide a brief discussion on the major implications of the strongly coupled dynamics of quarks and gluons as well as on effects due to their collective motion on the physics of the early universe and in astrophysics.
}



\begin{document}

\section{Introduction and Historical~Perspective}
\label{Sec:Introduction}

The strongly coupled dynamics of quarks and gluons has many important implications in particle physics, astrophysics, and~cosmology~\cite{Shuryak:1980tp, Rafelski:2003zz, Polyakov:1978vu, Olive:1980dy, Witten:1984rs, Ornik:1987up, Busza:2018rrf, Pasechnik:2016wkt, Lattimer:2015nhk, Shuryak:2014zxa, Yagi:2005yb, Boyanovsky:2006bf, Sanches:2014gfa, Zhitnitsky:2015dia, BraunMunzinger:2008tz, McInnes:2015hga}. The~fundamental theory of strong interactions, known as Quantum Chromodynamics (QCD), provides a successful description of a variety of observables in high-energy hadronic collisions~\cite{Campbell:2017hsr}, hadronic masses~\cite{Fodor:2012gf}, and, to~a lesser extent, also of the properties of phases of the QCD matter~\cite{Bazavov:2017dus, Philipsen:2019rjq}. While QCD is successful in the interpretation of short-distance phenomena (i.e., in the weakly coupled regime), a~long-standing theoretical problem is a dynamical description of the color confinement phenomenon. The~latter appears in the infrared (strongly coupled) regime of QCD and still remains the major unsolved problem of the Standard Model (SM) of particle physics~\cite{Shuryak:2018fjr,Pasechnik:2021ncb}. 

Due to confinement, color-charged particles do not exist as free states at large spacetime separations. They are instead bound together into colorless collective excitations that evolve into a gas of hadrons. No exact dynamical transition in spacetime between the fundamental (parton) and the composite (hadron) states of QCD is known to date despite the wealth of phenomenological information available from particle and heavy-ion collision experiments. Therefore, one usually resorts to a heuristic description using the concept of quark-hadron duality~\cite{Poggio:1975af, Shifman:2000jv} together with effective field theoretical (EFT) approaches~\cite{Romatschke:2017ejr}; this is used also in the framework of thermal field theory (for recent reviews of the latter, see also Refs.~\cite{Lundberg:2020mwu,Hofmann-book, Aarts:2015tyj}). On~the theory side, effective (typically, static or equilibrium) approaches, such as lattice QCD (LQCD)~\cite{Shuryak:2018fjr, Bazavov:2017dus}, are commonly being exploited while very little has been done on first-principle real-time evolution of QCD states~\cite{Pasechnik:2016wkt}.

The term ``quark matter'' was first used in 1970 by Itoh~\cite{Itoh:1970uw} in the context of neutron stars. Even before then, in~1965, Ivanenko and Kurdgelaidze~\cite{Ivanenko:1965dg} considered a star made of quarks. Since the mechanism for quark confinement was unknown at that time, they had to assume that the quark masses are much larger than the masses of ordinary baryons. A~few years later, however, following the realization that QCD exhibits asymptotic freedom~\cite{Gross:1973id,Politzer:1973fx}, several authors have suggested that the transition from a hadronic phase to a one dominated by quarks and gluons may be relevant to describe the state of matter in the early universe or inside the neutron stars with~a possibility to re-create such a condition also in the laboratory by colliding heavy ions~\cite{Collins:1974ky,Cabibbo:1975ig, Shuryak:1977ut, Shuryak:1978ij,Freedman:1976ub, Polyakov:1978vu, Kapusta:1979fh, Witten:1984rs}.

The terms ``hadronic plasma'' \cite{Shuryak:1977ut}  and ``quark-gluon plasma'' (or QGP) \cite{Shuryak:1978ij} were coined by Shuryak to describe a hypothetical state of matter existing at temperatures of order 100 MeV. The~corresponding phenomena were expected to occur at a characteristic energy density close to 1 GeV$/$fm$^3$. This makes a good analogy with a classical gaseous plasma in which electrically neutral gas at high enough temperatures turns into a statistical system of mobile charged particles~\cite{Ichimaru:1982zz}. While for such plasma the particle interactions obey the U(1)$_{\text em}$ gauge symmetry of Quantum Electrodynamics (QED), in~the former QCD case, the~interactions between plasma constituents are driven by their \mbox{SU(3)$_{\text c}$} color charges. For~an exhaustive collection of key references tracing the development of theoretical ideas on the QGP up to 1990, see e.g.,~Ref.~\cite{Rafelski:2003zz}. For~a summary of later developments, see more recent reviews~\cite{Shuryak:2014zxa, Braun-Munzinger:2015hba, Pasechnik:2016wkt}.  

Let us note that contrary to initial oversimplified expectations~\cite{Rafelski:2003zz}, strongly interacting multi-particle systems feature numerous emergent phenomena that are difficult to predict from the underlying QCD theory, just like in condensed matter and atomic systems where the interactions are controlled by QED. In~addition to the hot QGP phase, several additional phases of QCD matter were predicted to exist~\cite{BraunMunzinger:2008tz, Fukushima:2010bq}. In~particular, the~long-range attraction between the quarks in the color anti-triplet ($\bf \bar 3$) channel was predicted to lead to the color superconductivity (CSC) phase with condensation of $^{1}S_0$ Cooper pairs~\cite{Barrois:1977xd, Bailin:1983bm}. This result was anticipated, though~using a different reasoning, already in 1969 by Ivanenko and Kurdgelaidze~\cite{Ivanenko:1969gs}, who predicted that the superconducting quark phase may be relevant for the super-dense star interiors. At~high baryon density, an interesting symmetry breaking pattern \mbox{SU(3)$_{\text c} \times$ SU(3)$_{\text L} \times$ SU(3)$_{\text R} \times$ U(1)$_{\text B}$} $\rightarrow$ \mbox{SU(3)$_{\rm c+L+R} \times$ Z(2)} leading to the formation of quark Cooper pairs was found in QCD with three massless quark flavors (i.e., under  an assumption that~$m_u$ = $m_d$ = $m_s$ = $0$) \cite{Alford:1998mk, Fukushima:2010bq}.  This breaking of color and flavor symmetries down to the diagonal subgroup  \mbox{SU(3)$_{\rm c+L+R}$} implies a simultaneous rotation of color and flavor called the color-flavor locking (CFL). It is expected that CSC and CFL phases might play important role in the equation of state (EoS) of neutron stars~\cite{Becker-book}. 

Another interesting phase of QCD matter called quarkyonic matter situated in the QCD phase diagram between the chirally restored and the confined phases was proposed in Ref.~\cite{McLerran:2007qj}. The~quarkyonic matter is expected to exist at densities parametrically large compared to $\sim$$\Lambda_{\rm QCD}^4$ when the number of colors $N_c$ is large. Since gluons are in the adjoint representation of \mbox{SU(3)$_c$}, their effects are scaled as $\sim$$N_c^2$, and~so, they dominate all quark-induced $\sim$$N_c$ effects. This provides the  binding of gluons into quark-free states, so-called glueballs, and~so, the quarkyonic matter has only $N_c$ degrees of freedom (DoFs). This form of matter is expected to play some role in the structure of neutron stars~\cite{McLerran:2018hbz}. The existence of another peculiar form of hadronic matter --- the pion condensate --- was suggested by Migdal already in the 1970s~\cite{Migdal:1978az}. 

The rich phase structure of QCD at nonzero temperature and baryon chemical potential was recently reaffirmed by the proposed existence of phases with spatial modulations; see~\cite{Pisarski:2021qof} and references therein. Their moat-shaped  energy spectrum with a minimum of the energy over a sphere at nonzero momentum leads to a characteristic peak. In~heavy-ion collisions at low energy, these new QCD phases are expected to leave their imprints in particle spectra and their correlations. Their cosmological implications are so far~unexplored.

\begin{figure}[H] 

\includegraphics[width=.7\textwidth]{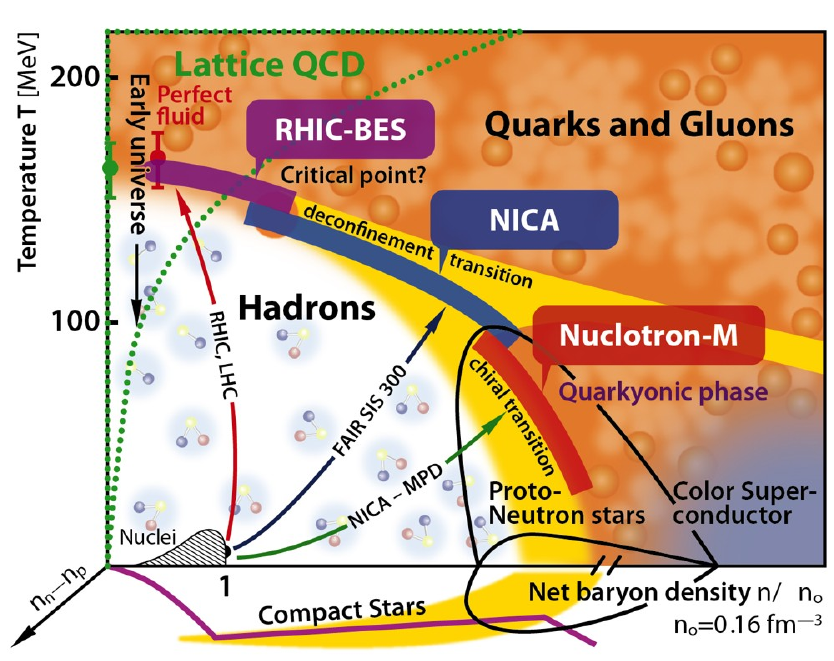}
\caption{
The schematic phase diagram for QCD matter in terms of the temperature $T$ and net baryon density $n$ normalized to the cold nuclei baryon density $n_o$. 
From \url{https://nica.jinr.ru/physics.php} (see also Ref.~\cite{Tejeda-Yeomans:2020spv}).
}
\label{fig:qcdpd}

\end{figure}

Our current knowledge of the QCD phase diagram is illustrated in Figure~\ref{fig:qcdpd}. Comparing this diagram to the phase diagram of water, see e.g.,~Ref.~\cite{Pasechnik:2016wkt}, one notices that (at least, theoretically) the complexity of the former approaches the latter.

\begin{figure}[H] 

\begin{adjustwidth}{-\extralength}{0cm}
\centering
\includegraphics[height=25em]{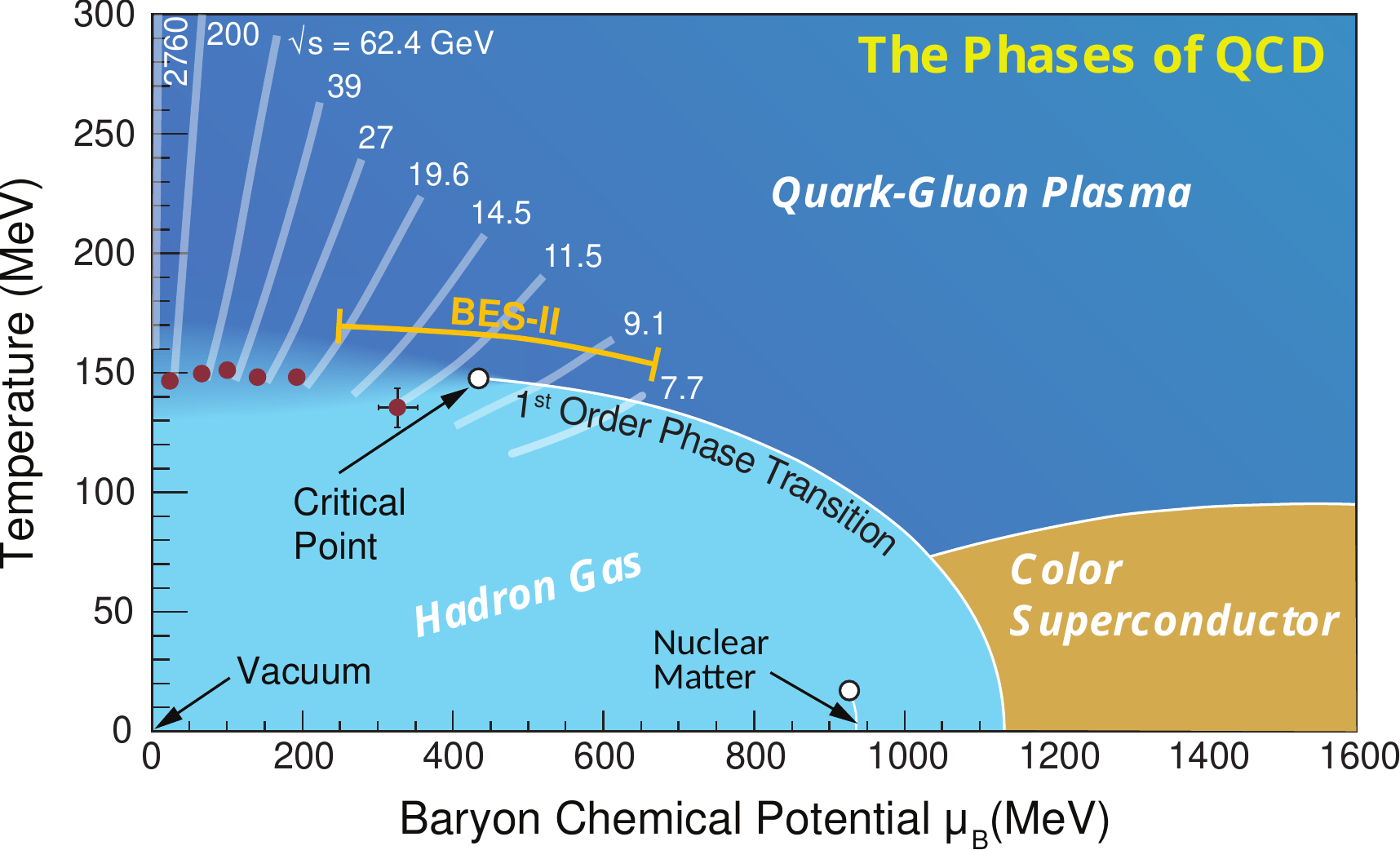}
\end{adjustwidth}
\caption{
A schematic QCD phase diagram in the thermodynamic parameter space spanned by the temperature $T$ and the baryonic chemical potential $\mu_B$. The~corresponding (center-of-mass) collision energy ranges for different accelerator facilities, especially the RHIC Beam Energy Scan (BES II) program, are indicated in the figure. Adapted from Ref.~\cite{Aprahamian} (see also Ref.~\cite{Bzdak:2019pkr}).}
\label{fig:besqcd}

\end{figure}

Experimental study of the QCD phase diagram at high temperatures, see Figure~\ref{fig:besqcd}, dates back to the CERN SPS fixed-target program with the lead ion beams in 1995-2000 and covers the domain of the baryon chemical potential $\mu_{\rm B} \!=\!200-400$ MeV~\cite{Pasechnik:2016wkt}. With~the advent of a first heavy-ion collider in 2000, the investigation of the $\mu_{\rm B} \!\simeq\!0$ region soon led to a discovery of the strongly interacting quark-gluon plasma (sQGP) at RHIC in 2005~\cite{BRAHMS:2004adc,PHOBOS:2004zne,STAR:2005gfr, PHENIX:2004vcz}. The~existence of this new phase of hot and strongly interacting QCD matter was five years later confirmed at order-of-magnitude higher energies of the LHC at CERN~\cite{Braun-Munzinger:2015hba, Pasechnik:2016wkt}. 

Starting from 2010, it became possible to explore systematically the phase structure of hot and dense matter at nonzero baryon density and, in~particular, to~search systematically for the critical endpoint (CEP) of the QCD phase diagram. The~CEP, a~postulated second-order phase transition point, is an expected endpoint of a line of the first-order phase transitions (FOPTs) that separates the low-temperature, low-density hadronic phase from a low-temperature, large-baryon number density QGP phase. Similarly to the water-steam transition where at the critical point, one finds bubbles of steam and drops of water intermixed at all length-scales from macroscopic, visible sizes down to atomic scales (with drops and bubbles near micron scale causing the strong light scattering called ``critical opalescence'' \cite{Wilson:1993dy}), several interesting phenomena are also expected to occur near the CEP of the QCD phase diagram~\cite{Stephanov:1998dy, Gupta:2011wh, Sumbera:2013kd, Adamczyk:2017iwn, Bzdak:2019pkr}. The~search for the CEP is conducted by the STAR collaboration at RHIC within its Beam Energy Scan (BES) program at the energies indicated in Figure~\ref{fig:qcdpd}. 

Current experimental and theoretical studies of the QCD phase diagram thus cover a wide region in temperature and baryon chemical potential $(T,\mu_{\rm B})$, particularly, at~small $\mu_{\rm B} \!\simeq\!0$ \cite{Bellwied:2015rza, Ding:2015ona, Bazavov:2017dus, Philipsen:2019rjq} and large $\mu_{\rm B}\!\simeq 100-600$ MeV~\cite{Fukushima:2010bq, Gupta:2011wh, Bzdak:2019pkr, Philipsen:2019rjq}, see Figure~\ref{fig:besqcd}. The~red and black full circles denote the critical endpoints of the chiral and nuclear liquid-gas phase transitions, respectively. The~(dashed) freeze-out curve indicates where the hadro-chemical equilibrium is attained at the final stage of the collision. The~nuclear matter ground-state at $T = 0$ and $\mu_{\rm B}$ = 0.93 GeV and~the approximate position of the QCD critical point at \mbox{$\mu_{\rm B}$ $\sim$ 0.4 GeV} are also indicated. The~dashed line is the chiral pseudo-critical line associated with the crossover transition at low~temperatures. 

The hot and dense QCD matter is considered to be a dominant ingredient of the early universe evolution in its first few microseconds. Physics of heavy-ion collisions (HIC), therefore, provides necessary means for theoretical understanding of the cosmological processes at those time scales. In~HIC theory, an~important progress has been made when relativistic viscous fluid dynamics was formulated starting from the first principles in an EFT framework, which was based entirely on the knowledge of symmetries and long-lived degrees of freedom, see e.g.,~Ref.~\cite{Romatschke:2017ejr} and Appendix \ref{App:B} of this review. However, for~proper understanding of the cosmological evolution, at~least in a vicinity of the QCD phase transition epoch, the~precise dynamical information on color-field media at finite temperatures is mandatory. Ongoing precision tests of QCD under extreme conditions, in~particular those at the Large Hadron Collider (LHC) at CERN and the Relativistic Heavy Ion Collider (RHIC) at Brookhaven National Laboratory (BNL), are currently pushing the energy, temperature and density frontiers, opening up new largely unexplored possibilities for understanding also the cosmological QCD phase transition. There is strong hope that the growing amount of  data and phenomenological concepts will eventually boost theoretical developments on infrared and finite-temperature dynamics of QCD. The~latter is particularly relevant for understanding the real-time evolution of its ground state and the associated phase transitions as well as hadronization processes relevant for the dynamics of the early~universe.

The QCD dimensional transmutation mechanism, breaking the conformal symmetry of the classical QCD action, has deep implications for the early universe evolution. Indeed, from~higher to lower primordial plasma temperatures, QCD crosses a phase transition to a chiral symmetry-breaking ground state related to the color confinement phase. Thus, an~attractive possibility is that the QCD vacuum energy may provide a source of universe acceleration and Dark Energy (DE) \cite{Pasechnik:2013poa,Addazi:2018fyo}. For~the current status of this problem, see also Ref.~\cite{Pasechnik:2016sbh} and references~therein.

At high temperatures above the confinement scale $\Lambda_{\rm QCD}$, i.e.,~during the first micro-seconds after the Big Bang, the thermal bath in the early universe was dominated by the primordial QGP~\cite{Bailin:2004zd,Yagi:2005yb,Boyanovsky:2006bf, Gorbunov:2011zzc}. When the temperature of the universe decreases down to $\Lambda_{\rm QCD}$, the~QGP dissolves out through collective hadronization phenomena. It is worth remarking that the QGP formation can be highly favoured under the very high-density conditions, where the matter chemical potential starts to be comparable to the QCD critical scale. Indeed, this can happen in high-density objects such as in the core of neutron stars as an effect of the gravitational potential~\cite{Lattimer:2015nhk,Oertel:2016bki, Baym:2017whm}. Nevertheless, the~issue of whether the QGP exists inside the neutron stars is still highly controversial and is under intense debate in the literature. In~fact, the~critical phenomena of QCD dynamics and hadronization and their connections to the QCD ground-state evolution in real time have paramount consequences on the whole cosmological scene, which may also shed light on the late-time cosmic acceleration~\cite{Pasechnik:2013poa}, the~formation of Dark Matter (DM) \cite{Bertone:2016nfn}, primordial black holes (PBHs) \cite{Green:2020jor} and could be imprinted in the spectra of primordial density perturbations and gravitational waves (GWs) \cite{Addazi:2018ctp}.

The main aim of our review is to provide a new critical sight on our current picture of quantum Yang-Mills (YM) field theories in the strong-coupling regime in a dynamical (i.e., non-stationary) spacetime background and in cosmology, in~connection to the empirical knowledge that comes from particle physics measurements and cosmological data. In~the following, unless~otherwise noted, we will mainly exploit the standard natural units $\hbar=c=k_B=1$, where $k_B$ is Boltzmann constant, $c$ is speed of light in the vacuum and $\hbar=h/(2\pi)$, with~$h$ is the Planck~constant. 

The review is organized as follows. In~Section~\ref{Sec:EUPT}, we discuss the nowadays standard scenario of the phase transitions in the early universe, making a connection to the  production of primordial black holes and to super-dense weakly interacting saturated QCD matter. We also discuss possible applications of the axion dynamics to the early universe and close with the possible role of non-perturbative QCD ground-state cosmological evolution. For~completeness, we also mention the possible role of the phase transitions in grand unified theories of particle interactions. In~Section~\ref{sec:Cosmoqgp}, we first introduce the basic notions of the hydrodynamical description of an expanding universe. There, we discuss simple models with constant speed of sound and then move on to a more complicated equation of states for the early universe. We also present current progress in the description of the dissipative effects in relativistic hydrodynamics. The~section is finalized by an overview of the problematics regarding the Cosmological Constant and the Vacuum Catastrophe. Section~\ref{sec:YMtheory} is devoted to a brief discussion of the real-time dynamics of the ground state in an effective action approach to quantum YM theories. We first discuss the YM ground state as time crystal; then, we develop an effective action approach providing the equation of state of the quantum ground state of the universe. 
The~section is closed with a discussion of cosmological attractors --- the solutions of the YM-Einstein equations using the Renormalization Group (RG) methods. Section~\ref{Sect:YMinflation} provides an overview of basic concepts of cosmic inflation models driven by YM dynamics in the early universe. Finally, a~short summary is given in Section~\ref{Sect:summary}.
   
\section{The Phase Transitions in the Early~Universe}
\label{Sec:EUPT}

\subsection{The Phase Transitions in the Standard~Model}
\label{Sec:SMPT}

In the SM of elementary particle interactions, the dynamics of fireball expansion is based on the asymptotic freedom  property of underlying non-Abelian gauge theories~\cite{Gross:1973id, Politzer:1973fx}. QCD is a quantum non-Abelian field theory, an~important part of the SM, that describes the fundamental interactions between colored quarks and gluons. The~generalization of classical electrodynamics to non-Abelian gauge theories was first studied and~exemplified in \mbox{SU(2)} by~Yang and Mills in 1954~\cite{Yang:1954ek}. The~classical Lagrangian density of an \mbox{SU($N$)} gauge theory reads,
\beql{classicalLagrangianDensity}
  \mathcal L_\text{cl}
  =
  -\frac{1}{4} F_{\mu\nu}^a F^{a \, \mu\nu} \,,
\eeq
in terms of the field strength tensor defined in terms of YM fields $A_\nu^a$ as {$F_{\mu\nu}^a = \partial_\mu A_\nu^a - \partial_\nu A_\mu^a +$ $g_\text{YM} f^{abc} A_\mu^b A_\nu^c$}, where $f^{abc}$ are the structure constants of the \mbox{SU($N$)} group. Throughout, $a, b,\ldots$ denote internal indices of \mbox{SU($N$)} in the adjoint representation. Here, the~parameter $g_\text{YM}$ is known as the YM coupling constant. Gauge theories based on \mbox{SU($N$)} are known as YM theories, and they became the target of a wider interest prompted by the discovery that massless particles may acquire a mass and a longitudinal polarization through spontaneous symmetry breaking (or Higgs) mechanism of a massless YM theory~\cite{Nambu:1961tp, Goldstone:1961eq, Goldstone:1962es}. The~latter is a vital part of the SM framework realizing the classical Higgs mechanism of EW symmetry breaking that has found an excellent confirmation through the discovery of the Higgs boson~\cite{ATLAS:2012yve,CMS:2012qbp}.

In the framework of the quantum field theoretical approach, the~YM field fluctuations are quantized around a given ground state through the first quantization procedure \`{a} la QED. However, due to self-interactions of the YM quanta, manifested via the term \mbox{$\propto$$A^2$} in the field strength tensor, the~quantum YM ground state acquires, in~general, a~very non-trivial structure. This structure is well understood in the weakly coupled (perturbative) regime of the theory, implying \mbox{$g_\text{YM} \ll 1$}, which is the case of the EW theory or in the UV regime of QCD with the so-called asymptotic freedom of color charges. The~strongly coupled (non-perturbative) regime, in~which \mbox{$g_\text{YM} \gtrsim 1$} that is realized in particular in the infrared limit of QCD, corresponds to the color confinement phenomenon and has remained the subject of active research over the last few decades. In~recent years, significant progress has been made in understanding of the quantum YM ground state in \mbox{SU($N$)} gauge theories at finite temperatures, see e.g.,~Refs.~\cite{Giacosa:2006hvm,Herbst:2004nk}.

\begin{figure}[H]

\includegraphics[trim= 0 10 0 5, clip, width=.45\textwidth]{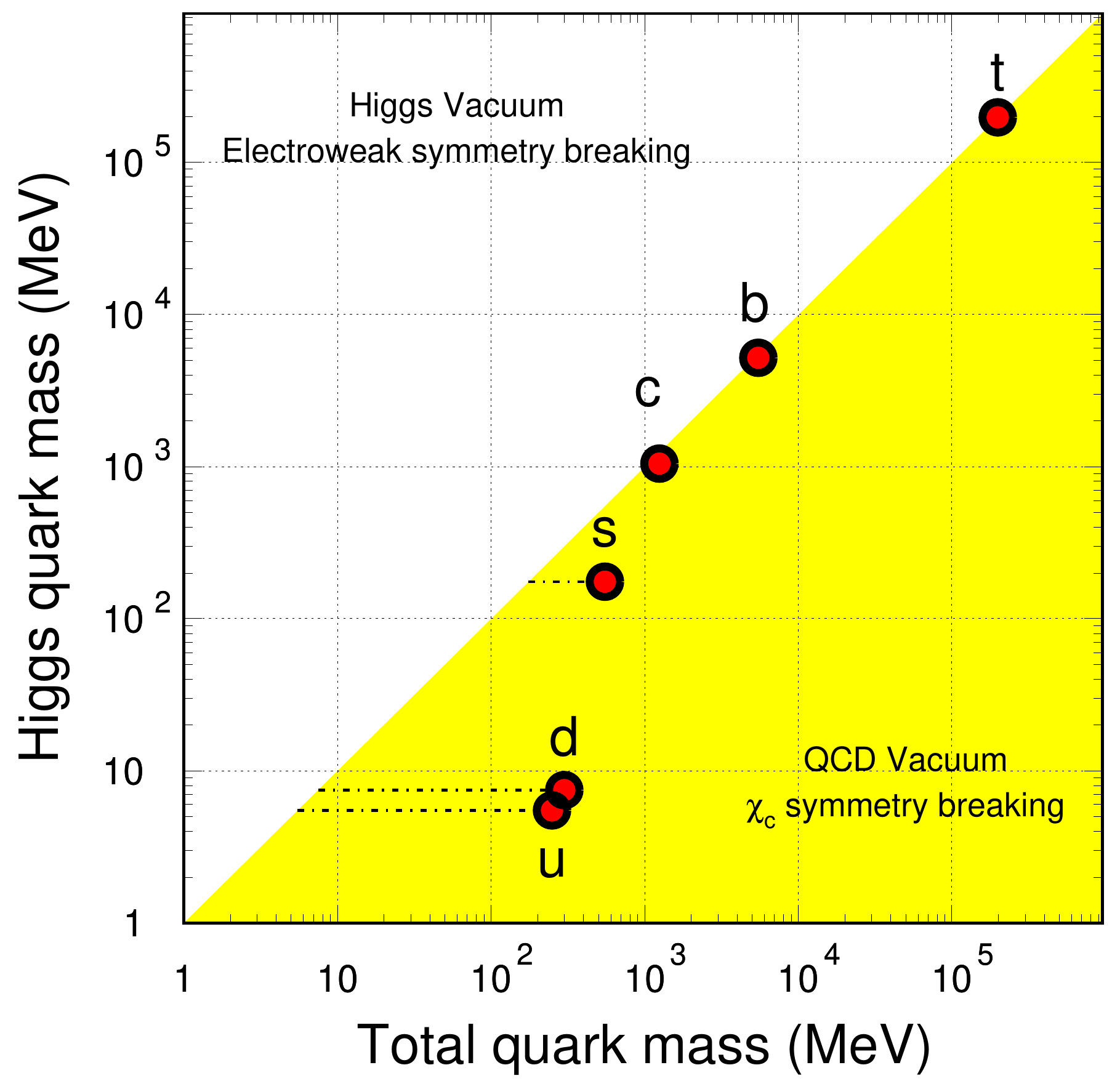}~~~~
 \includegraphics[trim= 0 12 0 10, clip, width=.46\textwidth, height=.425\textwidth]
{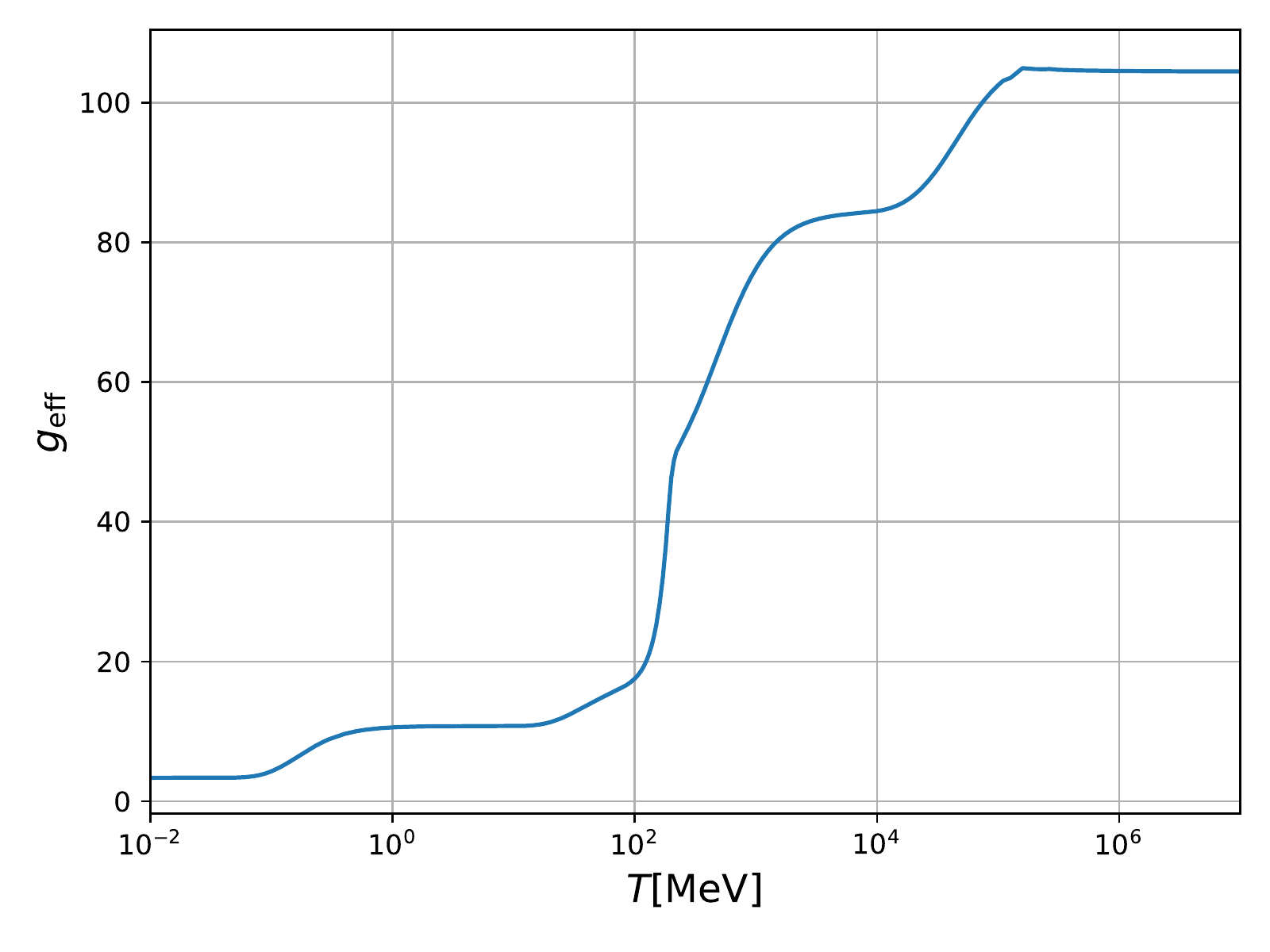}
\caption{(\textbf{Left}) Quark masses in the QCD vacuum and the Higgs vacuum. A~large fraction of the light quark $(u,d,s)$ mass is due the chiral ($\chi_c$) symmetry breaking in the QCD vacuum,
with numerical values from Ref.~\cite{ParticleDataGroup:2004fcd} (see also Ref.~\cite{Zhu:2006er}). (\textbf{Right}) The effective number of relativistic DoFs $g_{\rm eff}$ in the cosmological plasma in the SM as a function of temperature $T$, taking into account interactions between particles, obtained with both perturbative and lattice methods. From~Ref.~\cite{Hindmarsh:2020hop}.}
\label{fig:mass}

\end{figure}

As a result of a series of cosmological phase transitions that occurred in early universe during the first few microseconds after the Big Bang, new vacuum subsystems associated with breaking of the fundamental symmetries were formed. In~the early universe, the~SM predicts that cooling proceeds as a series of two phase transitions associated with the various spontaneous symmetry breakings of the corresponding gauge symmetries~\mbox{\cite{Linde:1978px, Bailin:2004zd, Boyanovsky:2006bf, Mukhanov:2005sc, Laine:2016hma, Mazumdar:2018dfl}}. One at the temperature $T_c^{\rm EW} = 160$ GeV~\cite{DOnofrio:2015gop} is responsible for spontaneous breaking of the EW symmetry providing masses to the elementary particles; see the left panel on Figure~\ref{fig:mass}. Due to the large value of its critical temperature $T_c$, it is not amenable to experimental study under the laboratory conditions. The~second and the only one accessible in the laboratory, QGP-to-hadronic matter phase transition happening at $T_c^{\rm QCD}\approx 160$ MeV~\cite{Bazavov:2018mes}, is related to the spontaneous breaking of the chiral symmetry and manifesting itself in the massless quark limit of the QCD Lagrangian. Since both phase transitions are considered to be analytic crossovers, the~bulk motion of the corresponding plasmas did not depart from thermal equilibrium. Therefore, such transitions, if~realized in nature, are not expected to generate cosmological relics~\cite{Borsanyi:2016ksw,Laine:2015kra, Mazumdar:2018dfl} or to be helpful for a baryogenesis~mechanism.

The QCD phase transition has occurred at characteristic temperatures of above $200 \,{\rm MeV}$ that correspond to a cosmological time-scale of above $10^{-5}\,{\rm s}$ and the Hubble length-scale of approximately $10\,{\rm km}$. The~nature of the QCD phase transition is still a matter of intense debates in the literature~\cite{Kogut:1987rz,Aoki:2006we,Boeckel:2010bey,Schettler:2010dp,Boeckel:2009ej,Ayala:2014jla, Cui:2017ilj,Burkert:2018bqq,Diamantini:2018mjg,Andronic:2017pug,Pang:2016vdc,Du:2019civ, Attems:2018gou, Addazi:2018ctp}, with~results derived so far heavily relying either on lattice field theory methods applied to QCD~\cite{Aoki:2006we,Boeckel:2010bey}, i.e.,~lattice QCD, or~on holographic analyses of QCD at early cosmology~\cite{Attems:2018gou}. For~a thorough review on various aspects of the QCD transition epoch, see e.g.,~Refs.~\cite{Boiko:1991uv,Bonometto:1993pj,Schwarz:2003du,Boyanovsky:2006bf,Castorina:2015ava,Tawfik:2008cd}.

It is undeniable that an abrupt QCD phase transition occurring reversibly in the early universe would lead to a promising cosmological scenario, according to which a large part of the quark excess would be condensed into invisible quark nuggets --- a possible explanation for DM only relying on QCD. As~Witten suggested in Ref.~\cite{Witten:1984rs}, this would happen only if quark matter retains an energy per baryon which is less than 938 MeV: then, neutron stars might generate a quark matter component for cosmic rays, and~detectable gravitational radiation could be produced during the QCD phase transition. Conversely, several recent studies drew a different conclusion, pointing toward the realization of second-order or crossover phase transition scenarios~\cite{Pang:2016vdc,Du:2019civ}.

In a hot and dense QCD matter, the~$u,d$ and to some extent, depending on the temperature, also the $s$ quarks become nearly massless, and the QCD Lagrangian acquires an approximate chiral symmetry \mbox{SU($N_F$)$_{\rm L}~\times$ SU($N_F$)$_{\rm R}$}, with~the  number of massless quark flavors $N_F=2$ ($u,d$) or $3$ ($u,d,s$). At~low $T<T_c^{\rm QCD}$, the QCD vacuum  becomes unstable, and this symmetry is spontaneously broken by $q\!-\!\bar{q}$ pairing. The~corresponding order parameter $\langle\bar{q}q \rangle\!=\!-(245 $ MeV$)^3$, known as the chiral quark condensate, gives rise to masses of light hadrons as well as to constituent masses of $u,d$ quarks and to some extent also to the $s$ quark; see the left panel of Figure~\ref{fig:mass}. Recent lattice calculations with $m_u=m_d=0$ and strange quark having its physical mass reveal that the chiral phase transition occurs at $T_{\chi}^{\rm QCD}\!=\!132^{+3}_{-6}$ MeV~\cite{Kaczmarek:2020sif}. 

Let us note that in~addition to the above scenario when the thermal history of the universe proceeds by a sequence of phase transitions from a more symmetric to a less symmetric state of matter, there is also a possibility of the reverse evolution when part of the zero-temperature unbroken gauge group of the SM or other gauge theory might have been broken in the early universe by thermal effects. As~first noted by Weinberg~\cite{Weinberg:1974hy} in the context of an $O(n)\times O(n)$ gauge theory, with~decreasing $T$, one may encounter a transition to a state of higher symmetry $O(n)\times O(n-1) \rightarrow O(n)\times O(n)$. Within~the minimal extensions of the SM containing an additional color triplet scalar field, the scenario in which the early universe underwent an epoch when SU(3)$_c$ was spontaneously broken but later restored was analyzed in Ref.~\cite{Patel:2013zla}. The attractiveness of such a multi-step phase transition scenario stems from the fact that it may generate the observed baryon asymmetry of the universe~\cite{Ramsey-Musolf:2017tgh}.

To describe the evolution of energy density $\epsilon(T)$ and entropy density $s(T)$ of the early universe, it is customary to normalize both quantities to their values $\epsilon_0(T)$  and $s_0(T)$ corresponding to an ideal massless Bose gas with a single degree of freedom (DoF)  \cite{Byrnes:2018clq, Hindmarsh:2020hop} 
\begin{eqnarray}
g_{\rm eff}(T) \equiv \frac{\epsilon(T)}{\epsilon_0(T)} \,, ~~~~
\epsilon_0(T)=\frac{\pi^2}{30} T^4  \,,
\label{eq:geff}
\\
h_{\rm eff}(T) \equiv \frac{s(T)}{s_0(T)} \,, ~~~~
s_0(T)=\frac{2\pi^2}{45} T^3 \,,
\label{eq:heff}
\end{eqnarray}
and call $g_{\rm eff}(T)$ and $h_{\rm eff}(T)$  the effective numbers of DoFs in energy and entropy, respectively. For~the particular case of a non-interacting gas consisting of $N_{\rm F}$ Dirac fermions, $N_{\rm V}$ massive vectors, $N_{{\rm V}0}$ massless vectors and $N_{\rm S}$ neutral scalars, the two functions are identical and read
\begin{equation}
g_{\rm eff}(T) = h_{\rm eff}(T) = \frac{7}{8} 4 N_{\rm F} + 3 N_{\rm V} + 2 N_{{\rm V}0} + N_{\rm S} \,, 
\label{eq:geffid}
\end{equation}
where the prefactors account for the DoF of each of the considered~particles. 

It is worth mentioning that in a generic case of interacting (non-ideal) gas $g_{\rm eff}(T)$ and $h_{\rm eff}(T)$, they are not identical and depend on temperature. Using the relationship $p=sT-\epsilon$ between the pressure $p$, the~entropy density $s$, the energy density $\epsilon$ and the generalized EoS parameter $w$,
\begin{equation}
p=w\epsilon \,.
 \label{eq:wein1}
\end{equation}
and the speed of sound $c_s$ can be expressed in terms of the effective DoF measures $g_{\rm eff}(T)$ and $h_{\rm eff}(T)$,
\begin{eqnarray}
w(T)=\frac{sT}{\epsilon}-1=\frac{4h_{\rm eff}(T)}{3g_{\rm eff}(T)}-1 \,,
\label{eq:wT}
\\
c^2_s(T)=\frac{dp}{d\epsilon}=
T\frac{ds}{d\epsilon}+s\frac{dT}{d\epsilon}-1= \frac{4}{3}\left[ \frac{4h_{\rm eff}(T)+T h^{'}_{\rm eff}(T)}
{4g_{\rm eff}(T)+T g^{'}_{\rm eff}(T)}\right]-1 \,,
\label{eq:csgh}
\end{eqnarray}
where the prime indicates differentiation with respect to temperature $T$ \cite{Byrnes:2018clq}. It is worth mentioning that the causality condition between the speed of sound and the speed of light $c_s\leq 1$ induces the inequality
\begin{equation}
\frac{h_{\rm eff}(T)}{g_{\rm eff}(T)} \leq \frac{3}{2} \,,
\label{eq:ghb}
\end{equation}
with an upper bound saturated for $w=1$ corresponding to the case of absolutely stiff~fluid.

Taking into account all permissible interactions in the SM, one can calculate either directly~\cite{Husdal:2016haj} or on a lattice the temperature dependence of $\epsilon(T)$ and $s(T)$ and extract corresponding DoFs. This is illustrated on the right panel of Figure~\ref{fig:mass} showing the temperature dependence of $g_{\rm eff}(T)$ in the SM. For~realistic values of both $g_{\rm eff}(T)$ and $h_{\rm eff}(T)$ for a wide temperature interval from 10 keV to 10 TeV, see Ref.~\cite{Husdal:2016haj}.

The simultaneous presence of the EW and the QCD matter in thermal equilibrium is one of the remarkable differences between the QGP produced in accelerator experiments and the primordial QGP in the early universe~\cite{Florkowski:2010mc, Guardo:2014rta, Sanches:2014gfa}. To~find out which form of matter prevails, let us use Equation~(\ref{eq:geff}) and compare the number of DoFs in an ideal massless gas consisting only of EW or QCD matter. Including only the particles which at the temperature $T\lesssim T_{\rm EW}$ can be considered as massless, we obtain $g_{\rm eff}^{\rm EW}=\tfrac{7}{8}(12+6)+2=17.75$ DoFs for the EW case. The~first term in the brackets corresponds to charged leptons $e,\mu,\tau$, and~the second term corresponds to neutrinos $\nu_e,\nu_{\mu}, \nu_{\tau}$. The~last term corresponds to photons. For~non-interacting or weakly interacting QCD matter, Equation~(\ref{eq:geffid}) reduces to 
\begin{equation}
g^{\rm QCD}_{\rm eff}=2\times 8 + \tfrac{7}{8}(3 \times N_F  \times 2 \times 2) \,,
\label{eq:g_qcd}
\end{equation}
where the first term accounts for the two spin and $N_c^2-1=8$ color DoFs of the gluons and the second term accounts for $N_c=3$ colors, $N_F$ flavors, two spin and two particle-antiparticle DoFs of the quarks. Including only the quarks with the mass $m_i/T\simeq 0,\, i=(u,d),\,s,\,c,\,b,$ we obtain successively $g_{\rm eff}^{\rm QCD}= (37,\, 47.5,\, 56,\, 68.5)$ DoFs. In~thermal equilibrium at the temperature $T\lesssim T_{\rm EW}$ and for $N_F$ active quark flavors, the QGP contains a factor of $g_{\rm eff}^{\rm QCD}/g_{\rm eff}^{\rm EW}\simeq 2 - 4$ more energy and pressure than those for the EW matter. For~temperatures $T \gg T_c^{\rm EW}$, deep inside the EW era, all six quarks $u,d,s,c,b,t$ can be considered to be massless, cf. the left panel of Figure~\ref{fig:mass}, and~$g_{\rm eff}^{\rm QCD}=79$. At~the same time, the EW matter acquires $g_{\rm eff}^{\rm EW} =\tfrac{7}{8}(12+6)+8+4=26.75$ DoFs, where $8=2\times(3+1)$ are the DoFs of massless gauge bosons, $W^{\pm},W^0,B^0$, and~the last term is due to the Higgs scalar doublet. For~this case, the QGP has a factor of $g_{\rm eff}^{\rm QCD}/g_{\rm eff}^{\rm EW}\simeq 3$ larger energy density and pressure than those of the EW matter. Hence, we conclude that the QGP was the most dense form of matter filling the early universe during both the QCD and EW~epochs.

\subsection{Creation of Primordial Black Holes during the Phase~Transitions}
\label{Sec:PBH}

According to inflationary theories, initially, very small inhomogeneities in the matter distribution were produced by the end of the exponential expansion regime. Such inhomogeneities filling the early universe are described by the metric perturbations $\delta g_{\mu\nu}$ which can be decomposed into three irreducible pieces --- scalar, vector and tensor ones, see, e.g.,~Refs.~\cite{Mukhanov:2005sc, Weinberg:2008zzc}. While the scalar part is induced by energy density fluctuations $\delta\epsilon$, the~vector and tensor perturbations are related to the rotational motion of the fluid  and to the gravitational waves, respectively~\cite{Mukhanov:2005sc}. Given the scope of this review, in~the following, we focus only on one spectacular phenomenon related to metrics fluctuations in the early universe --- the matter collapse into primordial black holes (PBHs). 

Whereas their existence was proposed already a half-century ago first by Zeldovich and Novikov~\cite{ZeldNov} and later by Hawking~\cite{Hawking:1971ei}, it was the detection of gravitational waves from mergers of tens of solar mass $M_{\odot}$ black hole binaries~\cite{Abbott:2016nmj} which has led to a surge of current interest in the PBHs as a Cold Dark Matter (CDM) candidate~\cite{Carr:2016drx,Carr:2020xqk,Green:2020jor,Biagetti:2021eep}. It can be shown that the creation of PBHs due to the gravitational collapse of hot and dense matter occurs for the density contrast $\delta=\delta\epsilon/\epsilon$ exceeding the critical threshold  $\delta_c(w[T]) \approx 0.3-0.45$, which generally depends on the EoS parameter $w$  \cite{Carr:2020xqk,Green:2020jor}. Such large values of $\delta$ can be generated, e.g.,~during a period of inflation in the very early universe~\cite{Biagetti:2021eep} or during an intermediate period dominated by long-lived massive particles (for recent work, see e.g.,~Ref.~\cite{Allahverdi:2021grt} and references therein) or when the universe in the course of the phase transition passes a local minimum in the pressure-to-energy density ratio $w=p/\epsilon$ \cite{Carr:2020xqk}.

For the PBHs forming from Gaussian inhomogeneities with root-mean-square amplitude $\delta_{\rm rms}$, the present CDM fraction for PBHs with a mass around $M$ is found as~\cite{Carr:2019kxo,Carr:2020xqk} 
\begin{equation}
f_{\rm PBH}(M)\approx 2.4 \beta(M)\sqrt{\frac{M_{\rm eq}}{M}} \,, ~~~ 
\beta(M)=\frac{2}{\pi}\int_x^{\infty}e^{-y^2} dy \,, ~~~ 
x=\frac{\delta_c(w[T(M)])}{\sqrt{2}\delta_{\rm rms}(M)} \,,
\label{eq:f_PBH}
\end{equation}
where $\beta(M)$ is the fraction of horizon patches undergoing collapse to PBHs when the temperature of the universe is $T$, $M_{\rm eq}=2.8\times 10^{17}M_{\odot}$ is the horizon mass at matter-radiation equality, and the numerical factor comes from the ratio of measured baryon $\Omega_{\rm b}$ and CDM $\Omega_{\rm CDM}$ abundances. In~Equation~(\ref{eq:f_PBH}), we have explicitly taken into account dependence of the critical overdensity $\delta_c$ on the EoS parameter $w(T)$. The~temperature depends on the PBH mass $M$ as $T \approx 200\sqrt{M_{\odot}/M}$ MeV. Note that the parameter $\delta_{\rm rms}(M)$ appearing in Equation~(\ref{eq:f_PBH}) can be always adjusted to counterbalance the theoretical uncertainties in the value of $\delta_c$ so that the current PBH DM fraction is preserved~\cite{Carr:2019kxo}. It is worth mentioning that in the scenarios where PBHs are formed during inflation, their abundance is larger than the Gaussian result by orders of magnitude, but~also the mass function has a more pronounced tail at larger masses~\cite{Biagetti:2021eep}.

In fact, there are a plethora of other mechanisms for PBHs formation (including besides the  already mentioned FOPTs, bubble collisions, and~the collapse of cosmic strings, necklaces, domain walls, non-standard vacua, etc., see e.g.,~the recent reviews~\cite{Carr:2020xqk,Green:2020jor}). In~the following, in~conformity with the topic of our review, we will concentrate on the softest point (SP) mechanism of creation of the PBHs discussed in Ref.~\cite{Carr:2019kxo}. Its virtue stems from the fact that by tracing the origin of PBHs to the SM phase transitions, it is capable of explaining the provenance of part, if~not all, of~the CDM in the universe~\cite{Carr:2016drx}.

The SP, corresponding to a local minimum in the pressure-to-energy density ratio $w=p/\epsilon$ as given in Equation~\eqref{eq:wT}, gives rise to elongation of the expansion time of the hot and dense matter. In~HICs, the~interest in locating the SP was started by the recognition that the longest-lived fireball might provide a clear signature of the QGP-to-hadron phase transition~\cite{Hung:1994eq}. Shortly after that, the formation of horizon-size PBHs due to a substantial reduction of pressure during adiabatic collapse in the course of the QCD transition was analyzed in the context of the early universe in Refs.~\cite{Jedamzik:1996mr,Jedamzik:1998hc}. Even though the previously used assumption of the first-order character of the phase transition was later on replaced by a crossover scenario, the~lattice calculations have found a local minimum in $w=0.145(4)$ at $T=159(5)$ MeV~\cite{Borsanyi:2010cj}.

To become acquainted with the influence of the SPs on the cooling of the universe during its radiation-dominated era, let us follow Ref.~\cite{Carr:2019kxo} and inspect the behavior of the function $g_{\rm eff}(T)$ shown on the right panel of Figure~\ref{fig:mass}. Let us focus on the temperatures when a part of the radiation matter transforms into a non-relativistic matter. Starting from $T\approx 200$ GeV downwards, this happens first to the top quark at $T \approx m_{\rm t} = 172$ GeV, which is followed by the Higgs boson at $125$ GeV and~the $Z$ and $W$ bosons at $92$ and $81$ GeV, respectively. The~fact that these particles become non-relativistic at nearly the same time of universe expansion induces a significant drop in the number of relativistic DoFs, from~$g_{\rm eff} = 106.75$ down to $g_{\rm eff} = 86.75$. Further changes at the $b$,$c$-quark and $\tau$-lepton thresholds are too small to be noticeable. Hence, further on, $g_{\rm eff}$ remains approximately constant until the QCD transition at around $160$ MeV. The~number of relativistic DoFs then falls abruptly to $g_{\rm} = 17.25$. A~little later, pions become non-relativistic, and~then muons, yielding $g_{\rm eff} = 10.75$. Thereafter, $g_{\rm eff}$ remains constant until $e^{+}e^{-}$ annihilation and neutrino decoupling at around $1$ MeV, when it drops down to $g_{\rm eff} = 3.36$ \cite{Carr:2019kxo}.

\begin{figure}[H]

\includegraphics[trim= 0 2 0 15, clip, width=.48\textwidth,height=.35\textwidth]{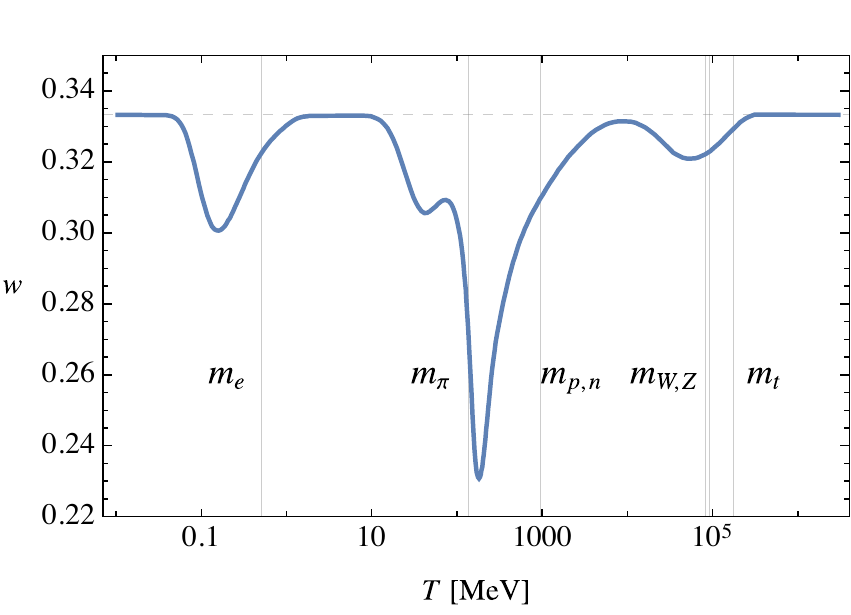}~~
\includegraphics[trim= 0 2 0 15, clip, width=.5\textwidth,height=.35\textwidth]{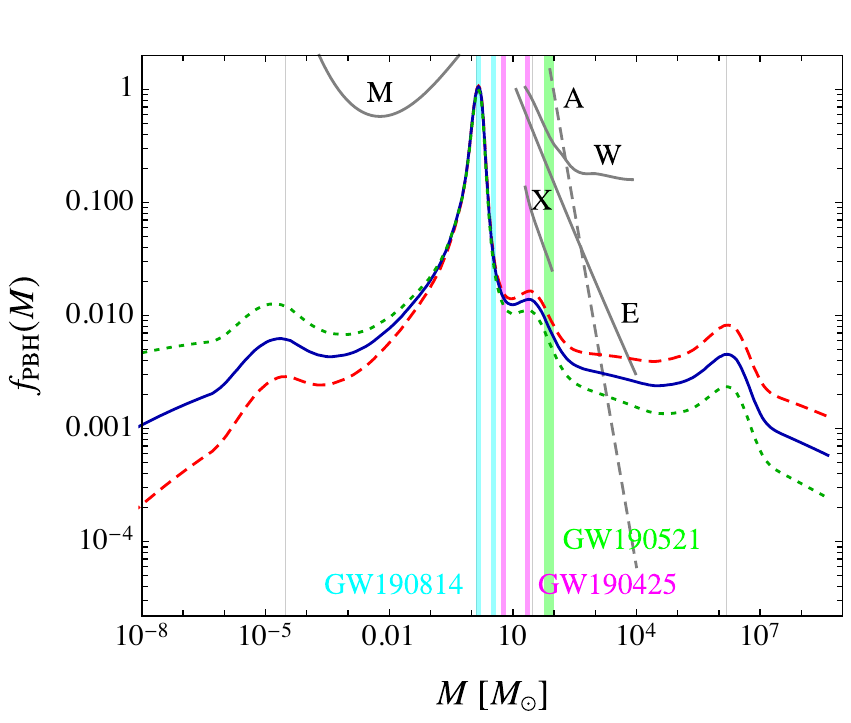}
\caption{({\bf Left}): EoS parameter $w$ as a function of temperature $T$. The~gray vertical lines correspond to the masses of the electron, pion, proton/neutron, $W,Z$ bosons and top quark, respectively. The~gray dashed horizontal line indicates value of $w = 1/3$. Adapted from Ref.~\cite{Carr:2019kxo}.
({\bf Right}): The mass spectrum of PBHs $f_{\rm PBH}(M)$ in solar mass units $M_{\odot}$. The~gray vertical lines correspond to the EW and QCD phase transitions and $e^{+}e^{-}$ annihilation. The~vertical colored lines indicate the masses of the three LIGO-Virgo events. Gray curves are constraints from microlensing (M), ultra-faint dwarf galaxies and Eridanus II (E), X-ray/radio counts (X), and~halo wide binaries (W).  The~accretion constraint (A) is shown dashed, as it relies on uncertain astrophysical assumptions. Adapted from Ref.~\cite{Carr:2019kxo}.}
\label{fig:w-T}

\end{figure}

Provided that total entropy is conserved, an~abrupt reduction of $g_{\rm eff}(T)$ leads to a sudden drop in the speed of sound $c_s(T)$, cf. Equation~(\ref{eq:csgh}), and~hence to a drop of  pressure, $p=w(T)\epsilon$, cf.~Equation~(\ref{eq:wT}). The~effect is clearly visible on the left panel of Figure~\ref{fig:w-T} showing the four periods in thermal history of the universe when $w(T)$ reaches its local minimum. After~each period, $w$ returns back to its relativistic value of 1/3, but each sudden drop modifies the probability of gravitational collapse of any large curvature fluctuations present at that time~\cite{Carr:2019kxo}. 

Consider one cooling period $T_1<T<T_2$ with $w(T)<w(T_{1,2})=1/3$ centered around the local minimum $w(T_{\rm SP})$ and define the quantity,
\begin{equation}
\Delta h_{\rm eff}(T)\equiv g_{\rm eff}(T) - h_{\rm eff}(T)\,,
\label{eq:Delta_h}
\end{equation}
measuring departure from the $w=1/3$ case; see Equation~(\ref{eq:wT}). At~the endpoints $\Delta h_{\rm eff}(T_2)=\Delta h_{\rm eff}(T_1)=0$ but for $w(T)<1/3$, it is always positive $\Delta h_{\rm eff}(T)> 0$, cf.~Equation~(\ref{eq:csgh}). Hence, the~initial drop in the entropy DoFs, $h_{\rm eff}(T)$, always precedes the jump in the energy density DoFs, $g_{\rm eff}(T)$. This leads to the following ``coarse-grained" scenario for the PBH formation: the reduction in $h_{\rm eff}(T)$ occurring for $T_2>T>T_{\rm SP}$ is followed by a fall in $g_{\rm eff}(T)$ for $T_1>T>T_{\rm SP}$. An~excess in entropy $\sim \Delta h_{\rm eff}(T_{\rm SP})$ lost by the radiation during its cooling is dumped into the collapsing matter, emerging eventually in the form of PBHs --- the matter with the largest entropy density in the universe~\cite{Bekenstein:1973ur}. This may explain why even at the present stage of the universe evolution, there is by a huge factor far more entropy in supermassive black holes (BHs) in galactic centers than in all other sources of entropy put together~\cite{Penrose:2018pyw}.

Assuming that the amplitude of the primordial curvature fluctuations is approximately scale-invariant~\cite{Carr:2019kxo}, one obtains from Equation~(\ref{eq:f_PBH}) the mass spectrum of PBHs $f_{\rm PBH}(M)$ shown on the right panel of Figure~\ref{fig:w-T}. The~peaks at $M \simeq 10^{-6}, 2, 30$ and $10^6 M_{\odot}$ correspond to the EW and QCD phase transitions, to~pions and muons becoming non-relativistic and to $e^{+}e^{-}$ annihilation, respectively. The~latter may also provide seeds for the supermassive BHs' formation in galactic nuclei. The~largest contribution to $f_{\rm PBH}(M)$ comes from the PBHs formed at the QCD transition epoch and that would naturally have the Chandrasekhar mass ($1.4\,M_{\odot}$)~\cite{Carr:2019kxo}. Moreover, the~peak in the range $1$-$10\,M_{rm \odot}$ could explain the LIGO/Virgo observations~\cite{Abbott:2016nmj}. The~latter favor mergers with low effective spins as expected for PBHs, but~it is hard to explain BHs of stellar origin~\cite{Wysocki:2017isg}.

The simple analytical models that describe the dynamical process of gravitational collapse which may be relevant for PBH formation were studied in Ref.~\cite{Adler:2005vn}. It is also worth noting that the gravitational collapse of large inhomogeneities during the quark-hadron transition epoch may also explain the baryon asymmetry of the universe~\cite{Garcia-Bellido:2019vlf}. The~asymmetry can be generated in local hot spots through the violent process of PBH formation at the QCD transition triggered by a sudden drop in the radiation pressure and the presence of large amplitude curvature fluctuations caused by the axion field --- the subject to be discussed in Section~\ref{Sec:Axions}.  

\subsection{Perturbative and Strongly Coupled Regimes of~QCD}
\label{Sec:pQCD}

\begin{figure}[H]
\includegraphics[trim= 0 0 0 0, clip, width=.48\textwidth, height=.44\textwidth]{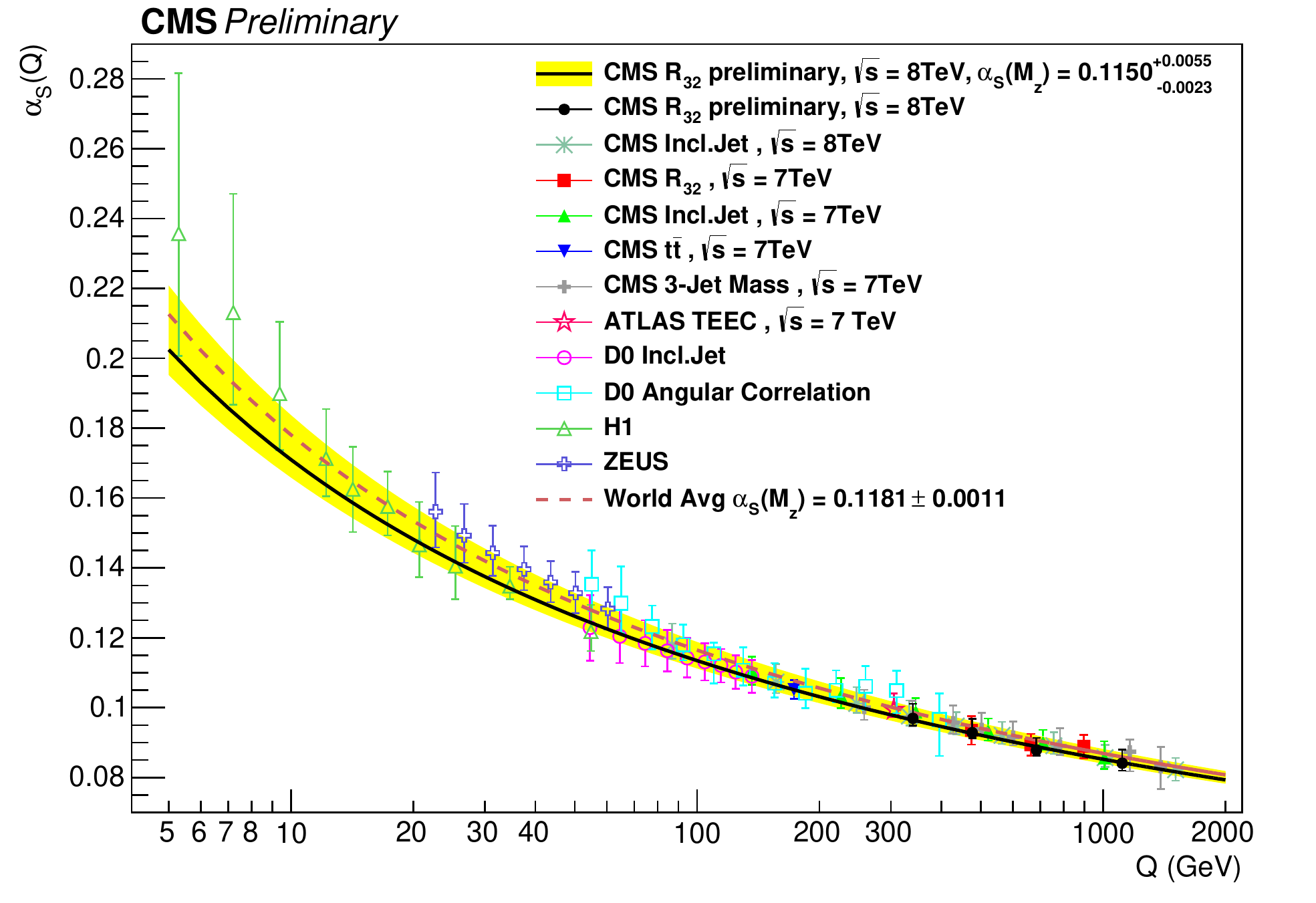}
\includegraphics[trim= 0 0 0 0, clip, width=.51\textwidth, ]{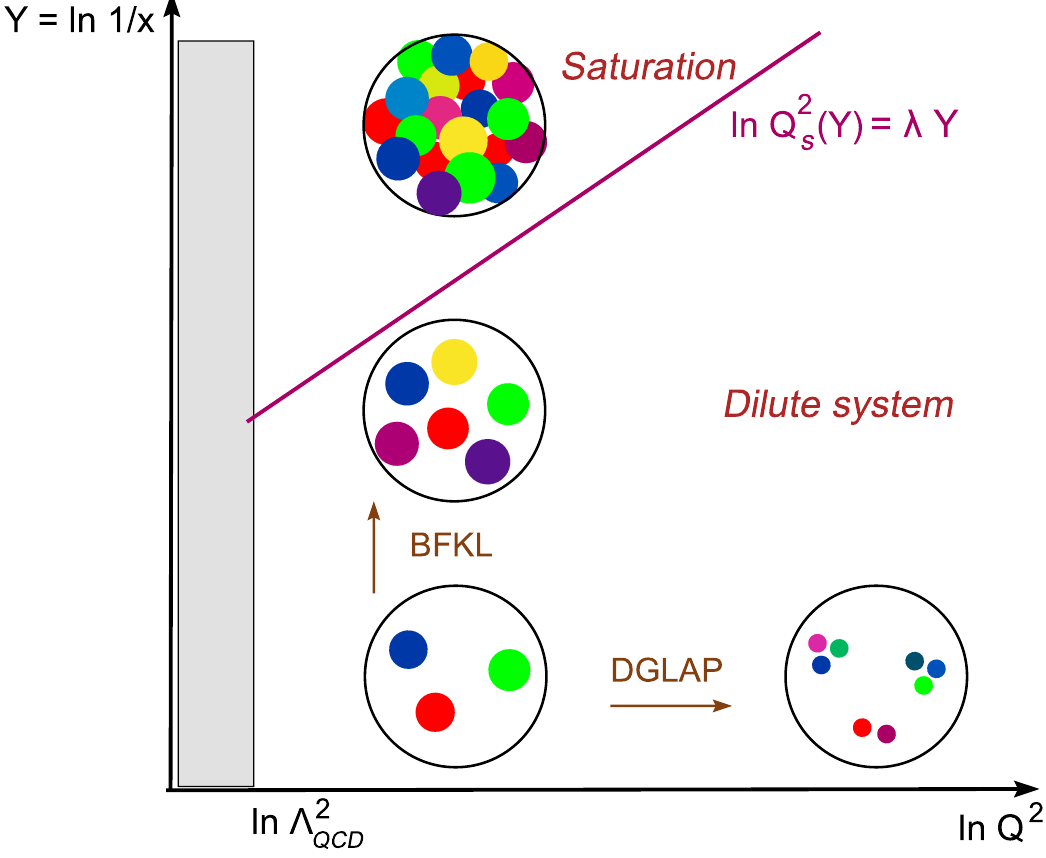}
\caption{\textbf{(Left)} The QCD coupling $\alpha_{\rm S}(Q)$ as a function of the momentum transfer scale $Q$  obtained by using the MSTW2008 NLO PDF set~\cite{Martin:2009iq}. 
Adapted from~Ref.~\cite{CMS:2017tvp}. \textbf{(Right)} The partonic phase diagram showing evolution of the partons' density and size as a function of their rapidity $Y = \ln(1/x)$ and the logarithm of momentum transfer squared $\ln Q^2$. Adapted from~Ref.~\cite{Gelis:2010nm}.}
\label{fig:alphas}
\end{figure}

An important contribution to the effective number of relativistic DoFs, $g_{\rm eff}$, comes from the hadron-to-QGP phase transition --- see a big jump in the interval $10^2 \lesssim T \lesssim 10^3$ MeV in Figure~\ref{fig:mass} (right panel). At~higher temperatures, in~the QGP region, the strength of the interactions between the quarks and gluons is set by the QCD coupling $\alpha_{\rm S}(Q)$ which at the one-loop order of perturbation theory takes the form,
\begin{equation} 
\alpha_{\rm S}(Q) \simeq  \frac{2\pi}{b_0 \ln(Q /\Lambda_{\rm QCD})}\,, \qquad 
b_0\!=\!11\!-\!\frac{2}{3}N_F \,,
\label{eq:alpha_S}
\end{equation}
where $Q$ is the momentum transferred during the interaction, $\Lambda_{\rm QCD}\simeq 200$ MeV is the characteristic QCD scale, and~$N_F$ is the number of active quark flavors. The~logarithmic decrease of $\alpha_{\rm S}(Q)$ with increasing $Q$, i.e.,~with decreasing distance among the quarks and gluons, is due to the fact that, in~contrast to the photon in QED, the~force carriers in QCD, the~gluons, have color charge. Their exchanges in higher-order processes involving both the quarks and the gluons occur more frequently with increasing $Q$ and lead to a color charge spread (or anti-screening). Indeed, the~gluon multiplicity increases at low momentum fractions corresponding to the limit of large energies. Dilution of the initial color charge is responsible for the weakening of $\alpha_{\rm S}$ at small distances $\ell \ll \Lambda_{\rm QCD}^{-1}$, i.e.,~when the quark experiences a large momentum transfer $Q$, see Equation~(\ref{eq:alpha_S}). This effect known as asymptotic freedom~\cite{Gross:1973id, Gross:1973ju, Politzer:1973fx} is illustrated on the left panel of Figure~\ref{fig:alphas} where the values of the $\alpha_{\rm S}(Q)$ extracted from proton-(anti)proton and lepton-proton collisions are shown~\cite{CMS:2017tvp}. In~agreement with Equation~(\ref{eq:alpha_S}), a slow logarithmic decrease from $\alpha_{\rm S}(Q_{\rm min}\!=\!5\,\rm{GeV})=0.22$ to $\alpha_{\rm S}(Q_{\rm max}\!=\!1500\,\rm{GeV})=0.08$ is~observed.

Before proceeding further, let us recall that quantum field theory (QFT) at finite temperature $T$ is often considered to be equivalent to Euclidean QFT in a space which is periodic, with~period $1/T$ along the ``imaginary time" axis (for a recent review of this subject, see e.g.,~Refs.~\cite{Ghiglieri:2020dpq, Lundberg:2020mwu} and references therein). Thus, in~order to formulate the theory at $T>0$ using its variant at $T=0$, one should replace zero components of all 4-momenta $k^{\mu}$ in the Euclidean integrals by the discrete Matsubara frequencies --- $2\pi n T$ for bosons and $(2 n + 1)\pi T$ for fermions, and~sum over $n\in \mathbb{Z}$ instead of integrating over $k^{\mu}$. Consequently, the~average momentum transferred during the interactions in the hot medium $Q$ can be related to the temperature as $Q=2\pi T$. In~particular, the~maximum value of the momentum transferred $Q_{\rm max}\!=\!1500\,\rm{GeV}$, which has been so far measured in $pp$ collisions at the LHC, roughly corresponds to the ``temperature'' $T_{\rm max}=Q_{\rm max}/(2\pi)\simeq 240\, {\rm GeV} \simeq T_c^{\rm EW}$.

The asymptotic freedom formula given by Equation~(\ref{eq:alpha_S}) is based on the applicability of QCD perturbation theory to the processes at high momentum transfers, as it is a well-known fact that the perturbative expansion is an example of asymptotic series. It looses its validity with decreasing $Q$ when the perturbative approximation breaks down. Interestingly, by~performing the matching of the fundamental theory onto the effective chiral Lagrangian at the infrared scale $Q \simeq 4\pi f_{\pi} \simeq 1$ GeV, at~which the ranges of validity of perturbative QCD and chiral perturbation theory (describing interactions among low-momentum hadrons) descriptions meet, one can infer the information about the behavior of $\alpha_{\rm S}(Q)$ at large distances. Such a matching implies that the QCD coupling in the infrared region is ``frozen" at $\langle\alpha_{\rm S} \rangle_{\rm IR} \simeq 0.56$ \cite{Fujii:1999xn}, incidentally at twice the upper scale value of $\alpha_{\rm S}(Q)$ shown on the left panel of Figure~\ref{fig:alphas}.

Let us note that evolution of the strong coupling parameter $\alpha_{\rm S}(Q)$ described at the leading order by Equation~(\ref{eq:alpha_S}) is valid only for DoFs that dominate the thermodynamical evolution, i.e.,~for the partons (quarks and gluons) with momenta of order $T\!=\!Q/2\pi$, and it does not apply to the long wavelength non-perturbative modes residing at the length scales of ${\ell > T^{-1}}$. Those modes are occupied by a liquid in which neighboring ``unit cells'' are tightly coupled to each other~\cite{Blaizot:2011ks}. This strongly coupled regime~\cite{Shuryak:2014zxa} makes the QGP behave as the ideal fluid~\cite{Lacey:2006bc, Heinz:2013th}. The fluidity of the QGP was first established in the collisions of ultra-relativistic nuclei at RHIC~\cite{PHOBOS:2004zne, STAR:2005gfr, PHENIX:2004vcz} and later confirmed at higher energies of the LHC~\cite{ALICE:2010suc,ATLAS:2011ah,CMS:2012tqw}. The~most prominent signals of the strong interaction in the deconfined bulk manifest in a collective flow of matter~\cite{Heinz:2013th, Pasechnik:2016wkt} and in a spectacular phenomenon of suppression of very energetic partons passing through the QGP medium~\cite{Gyulassy:1990ye, Bielcikova:2016lgh, Pasechnik:2016wkt}. Direct evidence for the non-perturbative character of deconfined matter comes from the low-momentum spectra of direct photons measured in Au+Au collisions at RHIC. The~temperatures obtained from the spectra $T\simeq 220$ MeV~\cite{Adare:2008ab} point to the initial temperatures $T_{\rm ini} \simeq 300-600$ MeV at early times of $\tau_0=0.6-0.15$ fm/$c$, which are way below the perturbative regime of~QCD.

By definition, plasma is a state of matter in which charged particles interact via long-range (massless) gauge fields~\cite{Shuryak:2008eq}. This distinguishes it from neutral gases, liquids, or~solids in which the inter-particle interaction is of short range. So, plasmas themselves can be gases, liquids, or~solids, depending on the value of the plasma parameter $\Gamma$, which is the ratio of interaction energy to kinetic energy of the particles forming the plasma~\cite{Ichimaru:1982zz}. For~a classical plasma of $N$ particles with charge $Ze$ occupying a volume $V$,
\begin{equation}
  \Gamma \equiv  \frac{(Ze)^2}{a k_B T} \,, \qquad 
  a(T)=\left (\frac{3V}{4\pi N} \right )^{1/3}  \approx 0.62 n(T)^{-1/3} \,,
\label{eq:Gamclas}
\end{equation}
where $n(T)=N/V$ is the temperature-dependent particle number density. While most plasmas are ideal with $\Gamma<10^{-3}$, a~strongly interacting plasma has $\Gamma \gtrsim 1$. For~plasma with $Z\simeq 1$ at the temperature $T\simeq 10^6$ K $\simeq 100$ eV, the number density $n$ must be as high as $10^{26}$ cm$^{-3}$ to make $\Gamma \simeq 1$ \cite{Ichimaru:1982zz}. Ion plasma in a white dwarf has $\Gamma=10-200$, in~transient plasmas produced in explosive shock tubes, the values of $\Gamma=1-5$ are found~\cite{Ichimaru:1982zz}. A~more down-to-earth example is table salt (NaCl), which can be considered as a crystalline plasma made of permanently charged ions Na$^+$ and Cl$^-$ \cite{Shuryak:2008eq}. At~temperatures of $T\approx 10^{3}$ K, still too small to ionize non-valence electrons, the~table salt transforms into a molten salt, which is a liquid plasma with $\Gamma \approx 60$.

Generalization of Equation~(\ref{eq:Gamclas}) to the QGP case was suggested in Ref.~\cite{Thoma:2005aw}
\begin{equation}
\Gamma \simeq 2\frac{C_{q,g} \alpha_{\rm S}}{a T} \,, \qquad C_q=\frac{N_c^2-1}{2 N_c}=\frac{4}{3} \,, 
\qquad C_g=N_c=3\,,
\label{eq:Thoma}
\end{equation}
where the strong coupling $\alpha_{\rm S}$ is a slowly varying function of temperature, $C_q$ and $C_g$ are the Casimir invariants of fundamental and adjoint irreducible representations of the color \mbox{SU(3)$_c$} group corresponding to quarks and gluons, respectively, and~\mbox{$a = a(T)$} is the average inter-parton distance at a given temperature $T$ as follows from Equation~(\ref{eq:Gamclas}). The~factor $2$ in Equation~(\ref{eq:Thoma}) takes into account the equal importance of chromoelectric (CE) and chromomagnetic (CM) interactions in ultra-relativistic systems. For~ideal massless QCD gas with $N_F$ active quarks and $d_F=g^{\rm QCD}_{\rm eff}$ degrees of freedom, see Equation~(\ref{eq:g_qcd}), the~particle number density reads
\begin{equation} 
n= d_F\frac{\zeta(3)}{\pi^2}T^3 \approx  d_F \left(\frac{T}{2}\right)^3 \,, \qquad 
d_F=2\times 8 + \tfrac{7}{8}(3 \times N_F  \times 2 \times 2)\, .
\label{eq:n_qcd}
\end{equation}

From Equations~\eqref{eq:Gamclas} and \eqref{eq:n_qcd}, it follows that $a \simeq 1.24 d_F^{-1/3} T^{-1}$, and so, the term $aT$, appearing in the denominator of Equation~\eqref{eq:Thoma}, depends on $T$ through the temperature-dependent number of active quark flavours $N_F(T)$ only. Consequently, for~an ideal massless QCD gas, the temperature-dependence of the plasma coupling parameter $\Gamma$ is driven by  $\alpha_{\rm S} d_F^{1/3}$. For~the QGP created in HICs at RHIC $T\approx200$ MeV and $\alpha_{\rm S}=0.3-0.5$ with $N_F=2$, Equations~\eqref{eq:n_qcd} and \eqref{eq:Thoma} yield $d_F=37$ and $\Gamma \simeq 2 - 8$ well inside the strongly coupled regime. At~much higher temperatures, say, at~$T\simeq T_{\rm EW}$, with~$\alpha_{\rm S}=0.08$ and $N_F=5$, we obtain $d_F=52.5$ and $\Gamma\simeq 0.5-1.5$ --- the value located in the vicinity of the strongly coupled regime. At~even higher temperatures, the number of active quark DoFs saturates at $N_F=6$ and the evolution of $\Gamma(T)$ becomes solely driven by the (logarithmically decreasing) QCD coupling $\alpha_{\rm S}(T)$, cf.~(\ref{eq:alpha_S}). Let us note that the ideal gas approximation serves only as a lower estimate of $\Gamma$ because it ignores the interactions in the partonic liquid. The~latter will slow down the temperature dependence of the average inter-parton distance $a$, thus weakening the strong coupling parameter dependence on $T$. 

\textls[-15]{A more in-depth approach to strongly coupled non-Abelian plasmas~\cite{CasalderreySolana:2011us} was expected to come from the gauge/string duality~\cite{Maldacena:1997re} --- a correspondence between $d$-dimensional conformal QFT and $(d+1)$-dimensional string or gravity theory. In~these theories, the graviton needs not live in the same spacetime as the QFT, but due to the holographic principle, the description of gravity within a volume of spacetime can be thought of as encoded on a lower-dimensional boundary to the region in the formalism of conformal field theory~\cite{Susskind:2005js, McGreevy:2009xe}. However, the inherently conformal character of the gauge QFT used in the duality with anti-de Sitter gravity (AdS) is at variance with QCD where the scale invariance is broken by the confinement scale (for a recent review of confinement dynamics, see Ref.~\cite{Pasechnik:2021ncb}), causing the running of the coupling. This limits the applicability of gauge/string duality~\cite{Maldacena:1997re} to temperatures $T \gg T_c^{\rm QCD}$ and hence to weak~couplings.}

\subsection{QCD at High Parton Densities and~Saturation}
\label{Sec:Glasma}

To proceed further, a different type of analysis of the QCD dynamics of partonic matter is needed. There are two independent paths along which the density of partons can evolve, and these are illustrated in the right panel of Figure~\ref{fig:alphas}. The~two together form the basis of our current understanding of high-energy scattering in~QCD.

The first path follows the development of partonic cascade in variable $Q$. For~partons that occupy a transverse area $1/Q^2$, the increase of $Q$ and hence of the temperature $T\sim Q$ leads to dilution of their density. The~process is controlled by the Dokshitzer–Gribov–Lipatov–Altarelli–Parisi (DGLAP) \cite{Altarelli:1977zs,Dokshitzer:1977sg,Gribov:1972ri} equations describing the evolution of partonic density as a function of evolution variable $\ln (Q^2/\Lambda_{\rm QCD}^2)$ \cite{Ioffe:2010zz, Campbell:2017hsr}.

The second path follows the development of a parton shower in variable $x=k^+/P^+$, which is a~fraction of the light cone momentum\endnote{The light cone four-vectors are related to Minkowski four-vectors in a standard way $k=(k^0,k_{\perp},k_z)=[k^-,\boldsymbol{k_{\perp}},k^+]$ where $k^{\pm}=k^0\pm k^z$. The~Minkowski dot product in light-cone coordinates is $k \cdot p = \frac{1}{2}(k^{+} p^{-}+k^{-} p^{+} )- \boldsymbol{k_{\perp}}  \cdot \boldsymbol{p_{\perp}}$.} $P^+$  of the parent parton, which has radiated a parton emerging with the light-cone momentum $k^+$. In~the $\boldsymbol{x_{\perp}}$ plane transverse to the direction of the fast-moving primary parton, the~partonic cascade initiated by the primary parton can be visualized as a Brownian motion-like trajectory developing from $x=1$ toward $x\rightarrow 0$. The~corresponding Gribov diffusion process is controlled by the so-called evolution parameter $Y=\ln (1/x)$, leading to a~difference in rapidity between the primary and radiated partons, with~a coefficient being the diffusion constant proportional to $\alpha_{\rm S}$. Its evolution in the $Y$ variable is described by the Balitsky-Fadin-Kuraev-Lipatov (BFKL) equation that is complementary to the DGLAP evolution realized in $\ln Q^2$~(see standard textbooks, e.g.,~Refs.~\cite{Ioffe:2010zz, Campbell:2017hsr} and references therein).

At fixed $Q$, the~radiated partons (mostly, soft gluons with $x\ll 1$) are typically of the same size. When a parton-parton interaction cross-section $\sim$$\alpha_{\rm S}/Q^2$ multiplied by $x G_A(x,Q^2)$ --- the probability to find at fixed $Q$ a parton carrying a fraction $x$ of the parent parton momentum --- becomes comparable to the geometrical cross section $\pi R_A^2$ of the object $A$ occupied by the gluons, the~partons ``overlap''. The~repulsive interactions among the gluons ensure that their occupation number $f_g$ (the number of gluons with a given $x$ multiplied by the area each gluon fills up divided by the transverse size of the object) saturates at $f_g \sim 1/\alpha_{\rm S}$. Note that this is a very generic behavior --- the same density scaling as the inverse interaction strength $\alpha^{-1}$ is characteristic of a number of condensation phenomena such as the Higgs condensate, see, e.g.,~Ref.~\cite{Cea:2020lea}, or~superconductivity~\cite{McLerran:2008es}. 

The phenomenon of saturation~\cite{Gribov:1984tu} is thus important for gluons with transverse momenta $k_{\perp} \leq Q_s$  \cite{Kharzeev:2002np,Berges:2020fwq}, where 
\begin{equation}
Q_s^2(x) = \frac{\alpha_{\rm S}(Q_s)}{2(N_c^2-1)} \frac{x G_A(x,Q_s^2)}{\pi R_A^2} \sim \frac{1}{x^{\lambda}}
\label{eq:Q_s}
\end{equation} 
is the $x$-dependent saturation scale representing a fixed point of the parton density evolution in $x$ or, equivalently, the~emergent ``close packing'' scale~\cite{Berges:2020fwq} --- see the right panel of Figure~\ref{fig:alphas} where the saturation line $\ln Q_s^2(Y) = \lambda Y$,  $Y=\ln x$ is also displayed. Such gluons form a highly coherent configuration called Color Glass Condensate (CGC) \cite{McLerran:1993ni, Gelis:2010nm, Berges:2020fwq}, or~glasma~\cite{Kovner:1995ja}, which due to the high occupation number $f_g$ has properties of QCD in the classical regime~\cite{Kharzeev:2002np}.

At high temperatures, one usually expects that quantum effects become less important. To~show that, for the CGC, we follow Ref.~\cite{Kharzeev:2002np} and write the gluonic part of the QCD action in terms of the gauge field potential ${\mathcal A}_\mu^a$ and field strength ${\mathcal F}_{\mu\nu}^a$ which are obtained by rescaling of their equivalents $A_\mu^a$ and $F_{\mu\nu}^a$ used in a more traditional approach when the coupling constant $g_s$ multiplies the interaction terms in the Lagrangian,
\begin{eqnarray}
\label{eq:rescaleNotation}
A_\mu^a \rightarrow {\mathcal A}_\mu^a  \equiv g_s A_\mu^a \,,
  \quad
F_{\mu\nu}^a \rightarrow g_s F_{\mu\nu}^a \equiv {\mathcal F}_{\mu\nu}^a
  =
  \partial_\mu {\mathcal A}_\nu^a - \partial_\nu {\mathcal A}_\mu^a + f^{abc} {\mathcal A}_\mu^b {\mathcal A}_\nu^c \,,
  \label{grescal1}
 \\
S_g=-\frac{1}{4}\int F_{\mu\nu}^a F^{\mu\nu,a} d^4x = -\frac{1}{4g_s^2}\int {\mathcal F^{\mu\nu,a}} {\mathcal F_{\mu\nu}^a} d^4x \,,
\label{eq:grescal2}    
\end{eqnarray}
where $f^{abc}$ with $(a,b,c)\in \{ 1,\ldots,8 \}$ are the \mbox{SU(3)} group structure constants. For~a classical configuration of gluon fields, by~definition, ${\mathcal F_{\mu\nu}^a}$ does not depend on the coupling, and~the action is large, $S_g \gg \hbar$. The~number of quanta in such a configuration is then
\begin{equation}
f_g \sim \frac{S_g}{\hbar}  \sim \frac{1}{\hbar g_s^2}\rho_4 V_4 \,,
\label{eq:f_g}
\end{equation}
where we rewrote Equation~\eqref{eq:grescal2} as a product of four-dimensional gluon condensate density $\rho_4\sim \langle {\mathcal F^{\mu\nu,a}} {\mathcal F_{\mu\nu}^a} \rangle$ and spacetime volume $V_4$. The~number of quanta $f_g$ in such a configuration depends only on the product of the Planck constant $\hbar$ and the strong coupling squared $g_s^2=4\pi\alpha_{\rm S}$. The~classical limit $\hbar \rightarrow 0$ is indistinguishable from the weak coupling limit $g_s^2 \rightarrow 0$ \cite{Kharzeev:2002np}. Thus, the~weak coupling limit of small $\alpha_{\rm S}$ corresponds to the semi–classical~regime. 

An equivalent argument employs the fact that the path integral formulation of the quantum theory in Minkowski space sums over all field configurations weighted with $\exp(-iS_g/\hbar)$. Since $g_s^2$ appears in the exponential in the same place as $\hbar$, cf.~Equation~\eqref{eq:grescal2}, this already suggests that for $g_s^2 \rightarrow 0$, the path integral is dominated by the classical configurations. Such configurations are believed to describe the matter inside incident nuclei during the initial stage of relativistic HICs at RHIC and the LHC~\cite{McLerran:1993ni, Gelis:2010nm}. 

Although in its original formulation, the QCD saturation is used for partons with fixed $Q$ in case of the macroscopic bodies, it is more relevant to consider the partons at a fixed temperature $T=Q/2\pi$, avoiding at the same time the quantum entanglement problem~\cite{Tu:2019ouv}, which is inevitably present in the description of microscopic objects. In~this generalized setting, an object $A$ filled with gluons may represent not only a fast-moving proton or nucleus but~also the interior of the expanding early universe. Moreover, as~follows from Equation~\eqref{eq:Q_s}, the saturation phenomenon is not necessary related to the growth of the gluon density at small $x$. For~a big fast-moving domain of space filled with deconfined quarks and gluons, with~radius $R\gg 1$ fm, and~hence also for the fast-expanding early universe itself characterized by the Hubble horizon $L_H \ggg 1$ fm, the~saturation limit can be reached even at $x \simeq 1$ \cite{Ioffe:2010zz}\endnote{In the early universe with $\epsilon\sim T^4 \sim a^{-4}$, the saturation scale  $Q_s^2(x)\sim \alpha_{\rm S}(T)R_A(T)\sim [T\ln T]^{-1}$ was extremely small.}. In~the extreme case, when the gluonic part of QCD matter completely decouples from the QCD fermionic fields and forms the vacuum condensate, the~first term in Equation~\eqref{eq:g_qcd} can be neglected, and we obtain $g_{\rm eff}^{\rm QCD}/g_{\rm eff}^{\rm EW}\simeq 8/3$, making the CGC a prevailing form of matter during both the QCD and EW~epochs.

Thus, in~the periods of cosmological evolution when $T \gg \Lambda_{\rm QCD}$, including a very hot QCD era, EW era and beyond, it is perfectly conceivable that the universe was dominated by the fully saturated gluonic matter with occupation number $f_g \sim \alpha_{\rm S}^{-1}$. If~during its subsequent cooling, the universe followed a trajectory in the $\left[\ln(1/x), \ln Q^2\right]$ plane staying still above the $\ln Q^2_s=\lambda Y$ line, see the right panel of Figure~\ref{fig:alphas}, the~CGC phase would be a prevailing form of matter down to the temperatures $T\approx (2-5)\times \Lambda_{\rm QCD}$. For~lower temperatures, the glasma is expected to fragment into a strongly interacting~QGP.

The issue of emergence of classical behavior in the cosmological history has drawn recently a great deal of attention because of its conceptual as well as practical importance, see e.g.,~Ref.~\cite{Ashtekar:2020mdv} and references therein. Although~the origin of the observed anisotropies in the cosmic microwave background (CMB) radiation is traced back to vacuum fluctuations of quantum fields in the very early universe~\cite{Mukhanov:2005sc, Weinberg:2008zzc, Vilenkin:1984wp}, there is a general expectation that the main characteristics of the universe can be described in classical terms even in its early history~\cite{Ashtekar:2020mdv, Martin:2015qta, Green:2020whw}. This is consistent with the fact that the initial conditions of the Hot Big Bang were determined by cosmic inflation driven by the so-called inflaton field~\cite{Liddle:1998ew, Mukhanov:2005sc, Mazumdar:2018dfl, Linde:2014nna}. Apart from the quantum fluctuations of this field, its very emergence may well be a quantum phenomenon, e.g.,~the axion condensation (for a more thorough discussion, see below).

Let us note that the word ``glass'' appearing in the acronym CGC is used in condense matter physics to describe a non-equilibrium, disordered state of matter acting like solids on short time scales but liquids on long time scales~\cite{Sethna2008-book}. In~the glasma case, there are two scales present --- the light cone time $\tau_{\rm wee}$ of low $x$ (or wee) partons and the light cone time $\tau_{\rm valence}$ of primary (or valence) partons. For~partons of transverse momentum $k_{\perp}$, we obtain 
\begin{equation}
 \tau_{\rm wee}=\frac{1}{k^-} = \frac{2k^+}{k^2_{\perp}} = \frac{2xP^+}{k^2_{\perp}} \ll 
 \frac{2P^+}{k^2_{\perp}}\approx \tau_{\rm valence} \,,
 \label{eq:tau}
\end{equation}
suggesting that the valence parton modes are static over the times scales over which the wee modes are probed~\cite{Berges:2020fwq}. It is quite tempting to identify $\tau_{\rm valence}$ with the quantum break-time discussed in Refs.~\cite{Dvali:2017eba,Berezhiani:2020pbv} defined as the time-scale after which true quantum evolution of the parton densities departs from the classical mean field~evolution.

Glasses are formed when liquids are cooled too fast to form the crystalline equilibrium state. The~fast cooling leads to an enormous number of possible configurations $N_{\rm gl}(T)$ into which the glasses can freeze and consequently to their large entropy $S=\ln N_{\rm gl}(T)$, which does not vanish, even at $T=0$, see e.g.,~Refs.~\cite{PhysRevLett.61.570,Sethna2008-book}. In~case of the CGC, the fast cooling is expected to take place in the Grand Unified Theory (GUT) era (see Section \ref{Sec:Couplings}) when about one-third of all gauge bosons are gluons. By~the end of that period, the excess of effective entropy DoF $h_{\rm eff}$ is almost completely absorbed by the saturated gluonic~matter. 

The gluon condensation into (many domains of) the saturated phase was also facilitated by the fact that the wee partons ``see'' the color charge of other gluons over very large distances given by their transverse wavelength $\lambda_{\rm wee}\sim 1/k^+=1/(xP^+)$. Since the glasma domains were formed in separate and completely different gluonic configurations, the~saturated gluonic matter occupying the early universe had a substantial excess in the entropy DoF, $h_{\rm eff}^{\rm QCD}$, over~the effective number of DoF in energy, $g_{\rm eff}^{\rm QCD}$. Consequently, the~value of the generalized EoS parameter $w$ was higher than that of the ideal massless gas; see Equation~\eqref{eq:wT}.

One possibility of how the EoS of a non-equilibrium matter comprising weakly interacting gluons can be approximated by the ideal massless gas of quasi-particles with $w(T)>1/3$ follows from Equation~\eqref{eq:troj5} discussed later in Section \ref{Ssec:simple}. The~glasma  with $w(T)\approx 1/D$ may be looked upon as either a two-dimensional sheet ($D=2$), a~one-dimensional string ($D=1$) or, more generally, a~fractal with the Hausdorff dimension $1\leq D < 3$. Recent investigations of the dynamics of expanding glasma show that the spatial asymmetry introduced by the initial geometry is effectively transmitted to the azimuthal distribution of the gluon momentum field, even at very early times~\cite{Carrington:2020ssh,Carrington:2021qvi}.

\subsection{The Running Couplings of the Standard Model and Their~Unification}
\label{Sec:Couplings}

\begin{figure}[H]
\includegraphics[trim= 0 0 0 0, clip, width=.45\textwidth]{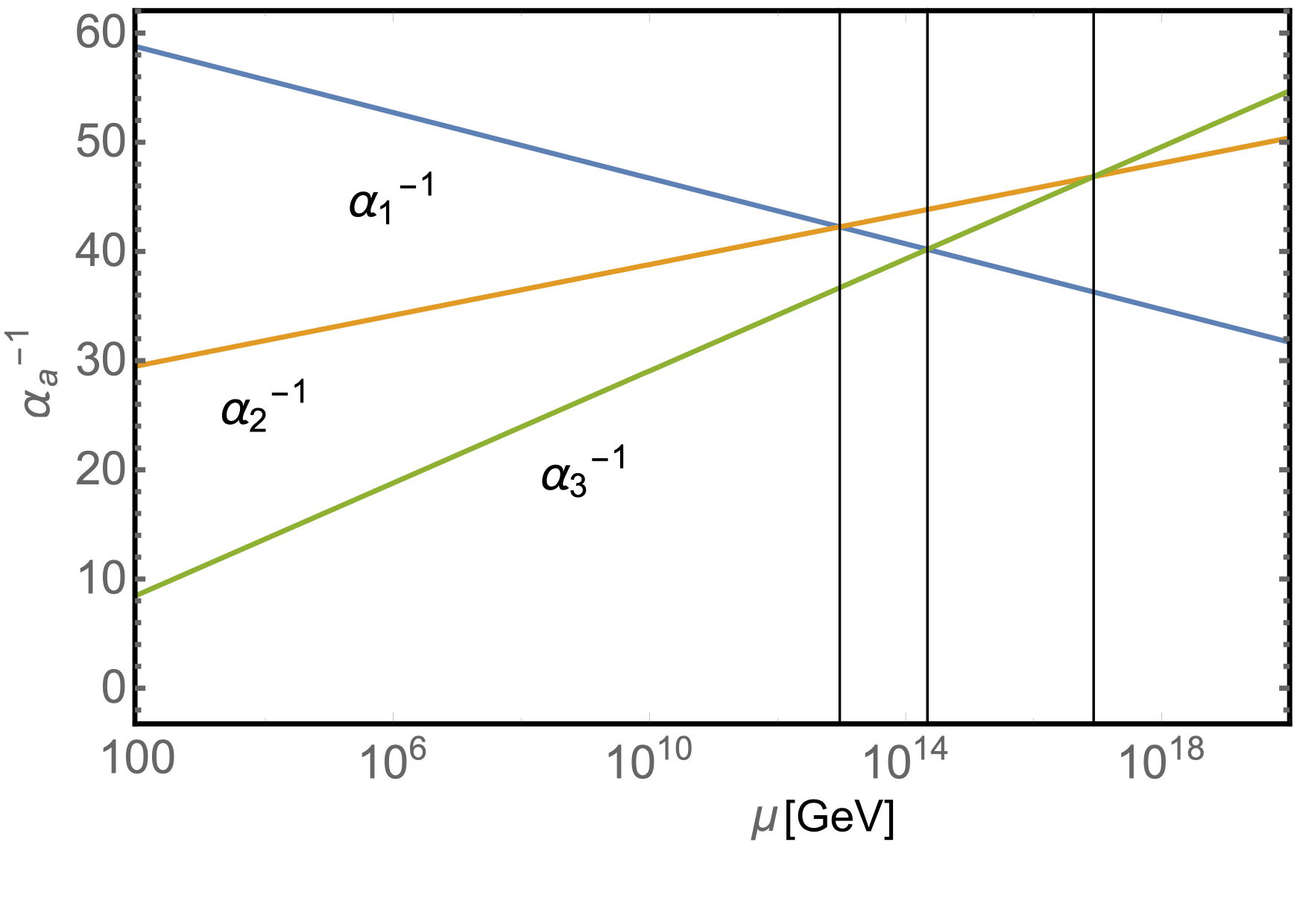}
\includegraphics[trim= 0 0 0 0, clip, width=.45\textwidth]{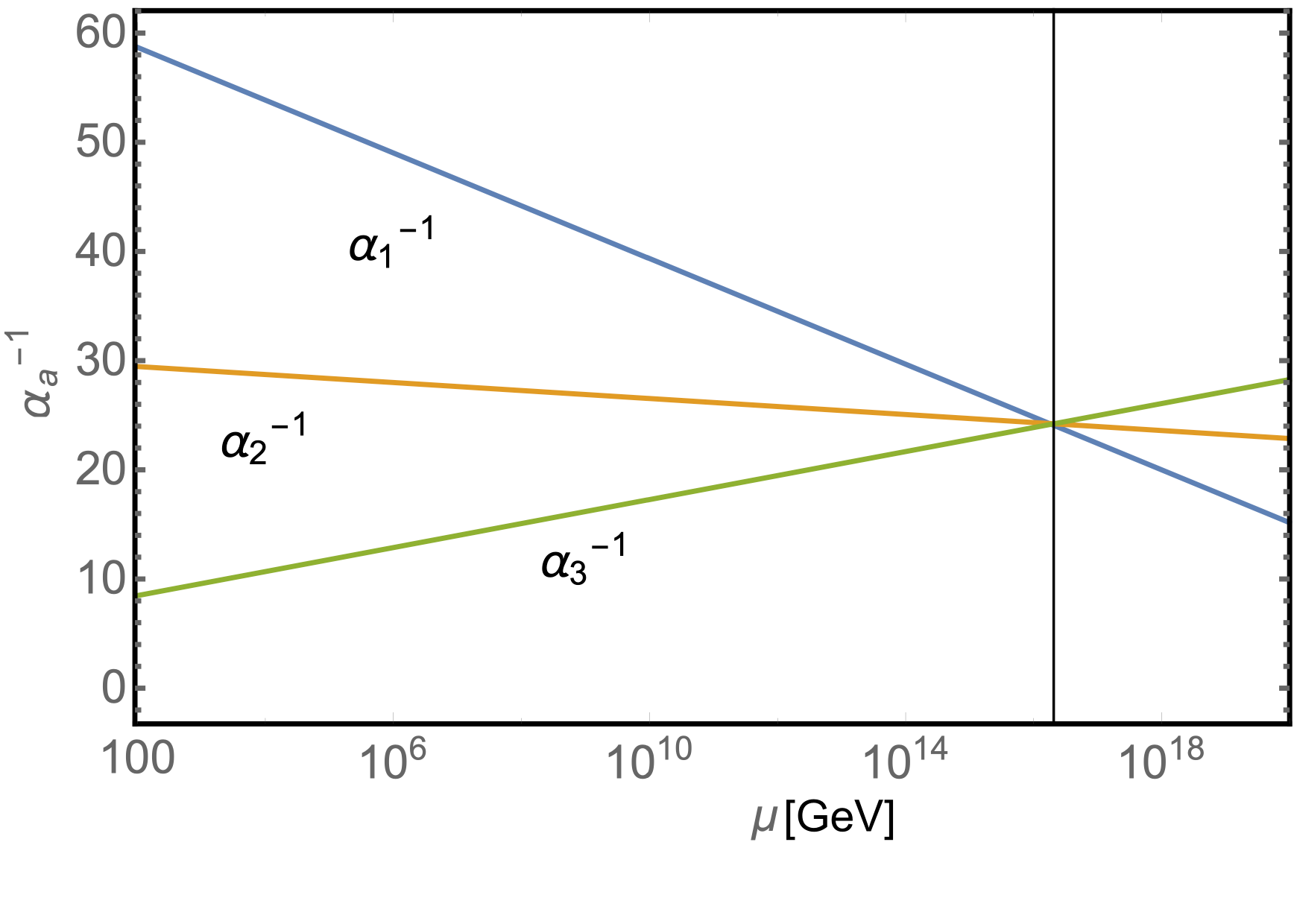}
\caption{(\textbf{Left}) RG flow of the inverse SM gauge couplings $\alpha_a^{-1}$ as functions of the renormalization scale parameter $\mu$. Index $a=1,2,3$ stands for QED ($a=1$), weak ($a=2$) and QCD ($a=3$) couplings. (\textbf{Right}) RG flow of the MSSM gauge couplings. Adapted from Ref.~\cite{Croon:2019kpe}.}
\label{fig:GUT}
\end{figure}

The importance of QCD interactions in the EW era of the universe evolution can be also seen by comparing the corresponding couplings --- the strong $\alpha_{\rm S}$, electromagnetic $\alpha_{\rm EM}$ and weak $\alpha_{\rm W}$ ones. This is illustrated on the left panel of Figure~\ref{fig:GUT} showing the RG flow in the scale $\mu=Q$ of the electromagnetic, weak and strong coupling parameters above $Q=100$ GeV. Note that due to the fact that the gauge group of SM interactions \mbox{SU(3)$_{\rm C}$} $\times$ \mbox{SU(2)$_{\rm L}$} $\times$ \mbox{U(1)$_{\rm Y}$} is not a simple Lie group, the~theory has not one but three coupling parameters, which are often denoted as $\alpha_1 \equiv \alpha_{\rm EM}$, $\alpha_2 \equiv \alpha_{\rm W}$ and $\alpha_3 \equiv \alpha_{\rm S}$ \cite{Weinberg:1996kr}.

To see how the values of the coupling parameters influence the state of early universe, let us compare the collision time among its constituents $t_c \sim 1/(\sigma n v)$ ($\sigma$ is the effective cross-section, $n$ is the particle number density and $v$ is their relative velocity) with the characteristic time-scale of the universe expansion $t_H \sim 1/H$ \cite{Mukhanov:2005sc}. Let us first restrict ourselves to the temperatures $T \gtrsim T_{\rm EW}$ when all particles of the SM are ultra-relativistic and the gauge bosons are massless. Then, the~cross-sections for strong and EW interactions have a similar energy dependence $\sigma \sim \alpha^2/T^2$, where $\alpha \simeq 10^{-1}-10^{-2}$ are the corresponding dimensionless running couplings $\alpha_{1,2,3}$ varying logarithmically with $T$; see Figure~\ref{fig:GUT}. Taking into account Equations~(\ref{eq:Fried1}) and (\ref{eq:sanch2}) for $n \sim T^3$, we find 
\begin{equation}
t_c \sim \frac{1}{\sigma n v} \sim \frac{1}{\alpha^2 T} \ll  t_H \sim \frac{1}{H} 
\sim \frac{1}{\sqrt{\epsilon}} \sim \frac{1}{T^2} \,.
\end{equation}

Thus, for~the temperatures $10^{15}-10^{17}$ GeV $\gtrsim T \gtrsim T_{\rm EW}$, the local equilibrium in the fluid is established before expansion of the universe becomes~relevant. 

At the lower temperatures down to $T_c^{\rm QCD}\approx 160$ MeV, the strong interactions prevail over the EW ones, and the local equilibrium is controlled by the interactions among the quarks and gluons in the strongly interacting QGP liquid. Even though the number density of the medium may be smaller\endnote{Let us recall that at the temperatures $T \gg T_c^{\rm QCD}$, most of the gluons are forming the condensate and are thus in the equilibrium but do not participate in two-particle scatterings.}, the big effective cross-section among the particles forming the medium guarantees that the condition for local equilibrium $t_c \ll t_H$ is satisfied. Thus, over~that whole range of the temperatures, the early universe is in the local equilibrium. It develops along the maximal possible entropy path, making it amenable to the hydrodynamical~description.

In spite of its success, the~SM cannot be the ultimate theory of particle physics. Such long-standing problems as the absence of a suitable DM candidate, no explanation of the observed baryon asymmetry in the universe, as~well as various hierarchy problems in the underlined mass spectra (such as the flavor problem, the~neutrino mass problem and the Higgs hierarchy problem) call for various bottom-up extensions of the SM framework as~well as continuous attempts to derive the SM structure from a top-down~perspective.

As was earlier discussed, e.g.,~in Ref.~\cite{Degrassi:2012ry}, the~Higgs boson quartic coupling in the SM turns negative at scales $\gtrsim$$10^{10}$ GeV, rendering the vacuum state of the theory unstable at high energies. The~current theoretical developments and experimental measurements suggest that the metastability of the Higgs vacuum is favored. This means a vacuum decay may occur with possibly catastrophic consequences for cosmology, since there are many catalysts that could trigger such a decay in the early universe. For~a comprehensive review on cosmological implications of the Higgs vacuum metastability, see, e.g.,~Ref.~\cite{Markkanen:2018pdo}.

Incidentally, almost immediately after the discovery of the asymptotic freedom, it was suggested~\cite{Georgi:1974sy} that at very high energies, the three gauge interactions of the SM are merged into a single force. The~model, a~first example of the Grand Unified Theory (GUT), is based on the smallest simple Lie group which contains the SM gauge groups \mbox{SU(5)} $\supset$ \mbox{SU$_c(3)$} $\times$ \mbox{SU(2)$_{\rm L}$} $\times$ \mbox{U(1)$_{\rm Y}$}. Among~its 24 gauge bosons, there are in addition to eight gluons of QCD and four EW gauge bosons $W^\pm,Z$ and $\gamma$ also 12 new ones called $X$ and $Y$. Their emission or absorption makes it possible to transform a lepton into a quark or~vice~versa. Hence, the~\mbox{SU(5)} GUT does not conserve baryon and lepton numbers separately, making it the first theory providing an explicit mechanism for the proton decay $p\rightarrow e^{+}\pi^{0}$, with~the half-time $\tau_p \simeq M^4_X/m^5_p \simeq 10^{30}-10^{31}$ years, where $M_X$ is the mass of \mbox{SU(5)} gauge boson at the scale of Grand Unification and $m_p$ is the mass of the proton. Although~it was later found to disagree with experimental lower limit of $\tau_p \geq 8.2\times 10^{33}$ years~\cite{Nishino:2009aa} \mbox{SU(5)}, unification is still considered an important example and a reference point of GUT~model-building.

The basic property of the \mbox{SU(5)} theory and its later GUT successors~\cite{Langacker:1980js, Nath:2006ut, Altarelli:2013aqa, Morais:2020odg,Croon:2019kpe} is that by virtue of the unification into a single (simple Lie) gauge group at very high energies, strict unification of the SM gauge couplings must take place. This is hinted at but not really achieved in the SM; see the left panel of Figure~\ref{fig:GUT}. First, $\alpha_1$ and $\alpha_2$ cross each other at $\mu\sim 10^{13}$ GeV; then, $\alpha_1$ crosses $\alpha_3$ at $\mu\sim 10^{14}$ GeV, and~finally, $\alpha_2$ and $\alpha_3$ cross at $\mu\sim 10^{17}$ GeV, providing a hint of unification close to the Planck scale. Thus, as~ultraviolet (UV) completions of the SM describing the physics at very high energies, GUTs are sometimes connected to theories of gravity such as string theory.

Let us now try to speculate on how the Grand Unification might have influenced the dynamics of the early universe at the temperatures when gluons decouple from quarks, forming a condensate. In~\mbox{SU(5)} GUT, with~increasing temperatures $T\gg T_c^{\rm EW}$, the gluon exchange between quarks (antiquarks) becomes overshadowed by the exchange of EW massless gauge bosons $W^{\pm},W^0,B^0$ to be finally superseded by the GUT super-heavy $X$ and $Y$ gauge bosons. The~latter also facilitate the conversion of quarks into leptons and vice~versa. Since the transformations occur in the thermal and chemical equilibrium, the number of quarks and leptons remains constant on~average.
 
One of the fundamental issues with the GUT models, which remains a challenge today, is the large hierarchy between the mass scale of Grand Unification and the EW scale. The~latter is a source of large loop corrections to the Higgs mass. The~problem is usually solved by extending the GUT with supersymmetry, the~hypothetical symmetry between fermions and bosons (for a recent review on the concepts of supersymmetric GUTs, see e.g.,~Ref.~\cite{Croon:2019kpe}), which is broken at lower energies. Some of its minimal realizations, such as the Minimal Supersymmetric Standard Model (MSSM) \cite{Dimopoulos:1981yj}, predict the unification of all three gauge couplings at the same scale; see the right panel of Figure~\ref{fig:GUT}.

Notwithstanding the drawbacks of GUTs, their cosmological signatures look quite promising. With~the critical temperature $T_c^{\rm GUT}$ of the GUT phase transition approximately of the same order of magnitude as the particle masses at that temperature, the phase transitions that take place in GUTs at $T \gtrsim 10^{14}$ GeV, as~a rule, prove to be FOPTs~\cite{Linde:1990flp}. Such transitions proceed via bubble nucleation~\cite{Gangui:2001wc}. While an isolated spherical bubble may produce GWs through sound waves in the plasma and magneto-hydrodynamics turbulence effects (see for example Refs.~\cite{Caprini:2001nb,Caprini:2009yp,Figueroa:2012kw,Hindmarsh:2016lnk} and references therein), the~process of bubble collision contributes to the GWs spectrum in the quadrupole approximation~\cite{Kamionkowski:1993fg, Mazumdar:2018dfl}. This contrasts with the SM phase transitions where the crossovers do not lead to a strong enhancement over the primordial GW spectrum. Moreover, the~FOPTs also generate a primordial magnetic field during the turbulence phase of the plasma and bubble collision~\cite{Durrer:2013pga,Mazumdar:2018dfl}, and in some instances, they may generate topological defects such as domain walls and strings~\cite{Vilenkin:2000jqa, Gangui:2001wc}.

Let us add that an FOPT in the EW sector though precluded in the SM is possible in many of its scalar sector extensions~\cite{Espinosa:1993bs}. In~the most exotic scenario, a very peculiar history of the universe may occur: a first-order QCD phase transition (with six massless quarks) triggers an EW FOPT, which is eventually followed by a low-scale reheating of the universe where hadrons (likely) deconfine again, before~a final, conventional crossover QCD transition to the current vacuum~\cite{Iso:2017uuu}.

\subsection{Axions} 
\label{Sec:Axions}

A promising avenue connecting the strong interactions with physics beyond the SM having at the same time far-reaching consequences in cosmology is provided by hypothetical ultra-light particles --- the axions~\cite{Weinberg:1996kr, Sikivie:2006ni, Peccei:2006as, Marsh:2015xka, DiLuzio:2020wdo}. The~QCD, unlike the EW interaction, is symmetric under time reversal and hence under a combined  charge conjugation \textsf{C} and parity \textsf{P} operation \textsf{CP}. In~principle, one can add to the QCD Lagrangian the term:
\begin{equation}
\mathcal{L}_Q=\theta \frac{g^2_s}{32\pi^2}F^a_{\mu\nu}\tilde{F}_a^{\mu\nu}\, ,   
\label{LQ}
\end{equation}
where $F^a_{\mu\nu}$ is the gluon field strength tensor (see Equation~\eqref{eq:rescaleNotation}), and $\tilde{F}_a^{\mu\nu}=\frac{1}{2}\epsilon^{\mu\nu\alpha\beta}F^a_{\alpha\beta}$ is its dual. For~a non-zero value of the parameter $\theta$, called the vacuum angle~\cite{tHooft:1976snw}, the~strong coupling permits violation of the \textsf{CP} symmetry. However, since $\mathcal{L}_Q$ can be written as a total derivative of $K^{\mu}$, the~Chern–Simons current, $\mathcal{L}_Q=\partial_{\mu}K^{\mu}$, the~new term does not produce any effects in perturbation theory and is therefore usually neglected. Nevertheless, classical configurations, topological in nature, do exist, one example being the instantons~\cite{Schafer:1996wv}, for~which this term cannot be ignored. For~instance, in~the semi-classical dilute instanton gas approximation (DIGA) \cite{Callan:1977gz}, the~QCD vacuum energy density $\epsilon_0$ depends on $\theta$ as 
\begin{equation}
\epsilon_0(\theta)=-2C e^{-S_{\rm inst}}\cos\theta \,, ~~ 
S_{\rm inst}=\frac{8\pi^2}{g^2_s} \,.
\end{equation}

Here, $C$ is a positive constant and $S_{\rm inst}$ is the QCD instanton action~\cite{Marsh:2015xka,DiLuzio:2020wdo}, \linebreak{cf. Equation~\eqref{eq:grescal2}}. Moreover, since $\mathcal{L}_Q$ in Equation~\eqref{LQ} preserves the charge conjugation \textsf{C}, it contributes directly to the neutron electric dipole moment $d_n\approx e m_q/m^2_n\theta$, where $e$ is the proton charge, $m_q$ denotes the mass of $u,d$ quarks, and~$m_n$ is the neutron mass. Current measurements~\cite{Abel:2020gbr} provide an upper bound on the \textsf{CP}-violation parameter, $|\theta| \lesssim 10^{-10}$.

Within the SM, the smallness of the $\theta$ parameter becomes a true fine-tuning problem. Since $\theta$ could acquire an $\mathcal{O}(1)$ contribution from the observed \textsf{CP}-violation in the EW sector (via the common quark mass phase, arg det$(M_q)$, where $M_q$ is the quark mass matrix), its not obvious why it becomes cancelled to a high precision by the (unrelated) gluon term~\cite{Marsh:2015xka}. To~solve this problem, the SM is augmented with an extra pseudo-scalar particle called axion $A$, whose only non-derivative coupling is to the \textsf{CP}-violating topological gluon density $F^a_{\mu\nu}\tilde{F}_a^{\mu\nu}$ that is suppressed by a large scale $f_A$. With~\mbox{$\theta \rightarrow \theta + \phi(x)/f_A$}, where $\phi$ is the angular DoF of spin-zero complex field, 
\begin{equation}
\varphi=|\varphi|e^{i\theta}= |\varphi|e^{i\phi/f_A} \,,
\label{eq:varphi}
\end{equation}
the minimum of the vacuum energy occurs when the coefficient $\theta + \phi/f_A$ in front of $F^a_{\mu\nu}\tilde{F}_a^{\mu\nu}$ vanishes. 

It is worth noting that interactions of the scalar field with ordinary matter is controlled by the factor $\partial_{\mu}\varphi/f_A$. Thus, even at its originally suggested value of $f_A \sim 250$ GeV at the EW breaking scale~\cite{Peccei:1977hh}, the~axions interact so weakly that they emerge without attenuation from reactor cores or stellar interiors. Present astrophysical constraints push $f_A$ to substantially higher values, which are somewhere between a few $10^8$ GeV and a few $10^{17}$ GeV.

In cosmology, since their introduction, the axion-like particles were considered to be potentially important candidates if not for all than at least for the main component of the DM --- a form of matter accounting for about one-quarter of its total energy density~\cite{Bertone:2016nfn}. In~order to fulfill their mission, the~axions must contribute a non-negligible amount to the energy density of the universe and should have not been in thermal equilibrium with the cosmological plasma at any time in the history of the universe. This, together with the smallness of axion mass, implies their large occupation numbers~\cite{Marsh:2015xka} --- the situation already encountered when we have discussed the properties of saturated gluon matter, cf.~Equation~\eqref{eq:f_g}. This implies that the axions over the whole history of the universe can be modeled by solving the classical field equations of a scalar condensate~\cite{Dvali:2017ruz}.

Starting from the Peccei--Quinn (PQ) scalar field $\varphi$ introduced in Equation~\eqref{eq:varphi}, the~Lagrangian invariant under the global ${\rm U}(1)_{\rm PQ}$ transformation reads~\cite{Kawasaki:2013ae}:
\begin{equation}
{\cal L} = \frac{1}{2} |\partial_\mu \varphi|^2 - V_{\rm eff}(\varphi,T) \,,~~
V_{\rm eff}(\varphi,T) = \frac{\lambda}{4} (|\varphi|^2 - f_A^2)^2 +
\frac{\lambda}{6}T^2|\varphi|^2\,.
\label{eq:axion-lag}
\end{equation}

\begin{figure}[H]
\begin{center} \hspace*{-.4cm}
\includegraphics[width=.45\textwidth]{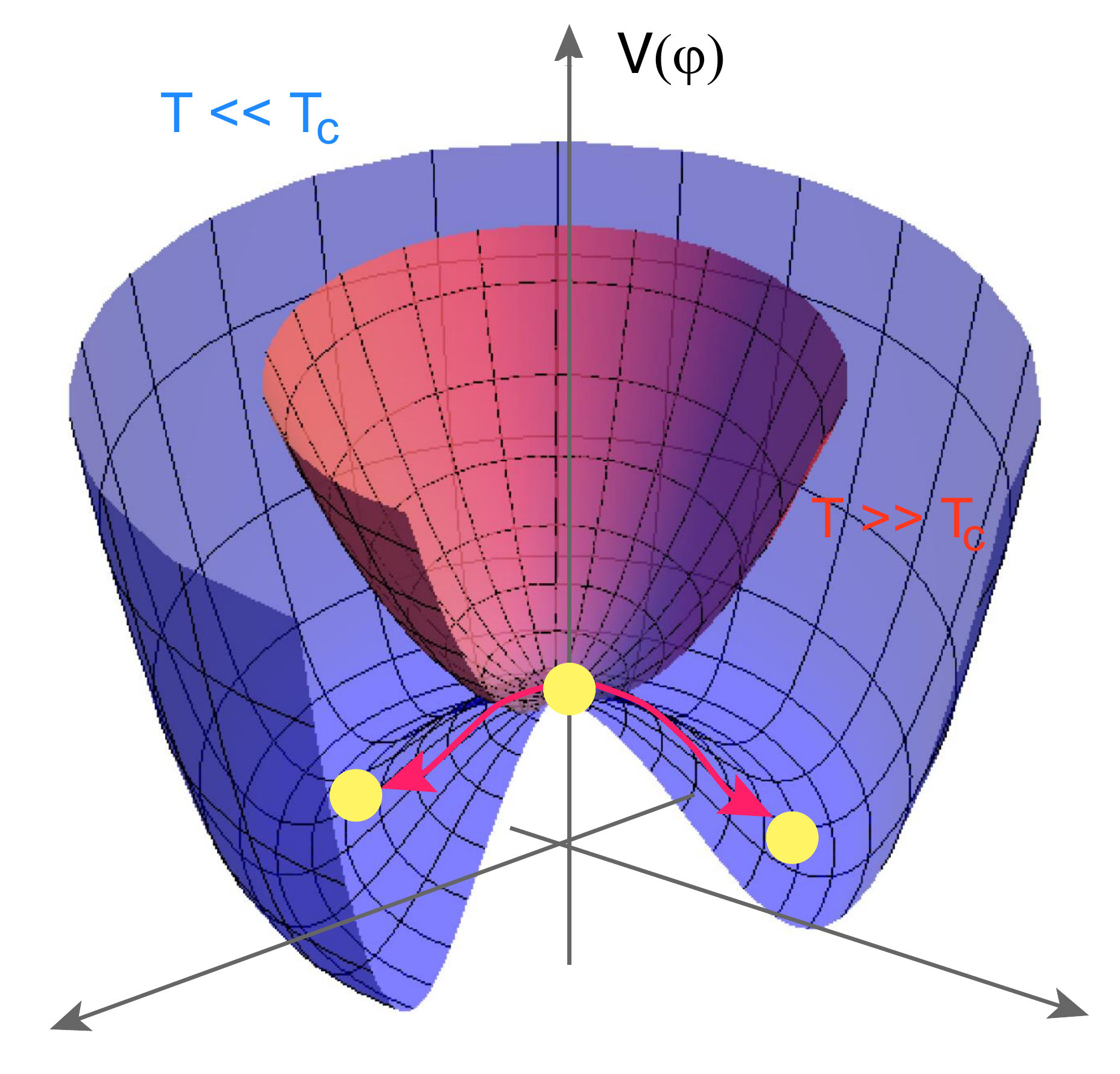}  
\includegraphics[width=.55\textwidth]{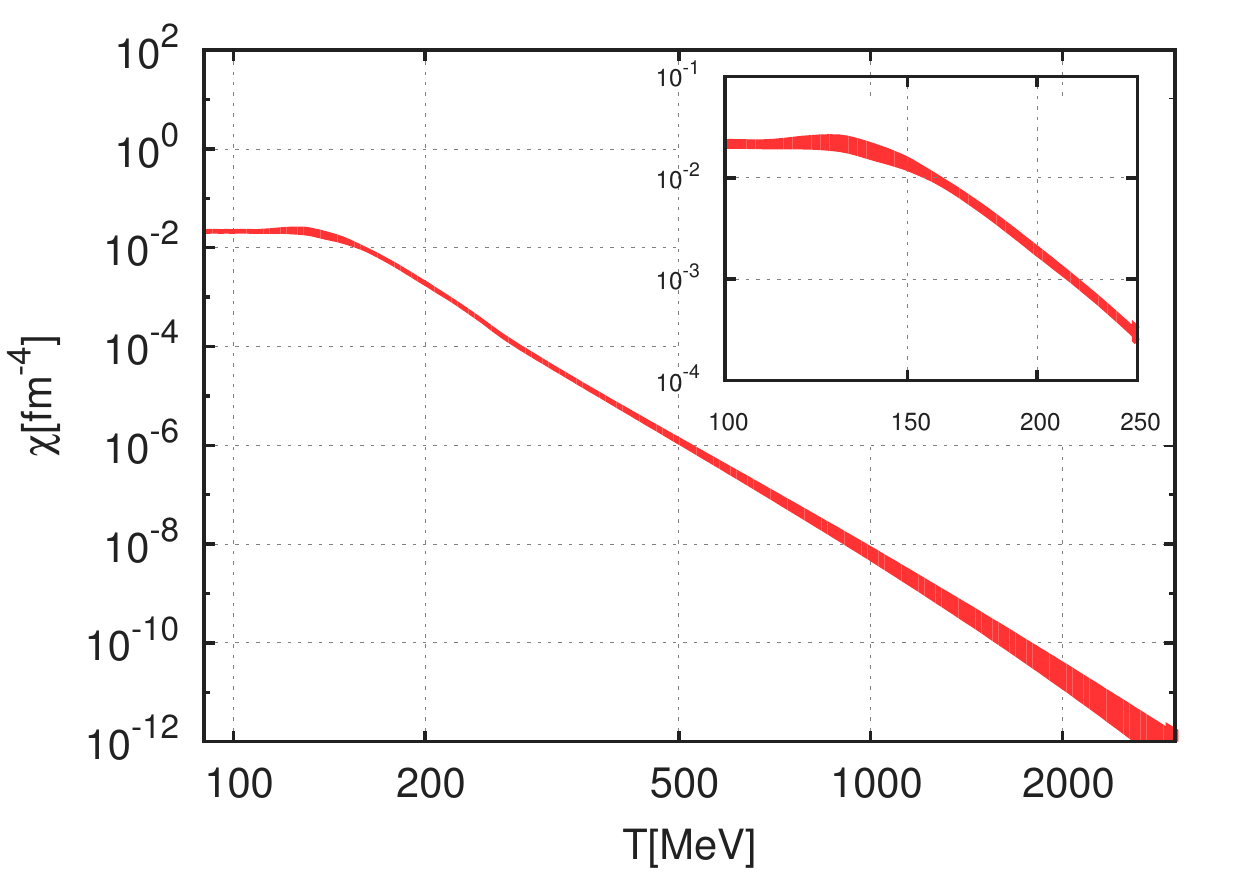}
\caption{(\textbf{Left}) Potential of the PQ scalar $V(\varphi)$ at different temperatures $T\gg T_c$ (pink) and $T\ll T_c$ (violet). The~yellow circles show the positions of the minimum. Adapted from Ref.~\cite{Kawasaki:2013ae}. (\textbf{Right}) Continuum limit of $\chi_{\rm top}(T)$ from LQCD. The~inserted sub-figure shows the behavior around the QCD phase transition temperature. Adapted from Ref.~\cite{Borsanyi:2016ksw}.}
\label{fig:axionpot}
\end{center}
\end{figure}

Focusing on the early universe, the evolution of the field  $\varphi$ passes the following milestones. At~high temperatures $T\gg T_c=\sqrt{3}f_A$, the~effective potential $V_{\rm eff}(\varphi,T)$ depicted on the left panel of Figure~\ref{fig:axionpot} has the ${\rm U}(1)_{\rm PQ}$ symmetric minimum at $\varphi=0$. With~increasing time $t$, the~universe cools, and the vacuum with $\varphi=0$ becomes unstable. Due to the misalignment mechanism, the field starts to roll down from $\varphi=0$, and the potential becomes tilted. At~$T\lesssim f_A$, the PQ symmetry is spontaneously broken --- the field acquires the vacuum expectation value $\langle \varphi \rangle = f_A$. Then, the axion --- the Nambu-Goldstone boson of the spontaneously broken ${\rm U}(1)_{\rm PQ}$ symmetry --- becomes a massless angular DoF at the minimum of the potential. The~detailed results depend on whether the PQ phase transition occurs before or after inflation. While in the former case, only one $\theta_0$ angle contributes (all other values are inflated away), in~a post-inflationary scenario, the initial value of the angle $\theta$ takes all values in the interval $\langle-\pi, \pi\rangle$. Eventually, when the universe cools to  the temperatures of a few GeV, the axion obtains a mass through the QCD non-perturbative instanton effect known as the axial anomaly~\cite{Adler:1969gk, Bell:1969ts, Weinberg:1996kr}.

At the leading order in $f_A^{-1}$, the axion mass $m_{A}(T)$ at some temperature $T$ can be extracted from the QCD-generating functional $\mathcal{Z}(\theta)$ in the presence of a theta term~\cite{DiVecchia:1980yfw},
\begin{equation}
m_{A}^2(T)=\frac{\delta^2}{\delta\phi^2}\ln \mathcal{Z}\left(\frac{\phi}{f_A} \right)\Big|_{\phi=0}=\frac{1}{f_A^2}\frac{d^2}{d\theta^2}\ln \mathcal{Z}(\theta)\Big|_{\theta=0}=\frac{\chi_{\rm top}(T)}{f_A^2}\, ,
\label{eq:axion-mass}
\end{equation}
where $\chi_{\rm top}(T)$ is the QCD topological susceptibility. This quantity is typically computed using the lattice methods developed by several groups, see e.g.,~Refs.~\cite{Alles:2000cg,Gattringer:2002mr,Bernard:2012fw,Bonati:2013tt,Berkowitz:2015aua,Kitano:2015fla,Borsanyi:2015cka,Bonati:2015vqz,Borsanyi:2016ksw,Taniguchi:2016tjc,Petreczky:2016vrs}, but~also in the framework of analytical approaches~\cite{DiVecchia:1980yfw,diCortona:2015ldu}. Its temperature dependence can be extracted either from the DIGA or from the LQCD calculations; see the right panel of Figure~\ref{fig:axionpot}. The~lattice simulations performed, for~instance, in~Ref.~\cite{Borsanyi:2016ksw} have revealed that for $T>T_{*}=150$ MeV, the susceptibility falls as $\chi_{\rm top}(T)\sim T^{-b}$ with $b=8.16$, extending thus the previous DIGA result $\chi_{\rm top}(T)=\chi(0) T^{-8}$ up to $T\approx 3$ GeV. At~the temperatures of $T=100-140$ MeV in the vicinity of the QCD chiral phase transition temperature $T_c$, see Section~\ref{Sec:SMPT}, $\chi_{\rm top}(T)$ flattens. Further analysis performed in Ref.~\cite{Borsanyi:2016ksw} exploiting the QCD EoS obtained therein has revealed that in the post-inflationary scenario, depending on the fraction of DM consisting of axions, $m_{A}=50-1500\,\upmu$eV. In~particular, for~the 50\% axion content of the DM, one obtains the axion mass scale of $m_{A}=50(4)\,\upmu$eV. Since at $T<T_c$, the chiral perturbation theory~\cite{Weinberg:1996kr} becomes applicable, it is worth making a comparison with the original formula for the axion mass,
\begin{equation}
m_{A}^2=\frac{m_u m_d}{(m_u+m_d)^2}\frac{m_{\pi}^2 f_{\pi}^2}{f_A^2}~~\Longrightarrow~~
    m_{A}\approx 5.7\left(\frac{10^{12}{\rm GeV}}{f_A} \right )\mu{\rm eV}\,,
    \label{eq:chiral-mass}
\end{equation}
where $f_{\pi}$ is the pion decay constant, and $m_d$ and $m_u$ are the down- and up-quark masses appearing in the QCD Lagrangian~\cite{Weinberg:1996kr, DiLuzio:2020wdo}. Let us note that if $m_{A} \gtrsim 20$ eV, the axions decay faster than the age of the~universe.

For temperatures above the chiral phase transition, the axion potential computed in DIGA reads~\cite{Callan:1977gz}:
\begin{equation}
   V(\phi,T) = \chi_{\rm top}(T)
   \left[ 1 - \cos\left(\frac{\phi}{f_A}\right) \right]\,.
   \label{eq:axion-pot}
\end{equation}

Expanding Equation~\eqref{eq:axion-pot} around $\phi/f_A=0$, we obtain $V(\phi)=m^2_A\phi^2/2$ at a finite $T$. Assuming the Friedmann-Lema\^itre-Robertson-Walker (FLRW) metric and classical axion field with this potential, the~corresponding unperturbed energy density $\bar{\epsilon}_A$ and pressure $\bar{p}_A$ due to the axion field read
\begin{equation}
    \bar{\epsilon}_A=\frac{1}{2}\dot{\phi}^2+\frac{1}{2}m^2_A\phi^2 ~~~ {\rm and} ~~~
    \bar{p}_A= \frac{1}{2}\dot{\phi}^2-\frac{1}{2}m^2_A\phi^2\, ,
    \label{eq:epspres}
\end{equation}
respectively. Substituting $\bar{\epsilon}_A$ and $\bar{p}_A$ from Equation~\eqref{eq:epspres} into the fluid Equation~(\ref{eq:Fried1}), we obtain
\begin{equation}
    \ddot{\phi}+3H\dot{\phi}+m^2_A=0\, .
    \label{eq:axeom}
\end{equation}

At early times $H(t)\gg m_A(t)$, we can neglect $m_A$ in Equation~\eqref{eq:axeom} to obtain the solution $\phi(t)=\phi_0$ --- the axion field is frozen at a constant value $\phi_0\sim f_A$.

With increasing time $t$, eventually, the~oscillating term proportional to $m_A^2(t)\equiv m_A^2(T(t))$ in the equation of motion of the axion field  (\ref{eq:axion-pot}) begins to contribute. At~the time $t_{\rm osc}$ defined implicitly as $m_A(t_{\rm osc}) \approx 3H(t_{\rm osc})$, the~universe is sufficiently large to host a sizeable fraction of one oscillation period --- the axion field starts to oscillate with an amplitude damped by the expansion rate. A~solution of Equation~(\ref{eq:axeom}) then reads~\cite{Weinberg:2008zzc},
\begin{equation}
\phi(t) = \phi_1 \left(\frac{a(t_1)}{a(t)}\right)^{3/2}
\cos\left(\int_0^t m_A(t)dt+\alpha\right)\,,
\end{equation}
where $t_1$ is the time at which $H(t_1)=m_{A}$, i.e.,~when the temperature drops below the QCD chiral phase transition temperature $T_c$, $\phi_1\sim f_A$ is the constant and $\alpha$ is the phase. In~particular, for~\mbox{$m_A(t)=m_{A}$} and a radiation-dominated universe, one obtains $\phi_1\approx 1.44\phi_0$ and $\alpha = -3\pi/8$ \cite{Weinberg:2008zzc}.

For the initial conditions at the onset of oscillations 
$\theta_i\equiv \theta(t_{\rm osc})$, $\dot{\theta}_i \equiv \dot{\theta}_i(t_{\rm osc})$, where $\theta_i$ is called the initial misalignment angle, we obtain from Equation~(\ref{eq:axeom})
\begin{equation}
    \theta_i=\theta_{\rm PQ}+\frac{\dot{\phi}_{\rm PQ}}{H_{\rm PQ}} ~~~~ {\rm and} ~~~~ \dot{\theta}_i =\dot{\theta}_{\rm PQ} \left(\frac{H(t_{\rm osc})}{H_{\rm PQ}} \right)^{3/2} \,,
\end{equation}
where $\theta(t_{\rm PQ}) \equiv \theta_{\rm PQ}$, $\dot{\phi}(t_{\rm PQ})\equiv \dot{\phi}_{\rm PQ}$, $H_{\rm PQ}\equiv H(t_{\rm PQ})$ and $a_{\rm PQ}\equiv a(t_{\rm PQ})$ are the values at the PQ symmetry breaking time $t_{\rm PQ}\ll t_{\rm osc}$. In~the second equation, we have also used $a \sim 1/T$ and \mbox{$H\sim T^2 G^{1/2}$} \cite{DiLuzio:2020wdo}.

While in the pre-inflationary scenario, inflation selects one patch of the universe within which the spontaneous breaking of the PQ symmetry leads to a homogeneous value of the initial misalignment angle $\phi_i$, in~the post-inflationary scenario, the PQ symmetry breaks with $\theta_i$, taking different values in patches that are initially out of causal contact; see the left panel of Figure~\ref{fig:axionpot}. However, today, they populate the volume enclosed by our Hubble horizon. In~the post-inflationary scenario, the initial misalignment angle $\phi_i$ takes all possible values on the unit circle. For~a quadratic potential, $V(\phi)=m^2_A\phi^2/2$, this is equivalent to an assumption that the initial condition reads $\phi_i\equiv \sqrt{\langle \phi_i^2 \rangle}  =\pi/\sqrt{3}$, where the angle brackets represent the value averaged over $\langle -\pi, \pi \rangle$ \cite{DiLuzio:2020wdo}.

Starting from $T_{\rm osc}$, the number of axions in a comoving frame becomes frozen, and~their number density evolves as
\begin{equation}
    n_A(T_{\rm osc})\approx m_A(T_{\rm osc})f^2_A\langle \phi_i^2 \rangle\, .
\end{equation}
For isotropic evolution, the~ratio of the number density $n_A$ to the entropy density $s$ in the comoving frame is conserved, i.e.,~$n_A(T)/s(T)=n_A(T_{\rm osc})/s(T_{\rm osc})$ leading for $T\ll T_c$ to the expression for the axion energy density,
\begin{equation}
 \epsilon_A^{\rm mis} = 
 m_{A} n_A(T_{\rm osc})\frac{h_{\rm eff}(T)}{h_{\rm eff}(T_{\rm osc})}\left(\frac{T}{T_{\rm osc}}\right)^3 =
 m_{A}f^2_A\langle \phi_i^2 \rangle \frac{h_{\rm eff}(T)}{h_{\rm eff}(T_{\rm osc})} \left(\frac{T}{T_{\rm osc}}\right)^3 T_c \,,
 \label{eq:axdene}
\end{equation}
where the effective number of DoFs of the entropy $h_{\rm eff}(T)$ is defined in Equation~(\ref{eq:heff}). Let us note that in contrast to Ref.~\cite{DiLuzio:2020wdo} in Equation~(\ref{eq:axdene}), the constancy of the axion mass for $T \lesssim T_c$ is already taken into~account.

After the spontaneous ${\rm U}(1)_{\rm PQ}$ symmetry breaking the axion field, $\phi$, being an angular variable, takes values in the interval $\langle 0, 2\pi f_A\rangle$, cf. Equation~(\ref{eq:varphi}). Consequently, the~axion potential $V(\phi)$ given by Equation~(\ref{eq:axion-pot}) is periodic in $\phi$ with period $\Delta\phi= 2\pi f_A/N_{\rm DW}$. In~the other words, $V(\phi)$ has an exact $Z_{N_{\rm DW}}$ discrete symmetry. The~axion acquires a periodic potential with $N_{\rm DW}$ equivalent minima. The~Kibble mechanism~\cite{Kibble:1976sj, Gangui:2001wc} then dictates that, depending on the homotopy group $\pi(\mathcal{M})$ of the manifold $\mathcal{M}$ of degenerate vacua, the~topological defects --- domain walls, strings or monopoles --- form each time the symmetry is broken~\cite{Vilenkin:2000jqa,Gangui:2001wc}. With~$\mathcal{M}={\rm U}(1)_{\rm PQ}$ and $\pi(\mathcal{M}=Z_{N_{\rm DW}})$, the production of axionic strings, which are vortex-like topological defects that form as soon as the symmetry is spontaneously broken, is possible. Those are not important when the PQ symmetry is broken before inflation --- they are inflated away --- but they play an important role in the post-inflationary scenario. When the Hubble parameter $H$ becomes comparable to the axion mass $m_A$, the~axion starts to roll down to one of the minima. Since the axion field settles into different minima in different places of the universe, domain walls are formed between the different vacua; see the left panel of Figure~\ref{fig:axionpot}. It is worth mentioning that this phenomenon is similar to ice formation on the surface of a pond or a puddle when the water begins to freeze in many places independently, and~the growing plates of ice join up in random fashion, leaving zigzag boundaries between them~\cite{Gangui:2001wc}.

As an example, let us consider a planar wall orthogonal to the $z$-axis $\phi=\phi(z)$. The~solution of the classical field equation with potential given by Equation~(\ref{eq:axion-pot}) reads~\cite{Saikawa:2017hiv}:
\begin{equation}
    \frac{\phi(z)}{f_A}=\frac{2\pi k}{N_{\rm DW}} + \frac{4 }{N_{\rm DW}}\tan^{-1}e^{m_A z} \,.
\end{equation}

This configuration interpolates between the two allowed vacua, \mbox{$\phi/f_A = 2\pi k/N_{\rm DW}$} at \mbox{$z \rightarrow -\infty$} and \mbox{$\phi/f_A= 2\pi(k+1)/N_{\rm DW}$}, which are separated by the wall of thickness $1/m_A$.

Astrophysical signatures of the axion can be broadly divided into its couplings to elementary/composite particles, i.e.,~photons, electrons, protons or neutrons, and~to the macroscopic objects in the universe such as BHs~\cite{DiLuzio:2020wdo}. In~the latter case, when the Compton length of the axions becomes of order of the BH size, they form gravitational bound states around it. The~phenomenon of superradiance~\cite{Penrose:1969pc, Zeldovich:1971} causes the axion occupation numbers to grow exponentially, providing a way to extract very efficiently energy and angular momentum from the BH. The~presence of axions could be inferred by observations of BH masses and angular momenta. Current measurements exclude the region of $6\times 10^{17}\, {\rm GeV} \leq f_A \leq 10^{19}$ GeV.

In particle physics signatures, the most important is the decay channel of axion into two photons with the decay width $\Gamma_{{\rm A}\rightarrow \gamma\gamma}=g^2_{{\rm A}\gamma\gamma} m_A^3/(64\pi)$ in terms of the model-dependent coupling constant $g_{{\rm A}\gamma\gamma}$. The~main uncertainty is due to the electromagnetic and color anomalies of the axial current associated with the axion. The~most relevant process induced by the $g_{{\rm A}\gamma\gamma}$ is the Primakoff process --- the conversion of thermal photons in the electrostatic field of electrons and nuclei into axions,
\begin{equation}
    \gamma+Ze\rightarrow A + Ze \,.
\end{equation}

A strong bound on exotic cooling processes in the sun is provided by the helioseismological considerations~\cite{Vinyoles:2015aba} giving the following constraint: $g_{{\rm A}\gamma\gamma}\leq 4.1 \times 10^{-10}\, {\rm GeV}^{-1}$. For~more details, consult Refs.~\cite{Sikivie:2006ni, Peccei:2006as, Marsh:2015xka, DiLuzio:2020wdo, Kawasaki:2013ae}. Another option is the inverse Primakoff scattering, which allows solar axions to coherently scatter off the atomic electric field and back-convert into photons in the detector volume, $A+Ze \rightarrow \gamma Ze$, proceeding through a $t$-channel photon exchange~\cite{Dent:2020jhf}.

The axion besides being a well-recognized DM candidate may provide also an interesting explanation of a wider range of phenomena related to the early universe dynamics. As~a first example, let us mention the SMASH model --- a minimal extension of the SM with additional particle content comprising three sterile right-handed neutrinos $N_i, i=1,2,3$, a~color triplet $Q$ and a complex SM-singlet scalar $\sigma$, whose VEV of $v_{\sigma} \sim 10^{11}$ GeV breaks the lepton number and the PQ symmetry simultaneously~\cite{Ballesteros:2016xej}. At~low energies, the~model reduces to the SM, which is augmented by seesaw-generated neutrino masses and mixing, plus the axion. In~this scenario, the~inflaton, a~scalar field driving cosmic inflation in the very early universe, is a mixture of $\sigma$ and the SM Higgs fields. The~reheating of the universe after inflation occurs via a mechanism known as a Higgs portal~\cite{Arcadi:2019lka} --- by the DM particles, which interact only through their couplings with the Higgs sector of the theory. The~model provides a consistent picture of particle physics from the EW scale to the Planck scale $M_{\rm PL}$ and of cosmology from inflation until today. In~particular, in~the SMASH model framework, the PQ symmetry is first broken and then restored non-thermally during preheating for $f_A = 4\times 10^{16}$ GeV.

The second example, the~Axion Quark Nugget (AQN) DM model, see, e.g.,~Ref.~\cite{Ge:2020xvf} and references therein, replaces the commonly accepted baryogenesis scenario with a charge separation process in which the global baryon number of the universe remains zero at all times. Similarly to Witten's idea of stranglets~\cite{Witten:1984rs}, the~AQN DM is composed of quarks and anti-quarks but now in a new high-density CSC phase. Initially, nuggets of both matter and antimatter are formed with equal probability as a result of the dynamics of the axion domain walls which at the same time provide the extra pressure needed to stabilize the CSC phase. Later on, due to the global \textsf{CP} violating processes associated with the initial misalignment angle $\theta_0\neq 0$ during the early formation stage, the~populations of the nuggets with the positive and negative baryon number become different. The~unobserved antibaryons hidden inside the DM would not participate in nucleosynthesis and, therefore, according to the usual definition would not contribute to the visible matter. However, since antimatter nuggets can interact with regular matter via annihilation leading to electromagnetic radiation, their existence has observational consequences~\cite{Ge:2020xvf}. 

It is worthwhile mentioning here an interesting generalization of the PQ mechanism where in addition to $\theta$ angle, also the strong coupling $\alpha_{\rm S}$ is promoted to a dynamical quantity. The~latter evolves through the VEV of a singlet scalar field that mixes with the Higgs field~\cite{Ipek:2018lhm,Berger:2020maa}. In~the resulting cosmic history, the~QCD confinement and EW symmetry breaking initially occur simultaneously close to the weak~scale.

To conclude this section, as~we have already noticed in several examples above, the~non-perturbative dynamics of QCD often exhibits very non-trivial and rather unexpected consequences at cosmological scales. An~attractive mechanism proposing that the QCD axion may emerge as a composite state has been discussed very recently in Ref.~\cite{Addazi:2021ivu}. In~particular, it was suggested that Majorana neutrinos, that combine into Cooper pairs, can form collective low-energy degrees of freedom. This motivates the existence of the QCD axion as a collective excitation of the neutrino condensate. Such a condensate can be produced after the QCD phase transition epoch in a cold coherent state by means of a misalignment mechanism, thus providing an alternative DM candidate. In~this case, a~QCD anomalous portal provides the necessary means for a tiny mass gap generation by neutrinos. Furthermore, the~Cosmological Constant emerges as a result of the spontaneously broken mirror symmetry of the QCD ground state triggered by the quantum gravity effects as suggested by Refs.~\cite{Pasechnik:2013poa,Addazi:2018fyo} (see also below). Hence, one concludes that QCD may be responsible for the dynamical generation of both the DM and DE components of the universe such that a complete knowledge of the QCD in the infrared regime may be absolutely critical for understanding of the cosmological evolution since the latest QCD transition epoch and for eventual formation of the current state of the~universe.

\section{Dynamics of the Early~Universe}
\label{sec:Cosmoqgp}

\subsection{Simple Models with Constant Speed of~Sound}
\label{Ssec:simple}

In accord with the observations, the Standard Cosmological Model (SCM) postulates that the cosmic matter at scales larger than 100 Mpc is homogeneously and isotropically distributed. Consequently, thermodynamic pressure $p$ at early times of its evolution depends on temperature $T$ and various chemical potentials $\mu_i, i=B, Q, L, \ldots$ corresponding to baryon number $B$, electric charge $Q$, lepton number $L$ etc., only via the energy density $\epsilon$. Solution of the Einstein equations of general relativity
\begin{equation}
R_{\mu\nu} -\tfrac{1}{2}g_{\mu \nu}(R-2\Lambda)  = -8\pi G T_{\mu \nu} \,,
\label{eq:Eeq}
\end{equation} 
where $R_{\mu\nu}$ is the Ricci tensor, $R=R_{\mu\nu}g^{\mu\nu}$ is the scalar curvature, $g^{\mu\nu}$ is the metric tensor, and $\Lambda$ and $G$ are the cosmological and gravitational constants; preserving the homogeneity and isotropy of space under its time evolution is a spacetime of constant curvature parameter $k=\{+1,0,-1 \}$. It is described by a single function --- the time-dependent scale factor $a(t)$ \cite{Mukhanov:2005sc, Weinberg:2008zzc, Zee:2013dea} which connects the Lagrangian (or comoving) coordinates $r$ with the physical Euler coordinates  $\hat{r}(t)=a(t)r$. The~metric tensor $g_{\mu\nu}$ in the preferred coordinate system where these symmetries are clearly manifest reads\endnote{It is worth mentioning that even though a fluid filling a FLRW universe (\ref{eq:FLRW}) homogeneously is static in the comoving frame $u^{\mu}=(1,0,0,0)$, the~expanding geometry induces a nonzero fluid expansion rate $\partial_{\mu}(\sqrt{-g})u^{\mu}/\sqrt{-g}=3H(t)$, where $g=-a^6(t)$ is the determinant of the FLRW metric tensor $g_{\mu\nu}$ with $k=0$.}
\begin{equation}
ds^2 \!= \!g_{\mu\nu}x^{\mu}x^{\nu}\!=\!dt^2 \!- \!a^2(t)\Big [ \frac{dr^2}{1\!-\!k r^2}\! +\! r^2 (d{\theta}^2 \!+ \!\sin^2{\theta}d{\phi}^2)\Big ]\,, \qquad k\!=\!{\rm const} \,.
\label{eq:FLRW}
\end{equation}

Using this metric in Equations~\eqref{eq:Eeq} and \eqref{eq:FLRW} and neglecting the dissipative terms yields the Friedmann equation for the time evolution of $a(t)$ and the fluid equation for the time evolution of $\epsilon(t)$, respectively (see e.g.,~Refs.~\cite{Mukhanov:2005sc, Liddle:1998ew}),
\begin{eqnarray}
H^2(t) \equiv \Big (\frac{\dot{a}}{a}\Big) ^2 =\frac{8\pi G}{3}\epsilon\! -\! \frac{k}{a^2} \!+\! \frac{\Lambda}{3} \,,
\label{eq:Fried} \qquad
\dot{\epsilon}+3(\epsilon+p )H(t) = 0 \,,
\label{eq:Fried1}
\end{eqnarray}
where $H(t)$ is the Hubble parameter. From~Equation~(\ref{eq:Fried1}), it follows that the expanding universe is characterized by a natural time-scale $H^{-1}=a/\dot{a}$. Any particle species will remain in thermal equilibrium with the cosmic fluid so long as the mean interaction time $t_c$ allows rapid adjustment to the falling temperature provided that $t_c<H^{-1}$.

In the period of the universe evolution when $\epsilon\gtrsim1$~GeV fm$^{-3}$, the terms containing constants $\Lambda$ and $k$ in Equation~(\ref{eq:Fried}) can be safely neglected,  transforming the above two equations into a single one describing the time evolution of the energy density~\cite{Ornik:1987up},
\begin{equation}
 -\frac{d\epsilon}{3 \sqrt{\epsilon}(\epsilon + p)}=\sqrt{\frac{8\pi G}{3}}dt \,.
 \label{eq:ornik}
\end{equation}

For the time-independent speed of sound $c_s$ and for the energy densities negligible compared to the initial density $\epsilon(t>0)\ll \epsilon(t=0)$ integration of Equation~(\ref{eq:ornik}), the calculations yield~\cite{Sanches:2014gfa} 
\begin{equation}
 \epsilon(t)=\frac{1}{6\pi G(1+c^2_s)^2 t^2}\,, \qquad c^2_s\equiv \frac{dp}{d\epsilon}= {\rm const}\, .
 \label{eq:sanch1}
\end{equation}

Substituting this equation into the fluid Equation~(\ref{eq:Fried}), we obtain the expansion rate of the early universe 
\begin{equation}
\dot{a}\sim t^{-\alpha} \,, \qquad \alpha=\frac{1+3c^2_s}{3(1+c^2_s)} \,.
 \label{eq:sanch2}
\end{equation}

In particular, for~the massless non-interacting gas with $c_s^2=1/3$, one obtains $\dot{a}\sim t^{-1/2}$ and hence $a(t)\sim t^{1/2}$.

It is worth mentioning that in the cosmological literature, see, e.g.,~Ref.~\cite{Weinberg:2008zzc}, it is customary to express the EoS in terms of the parameter $w$, Equation~(\ref{eq:wein1}). For~the constant speed of sound case, i.e.,~for $w={\rm const}$, the solution of the Friedmann Equation~(\ref{eq:Fried1}) yields 
\begin{equation}
\epsilon\sim a^{-3(1+w)} \,.
\label{eq:wein2}
\end{equation}

For the non-relativistic matter (dust) with $p=0$, we obtain $\epsilon\sim a^{-3}$, for~the massless non-interacting gas, $\epsilon\sim a^{-4}$, for~the EoS $p=\epsilon$ corresponding to absolutely stiff fluid~\cite{Zeldovich:1961sbr}, $\epsilon\sim a^{-6}$, and~for the vacuum energy with the EoS $p=-\epsilon$, Equation~(\ref{eq:wein2}) gives $\epsilon={\rm const}$.

Let us now follow Ref.~\cite{Trojan:2011ma} and consider an ideal gas of free particles in $D$-dimensional space. Its particle number density $n$, energy density $\epsilon$ and pressure $p$ expressed in terms of single-particle statistical sum $f(E,T,\mu)$ of particle with energy $E$, momentum $P$ and spin $s$ reads (see e.g.,~Ref.~\cite{Kapusta:1989tk}):
\begin{eqnarray}
f(E,T,\mu)=\frac{1}{\exp[(E-\mu)/T]\pm 1}\,, \qquad 
n=\gamma\int f(E,T,\mu)d^DP \,, 
\label{eq:troj1}\\
\epsilon=\gamma\int f(E,T,\mu) E(P)d^DP=
\gamma S(D)\int_0^{\infty} f(E,T,\mu) E(P)P^{D-1}dP \,,
\label{eq:troj2}\\
p=-\frac{T}{V}\ln Z =-\gamma T\int \ln f(E,T,\mu) d^DP=
\gamma\frac{S(D)}{D}\int_0^{\infty} f(E,T,\mu) \frac{\partial E(P)}{\partial P} P^D dP \,,
\label{eq:troj3}
\end{eqnarray}
where $d^D P=[D\pi^{D/2}]/[\Gamma(D/2+1)]P^{D-1}dP\equiv S(D)P^{D-1}dP$ is a volume element of the $D$-dimensional hypersphere and $\gamma\equiv (2s+1)(2\pi)^{-D}$.  When evaluating the first integral in Equation~\eqref{eq:troj3}, we have performed integration by parts assuming that the particle energy $E(P)$ is some generic function of $P$. The substitution of Equations~\eqref{eq:troj2} and \eqref{eq:troj3} into Equation~\eqref{eq:wein1} constrains $E(P)$ to satisfy the differential equation
\begin{equation}
\frac{P}{D}\frac{\partial E(P)}{\partial P}=w E(P) \,,
\label{eq:troj4}
\end{equation}
whose solution for $w={\rm const}$ reads  
\begin{equation}
E(P)=\xi P^{w D} \,,
\label{eq:troj5}
\end{equation}
with $\xi$ some arbitrary constant. Hence, the~medium with constant speed of sound squared $c_s^2=w$ in ordinary three-dimensional space can be equivalently described as an ideal gas of quasi-particles with energy $E$ and momentum $P$ satisfying the dispersion relation (\ref{eq:troj5}) in $D$-dimensional space~\cite{Trojan:2011ma}.  It is worth mentioning that at some instances, the dispersion relation (\ref{eq:troj5}) can be satisfied by the real particles.  The~case of $w=0$ corresponds to non-relativistic particles, while $w D=1$ corresponds to massless particles in $D$-dimensional space with the EoS $p=\epsilon/D$ and hence with the sound velocity $c_s=D^{-1/2}$. 

Unfortunately, this is not the case of the absolutely stiff fluid, as first discussed by Zeldovich~\cite{Zeldovich:1961sbr}. Notwithstanding that its EoS $p=\epsilon$ can be used to describe a large variety of systems such as phonon-like excitations in a thin channel ($D=1$) \cite{Trojan:2011ma}, thin film ($D=2$) of non-relativistic quasi-particles~\cite{Trojan:2011ma}, interiors of neutron stars~\cite{Rhoades:1974fn,Blaschke:2020vuy}, Big Bang nucleosynthesis~\cite{Dutta:2010cu} or warm self-interacting DM component~\cite{Stiele:2010xz}. The cosmology and thermodynamics of the FLRW universe with bulk viscous stiff fluid was studied in Ref.~\cite{Mathew:2014gpa}. The~fact that the stiff fluid saturates the holographic covariant entropy bound was used in  Ref.~\cite{Banks:2001px} to describe a cosmology of the very early universe. Last but not least, a~stiff perfect fluid is energetically equivalent to a time-like massless scalar field $\phi$, see, e.g.,~Ref.~\cite{Miguelote:2003ig}. From~its energy-momentum tensor,
\begin{equation}
 T^{\phi}_{\mu\nu}=\partial_{\mu}\phi\partial_{\nu}\phi -
 \frac{1}{2}g_{\mu\nu}\partial_{\alpha}\phi\partial^{\alpha}\phi \,, \qquad
\partial_{\alpha}\phi\partial^{\alpha}\phi>0 \,,
 \label{eq:migu1}
\end{equation}
we obtain that $p\equiv T^{\phi}_{kk}=\epsilon \equiv T^{\phi}_{00} =
\partial_{\alpha}\phi\partial^{\alpha}\phi$.

\subsection{Equation of State of the Early~Universe} 
\label{Sec:EoS}

As already discussed in Section~\ref{Ssec:simple}, the only way how to write the EoS compatible with homogeneity and isotropy of the universe is through the pressure expressed as a function of the energy density. Thus, given the barotropic EoS $p(\epsilon)$ of the expanding matter, we can for instance use Equation~(\ref{eq:ornik}) to predict the temporal evolution of the energy density $\epsilon(t)$ and the temperature $T(t)$. For~the ideal gas EoS, Equation~(\ref{eq:sanch1}) yields the relation between the temperature $T$ of the early universe and the time $t_{\rm sec}$ elapsed from the Big Bang, $T_{\rm MeV}\!=\!\mathcal{O}(1)/\sqrt{t_{\rm sec}}$ \cite{Mukhanov:2005sc}, i.e.,~the same time-dependence as for the time derivative of the scale factor $\dot{a}(t)$; see Equation~(\ref{eq:sanch2}) and the discussion below it. For~the QGP-to-hadronic matter transition, this leads to $t_{\rm sec}^{\rm QGP}\!\sim\!10^{-5}\, s$, and~for the EW phase transition, this leads to $t_{\rm sec}^{\rm EW}\!\sim\!10^{-11}\, s$. 

More sophisticated EoS can be based either on the phenomenological models or on the microscopic theory. In~the latter case, a~quantum physics formulation of statistical mechanics in terms of the $S$ matrix, which describes the scattering processes taking place in the thermodynamical system of interest, is available --- see Ref.~\cite{Dashen:1969ep}. It provides a simple prescription for calculating the grand canonical potential $Z(T, \mu)$ of any gaseous system given the free-particle energies and $S$-matrix elements. The application of $S$ matrix formulation to study the thermal properties of an interacting gas of hadrons can be found in Ref.~\cite{Welke:1990za}. It can be also used to show how the hadron resonance gas model emerges from the $S$-matrix framework~\cite{Lo:2017sde}. However, this approach is based on the perturbative expansion of the $S$-matrix and is therefore not applicable in the strong coupling regime of QFT. There, one resorts to direct calculation of the grand canonical potential on the~lattice.

Our first example of the EoS used in a description of the evolution of the early universe is the Bag Model (BM) EoS~\cite{Baacke:1976jv, Shuryak:1980tp, Fogaca:2009wf} based on the phenomenological description of the mass spectrum of the hadron states~\cite{Chodos:1974je,DeTar:1983rw, Yagi:2005yb} in terms of gas of massless color objects --- quarks and gluons --- moving inside the confining potential --- the bag,
\begin{equation}
\epsilon_q(T)=\sigma_q T^4 + \mathcal{B} \,, \qquad
p_q(T)=\frac{\sigma_q}{3}T^4 -\mathcal{B} \,, \qquad
p_q(\epsilon)=\frac{1}{3}\left (\epsilon-4\mathcal{B}\right)\, .
\label{MIT}
\end{equation}

In Equation~(\ref{MIT}), $\sigma_q=\frac{\pi^2}{30}g^{\rm QCD}_{\rm eff}$ is the Stefan-Boltzmann constant with $g^{\rm QCD}_{\rm eff}$ given by Equation~(\ref{eq:g_qcd}). The~BM EoS (\ref{MIT}) incorporates color confinement through the bag constant $\mathcal{B} = \epsilon_{\rm bag} - \epsilon_{\rm vac} > 0$, indicating the~difference between the energy densities of the physical vacuum and the ground state for quarks and gluons in the medium. The~latter can be interpreted as the energy needed to create a bubble in the vacuum in which the non-interacting quarks and gluons are confined. While the fit to hadron masses made in the original BM predict $\mathcal{B}^{1/4}\approx 140$ MeV, the value of $\mathcal{B}^{1/4}\approx 220$ MeV is frequently quoted in works dealing with the vacuum structure of QCD, see, e.g.,~Ref.~\cite{Yagi:2005yb}. 

Let us note that BM EoS represents a bare-bones model of hadron-to-QGP phase transition. At~small energy densities, hadrons --- the bubbles inside the non-perturbative vacuum --- occupy only a small fraction of the total considered volume $V$. An increase of $\epsilon$ leads to the coalescence of several bubbles into larger ones. For~$\epsilon\geq 4\mathcal{B}$, the volume $V$ is filled with one large bubble, cf.~Equation~(\ref{MIT}), whose surface coincides with the enclosing walls.  Hence, there is no longer free surface against the vacuum,  and the new de-confined phase of matter is created~\cite{Baacke:1976jv}. The~critical temperature $T_c$ of the phase transition can be estimated using Gibbs criteria. Equating the BM pressure (\ref{MIT}) with the pressure $p_h$ of the hadron (pion) gas with 3 DoF and hence with $\sigma_h=3\pi^2/30$, we obtain
\begin{equation}
p_q(T_c)=\frac{\sigma_q}{3} T_c^4 -\mathcal{B}=p_h(T_c)= \frac{\sigma_h}{3}T_c^4 \,, \qquad
T_c=\left (\frac{3\mathcal{B}}{\sigma_q-\sigma_h} \right)^{1/4} \,.
 \label{T_c}
\end{equation}

For $N_F=3$ active quark flavors $u,d,$~and~$c$, Equation~(\ref{eq:geff}) yields  $T_c\approx  0.67 \mathcal{B}^{1/4} and \approx 150$ MeV. Let us add that Equation~(\ref{T_c}) describes the first-order phase transition  with energy density discontinuity,
\begin{equation}
\Delta\epsilon=\epsilon_q(T_c)-\epsilon_h(T_c) =3p_q(T_c)+4\mathcal{B}-3p_h(T_c)=4\mathcal{B}\,.
 \label{eps_c}
\end{equation}

Last but not least, expressing the BM EoS (\ref{MIT}) in terms of the dimensionless interaction measure --- the trace anomaly describing the thermal contribution to the trace of the energy-momentum tensor $\mathcal{T}^{\mu\nu} \equiv T^{\mu \nu}$
\begin{equation}
 \Theta \equiv \frac{ \mathcal{T}^{\mu\mu}(T)}{T^4} = \frac{\epsilon_q-3p_q}{T^4}=
 \frac{4\mathcal{B}}{T^4} =
 \frac{4\sigma_q\mathcal{B}}{\epsilon_q -\mathcal{B}} \,,
 \label{tran}
\end{equation}
we observe a monotonous weakening of the interaction strength with increasing $\epsilon_q$ \cite{Collins:1974ky}, leading ultimately to a Stefan-Boltzmann (SB) value of $\Theta=0$. At~the same time, the entropy density $s_q=(\epsilon_q+p_q)/T=4/3\sigma_q^{1/4}(\epsilon_q-\mathcal{B})^{3/4}$ converges to its SB limit from below. The~speed of sound derived from the BM EoS (\ref{MIT}) is energy-density-independent and coincides with that of the ideal gas of massless particles $c_s^2=dp_q/d\epsilon_q=1/3$. 

More sophisticated EoS can be constructed, e.g.,~by adding the term $\sim T^2$ to expressions for $\epsilon(T)$  and $p(T)$ in Equation~(\ref{MIT}) with $\sigma=\sigma_q$
\begin{equation}
\epsilon(T)=\sigma T^4 - CT^2 + \mathcal{B} \,, \qquad
p(T)=\frac{\sigma}{3}T^4 - DT^2 -\mathcal{B}  \,.
\label{FZB1} 
\end{equation}

The motivation for such a modified BM EoS comes from the observation~\cite{Pisarski:2006hz, Megias:2007pq} that in the case of a pure gauge theory up to temperatures a few times the transition temperature $T_c$, the dominant power-like correction to the pQCD high-temperature behavior is $\mathcal{O}(T^{-2})$ rather than $\mathcal{O}(T^{-4})$. Moreover, the~quadratic thermal terms in the deconfined phase can be also obtained from gauge/string duality~\cite{Zuo:2014vga}.

In its original setting with $C=D>0$, Equation~(\ref{FZB1}) represents the LQCD-motivated ``fuzzy'' BM EoS of Ref.~\cite{Pisarski:2006yk}. On~the other hand, for $C=-D<0$, it represents a particular case of gas of gluonic quasi-particles EoS with a temperature-dependent bag function $\mathcal{B}(T)=-CT^2+\mathcal{B}$; see e.g.,~Refs.~\cite{Schneider:2001nf, Giacosa:2010vz}. In~the following discussion, we will keep the values of the constants $C$ and $D$ in Equation~(\ref{FZB1}) unrestricted.

By inverting $\epsilon(T)$ in Equation~(\ref{FZB1}) with respect to temperature squared $T^2$,
\begin{equation}
T^2(\epsilon)=\frac{C+ \sqrt{C^2+4\sigma(\epsilon-\mathcal{B})}}{2\sigma}>0 \, ,
\label{FZB3}
\end{equation}
and substituting for $T^2(\epsilon)$ into Equation~(\ref{FZB1}), we obtain the barotropic form of the EoS (\ref{FZB1}):
\begin{equation}
p(\epsilon) = 
\frac{1}{3}\left (\epsilon - 4\mathcal{B}\right) - \frac{1}{3}{\rm sgn}(A)|A| T^2(\epsilon)~~~A=3D-C \,.
\label{FZB2} 
\end{equation}

The corresponding sound velocity squared reads
\begin{equation}
c^2_{s}(\epsilon)= \frac{d p(\epsilon)}{d \epsilon}= 
\frac{1}{3}\left (1 - \frac{{\rm sgn}(A)|A|}{\sqrt{C^2+4\sigma(\epsilon-\mathcal{B})}}\right) \,.
\label{FZB4}
\end{equation}
Note that for $A=0$, both Equations~(\ref{FZB2}) and (\ref{FZB4}) degenerate to the corresponding result from the BM EoS; see Equation~(\ref{MIT}). In~the context of Equation~(\ref{FZB1}), the situation with $C=3D=\frac{3}{2}N_F\mu_B^2$ describes the EoS of ideal QGP with non-zero baryon chemical potential~\cite{Shuryak:2004pry}. It is worth mentioning that by tuning the phase transition temperature to $T_c=160$ MeV, this EoS predicts a six times higher value of the bag constant $\mathcal{B}$ than the original BM~\cite{Shuryak:2004pry}.

Therefore, in the following, we discuss only the non-trivial case of $A\neq 0$. First, we check which values of the constants $C$ and $D$ in Equation~(\ref{FZB4}) are compatible with the condition $c_s^2>0$.  While for $A<0$ (i.e., $C>0$ and $-C< D <C/3$), Equation~(\ref{FZB4}) is  always  satisfied  for $A>0$ ($C\geq 0$ and $D >C/3$  or  $D>-C>0$), there exists a lower bound $\epsilon_0$ on energy density 
\begin{equation}
 \epsilon > \epsilon_0 = \frac{A^2-C^2}{4\sigma}+\mathcal{B} \, .
\label{FZB5}
\end{equation}

Thus, in both cases ($A>0$ or $A<0$), Equation~(\ref{FZB3}) represents the genuine non-trivial EoS of non-ideal high-density matter each with its own sound velocity approaching for $\epsilon \to \infty$, either from below or from above, the~SB limit of $\sqrt{1/3}$. However, only for $A<0$, the second term on the right-hand side of Equation~(\ref{FZB2}) with $-\frac{1}{3}AT^2(\epsilon)$  represents independent pressure. It is also worth mentioning that for the latter case ($A<0$), the trace anomaly
\begin{equation} 
\Theta(\epsilon) = \frac{4\mathcal{B}}{T^4(\epsilon)}+\frac{{\rm sgn}(A)|A|}{T^2(\epsilon)} 
  \, ,
  \label{tran2} 
  \end{equation}
with $T^2(\epsilon)$ defined in Equation~(\ref{FZB3}), acquires a peak at the energy density
\begin{equation}
 \epsilon_p = \frac{2\mathcal{B}\pm \sqrt{2 \mathcal{B} |A|}C}{|A|\sigma} -\mathcal{B}
  \, .
 \label{tran3}
\end{equation}

Standard explanation of this phenomenon within the \mbox{SU(3)$_c$} gauge theory, see, e.g., Ref.~\cite{Castorina:2011ja}, relies on the fact that in the region around and just above the critical temperature $T_c$ of hadron-to-QGP phase transition, the~energy density rises much more rapidly than the pressure, leading to the observed rapid increase of $\Theta$. Since asymptotically $\epsilon/T^4$ and $3p(T)/T^4$ converge to their common Stefan-Boltzmann value of $\sigma$, see Equation~(\ref{FZB1}), there must be some temperature $T_p$ (and hence also some energy density $\epsilon_p=\epsilon(T_p)$) at which the growth rates change roles, with~the pressure now increasing more rapidly. The further decrease of $\Theta$ is in good approximation given by $T^{-2}$, so that $T^2\Theta(T)$ becomes approximately constant very soon above $T_c$ and~up to about $5T_c$ \cite{Pisarski:2006yk}.

Note that Equation~(\ref{eq:ornik}) with the initial condition $\epsilon_0(t_0)\!=\!10^4$ GeV fm$^{-3}$ at $t_{\rm sec}^{\rm QGP} \ll t_0\!=\!10^{-9}$ s $\ll t_{\rm sec}^{\rm EW}$ was used in Ref.~\cite{Sanches:2014gfa} to study the sensitivity of dilution and cooling of the early universe to the changes in the primordial QGP EoS. The~latter might modify the pattern of the emission of GWs or the generation of baryon number fluctuations. No dramatic changes between different EoS, including Equations~(\ref{MIT}) and (\ref{FZB1}) (with $C=D$) and others, were found in the whole considered time interval. This finding seems to be supported by the LQCD calculations which show that the transition from primordial QGP matter to hadronic matter proceeded as a continuous crossover~\cite{Aoki:2006we}. The~latter does not introduce any fluctuations on length scales much longer than the natural length scales of QCD $\sim\Lambda_{\rm QCD}^{-1}$, so it has probably left no imprint in the microseconds-old universe that survived so as to be visible in some way today~\cite{Busza:2018rrf}.

A different conclusion was, however, reached in Ref.~\cite{Addazi:2018ctp} where a novel mechanism for the production of GWs during the QCD phase transition has been proposed. It was found that while the energy density of the homogeneous gluon condensate is smoothly decaying in cosmological time $t_{\rm sec}$, the pressure of the condensate undergoes a sequence of violent oscillations at the characteristic QCD time scales --- the process called relaxation --- the basic feature of the QCD transition. Such relaxation processes generate a very specific multi-peaked GWs signature in the domain of radio frequencies. In~particular, gravitational echoes of the QCD transition potentially accessible by the FAST~\cite{FAST:2020flt} and SKA~\cite{SKA:2018ckk} GW telescopes are sourced by the formation of domain walls in the QCD vacuum (also responsible for color confinement in the infrared regime). More details about the dynamical theory of the YM vacuum will be provided below in Section~\ref{sec:YMtheory}.

Recently, the~stochastic GWs induced by the scalar perturbations of the metrics were investigated in Ref.~\cite{Abe:2020sqb}  as a cosmological probe of the sound speed $c_s$ during the QCD phase transition. Although~the GW propagation itself does not depend on $c_s$, the~sound speed value affects the dynamics of primordial density. Induced stochastic GWs can thus be an indirect probe of both the EoS parameter $w=p/\epsilon$ and $c_s=\sqrt{dp/d\epsilon}$. In~particular, similarly to the conclusions of Ref.~\cite{Addazi:2018ctp}, the~GW frequency $\sim 10^{-8}$ Hz corresponding to the Hubble scale during the QCD phase transition appears to be in the range of the planned GW~detectors.

Let us now turn to a description based on the fundamental theory. Using the Lagrangian of the SM, one can extract the thermodynamical quantities either directly from the lattice calculations~\cite{Borsanyi:2016ksw}, deduce them from lattice simulations using a dimensionally-reduced EFT~\cite{Laine:2015kra} or use the perturbation theory. With~the temperature $T$ depending on the lattice spacing $a$ and the number of lattice points in the temporal direction $N_t$ as $T=(aN_t)^{-1}$, the~reliable Monte Carlo simulations of QFT in the high-temperature regime need very fine lattices. At~the same time, as the lattice spacing $a$ is reduced, the~autocorrelation times for zero temperature simulations rise, and~the costs of these simulations explode beyond feasibility~\cite{Borsanyi:2016ksw}. This makes the LQCD simulations in the region of temperatures higher then the few GeVs practically impossible. Alternatively, one can vary the gauge coupling $g_s$, which leads to changing $T$ as well, although~the spacial and temporal dimensions do not~\cite{Laine:2015kra}. In~addition to that, even at the temperatures of the EW phase transition $T_c^{\rm EW}$, the~dynamics needs to be treated with lattice methods. This is due to the fact that when the particle momenta in the range $k \sim g^{2}T/\pi$ are considered, then the dynamics of the system is non-perturbative~\cite{Linde:1990flp,Linde:1980ts,Gross:1980br}. Values much below and far above this value $p/T^4$ can be determined by a direct perturbative computation~\cite{Laine:2015kra}.

In the SM framework, the~basic thermodynamical observable is the pressure~\cite{Gynther:2005dj}, which can be formally defined through the grand canonical partition function $\mathcal{Z}$ as
\begin{equation}
\mathcal{Z}\equiv \exp \left[\frac{p_B(T)V}{T}  \right] \,, ~~~
p_B(T)=p_E(T)+p_M(T)+p_G(T) \,,
\label{eq:pfromZ}
\end{equation}
where $p_B$ denotes the ``bare'' result related to the physical (renormalized) pressure as $p(T)=p_B(T)-p_B(0)$. The~pressure terms $p_E(T),p_M(T)$ and $p_G(T)$ appearing in Equation~(\ref{eq:pfromZ}) collect the contributions from the momentum scales $k\sim\pi T$, $k \sim gT$, and~$k \sim g^2 T/\pi$, respectively. The~couplings $g$ relevant in the SM are $g\in\left\{h_t, g_1,g_2,g_3\right\}$, where $h_t$ denotes the Yukawa coupling between the top quark and the Higgs boson, and~$g_1,g_2,g_3$ are the SM couplings related to \mbox{U(1)$_Y$}, \mbox{SU(2)$_{\rm L}$} and \mbox{SU$_c(3)$} gauge groups, respectively. Note that in Equation~(\ref{eq:pfromZ}), the thermodynamic limit $V\rightarrow \infty$ is~implied.

Using this technique, the SM calculations of the dimensionless function $p(T)/T^4$ and of the trace anomaly $\Theta(T)$ (\ref{tran}) up to $\mathcal{O}(g^5)$ were performed in Ref.~\cite{Laine:2015kra}. It was found that similarly to EoS Equation~(\ref{FZB1}), Higgs dynamics induces a peak in heat capacity $c(T)=d\epsilon/dT$  occurring around $T_c^{\rm EW}\approx 160$ GeV. This  leads to a short period of slower temperature change, and~correspondingly, a mildly increased abundance of produced particles. However,~in general, the largest radiative corrections originate from QCD effects, reducing the energy density by a couple of percent from the free value even at $T > 160$ GeV~\cite{Laine:2015kra}. It is worth emphasizing that the above-mentioned effects do not exhaust all possible phenomena relevant for dynamics of the early universe. In~particular, as~already discussed in Section~\ref{Sec:Glasma}, at~high temperatures, the saturation phenomena become ubiquitous. In~this case, the gluon DoFs are frozen in the classical field (condensate) and do not contribute to the density of the QGP. To~shed more light on the microscopic theory in the canonical picture, the~dynamical aspects of relativistic gluon and hot meson plasmas, both coupled to the homogeneous condensate, are discussed in detail in Refs.~\cite{Prokhorov:2016gei,Pasechnik:2014gqw,Prokhorov:2013xba,Prokhorov:2017nst}, respectively.

\begin{figure}[H]
\begin{adjustwidth}{-\extralength}{0cm}
\centering
\includegraphics[trim= 90 70 50 60, clip, width=.5\textwidth, height=.42\textwidth]{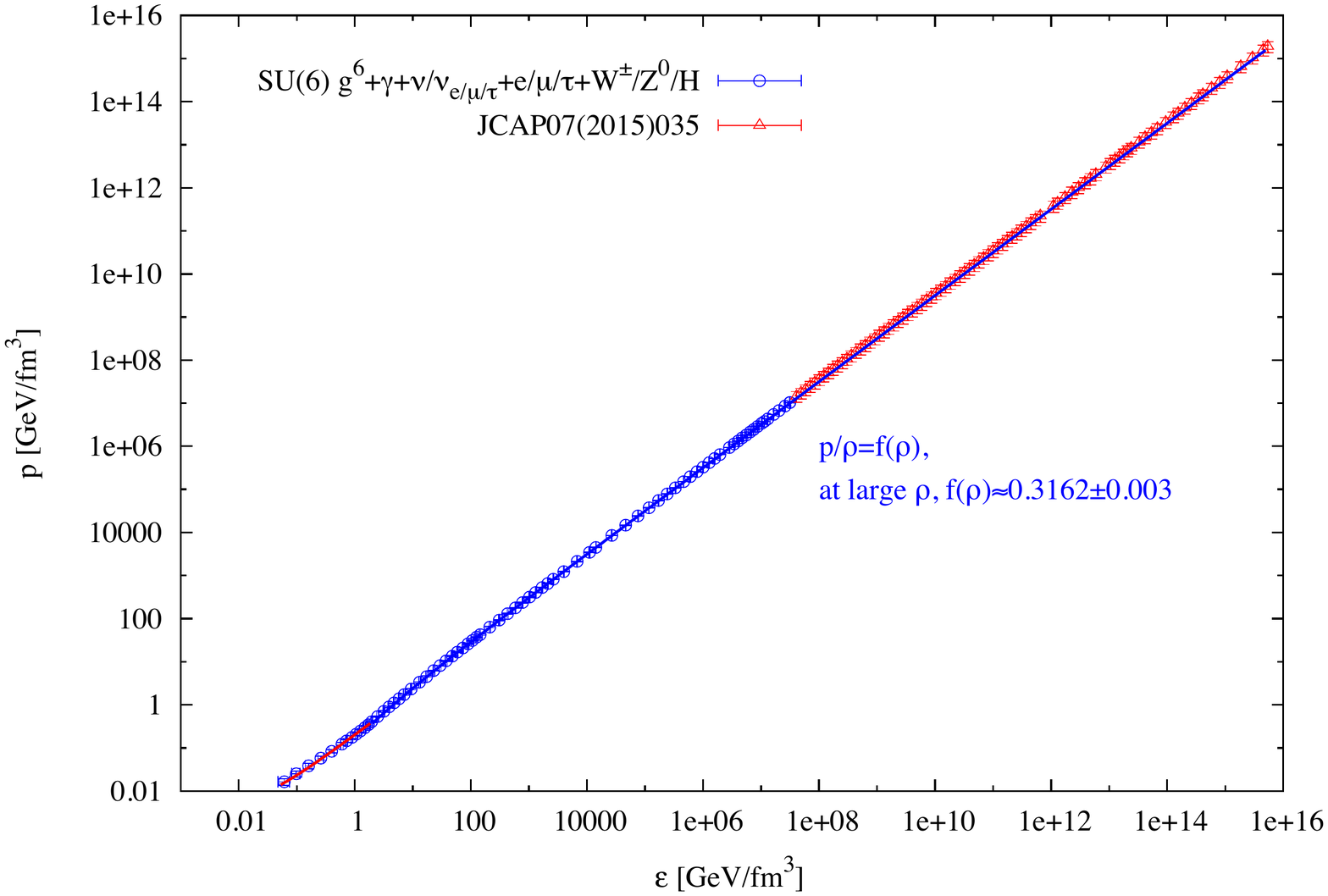}~~~~
 \includegraphics[trim= 90 60 80 60, clip, width=.51\textwidth, height=.425\textwidth] {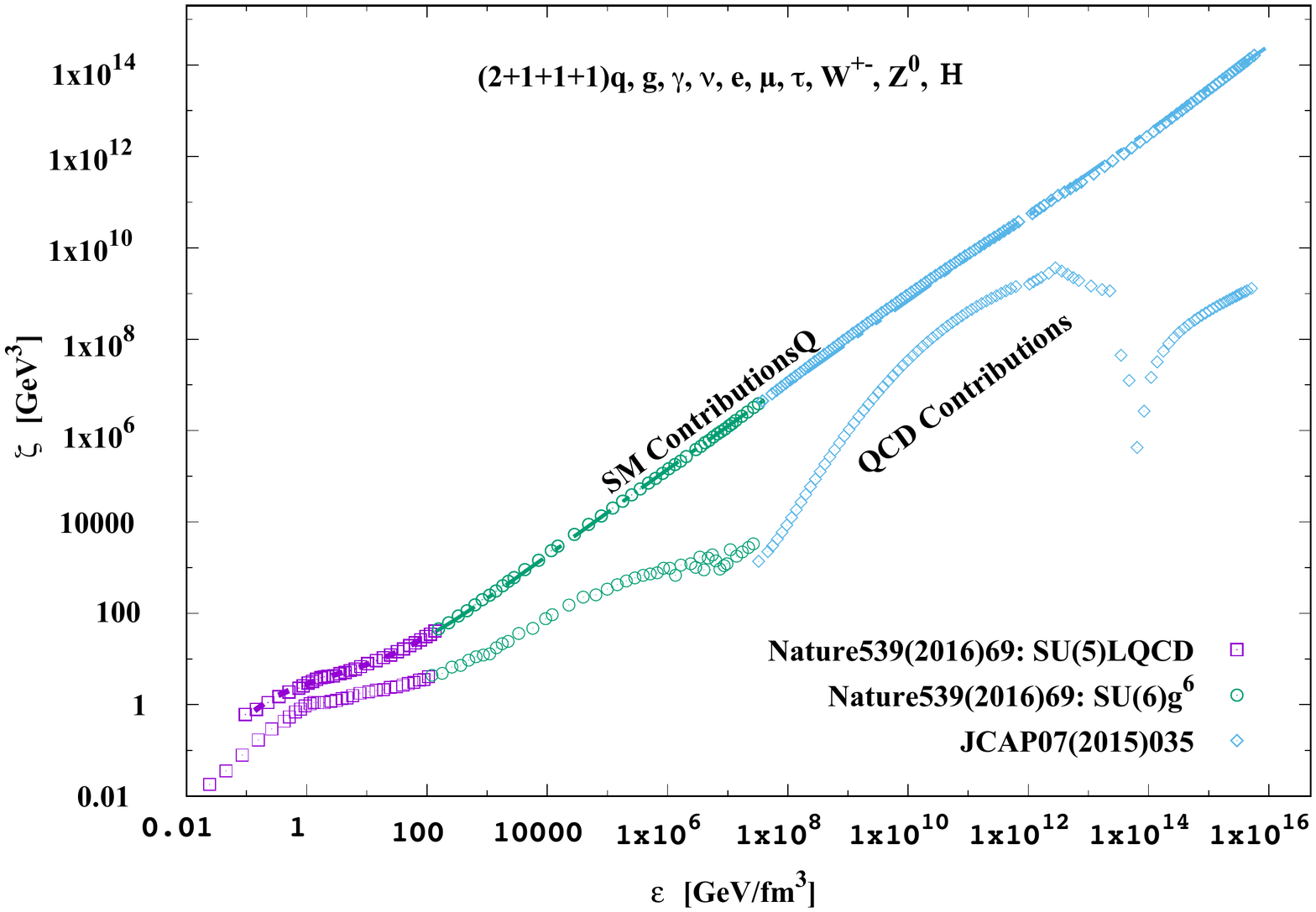}
 \end{adjustwidth}
\caption{(\textbf{Left}) The combined EoS $p(\epsilon), \epsilon\equiv\rho$ of QCD and EW matter, using non-perturbative results~\cite{Borsanyi:2016ksw} extended to include other DoFs such as $\gamma$, neutrinos, leptons, EW, and~Higgs bosons as well as perturbative results~\cite{Laine:2006cp, Laine:2015kra}. Adapted from Ref.~\cite{Tawfik:2019jsa}. (\textbf{Right}) Bulk viscosity $\zeta$ at $\mu_B=0$ as a function of the energy density $\epsilon$. The~top symbols stand for the SM contributions, while the bottom ones stand for the QCD contributions, only. Adapted from Ref.~\cite{Tawfik:2019qyd}.}
\label{fig:eos}

\end{figure}

In Ref.~\cite{Tawfik:2019jsa}, the results from Refs.~\cite{Borsanyi:2016ksw, Laine:2006cp, Laine:2015kra} were used to analyze the early universe EoS $p(\epsilon)$. The~extracted EoS depicted on the left panel of Figure~\ref{fig:eos} covers the broad interval in energy density $10^{-2} \leq \epsilon \leq 10^{16}$ GeV fm$^{-3}$  and corresponds to the evolution periods down from the GUT era, through the EW era and the QGP era, into~the hadron era. The~apparent smoothness of the function $p(\epsilon)$ hides in the complicated temperature dependence $\epsilon(T)=g_{\rm eff}(T)\epsilon_0(T)$ depicted on right panel of Figure~\ref{fig:mass}. 

While the GUT EoS $p_{\rm GUT}=(0.330\pm0.024)\epsilon$ valid for $10^{8} \lesssim \epsilon \leq 10^{16}$ GeV$\cdot$fm$^{-3}$ (the red triangles in Figure~\ref{fig:eos}) appears only slightly below the ideal gas limit, the~hadronic-era EoS $p_h=(0.003\pm0.002)+(0.199\pm0.002)\epsilon$ characterizes the region $\epsilon \lesssim 1$ GeV$\cdot$fm$^{-3}$. Most interesting is the intermediate region containing both the QCD and the EW epochs, which can be described by a single function  $p_{\rm SM}=a+b\epsilon+c\epsilon^d$ with $a=0.048\pm0.016, b~=~0.316~\pm~0.031, c=-0.21\pm0.014, d=-0.576\pm0.034$ \cite{Tawfik:2019jsa}. It is worth noting that the critical energy density $\epsilon_c$ defined implicitly by the equality of the pressures in hadronic and QGP phases $p_h(\epsilon_c)=p_{\rm SM}(\epsilon_c)$ reads: $\epsilon_c\simeq (1.2\pm0.2)$ GeV$\cdot$fm$^{-3}$. 

As can be seen from the parametrization of the fit,  
\begin{adjustwidth}{-\extralength}{0cm}
\begin{equation}
p_{\rm SM}=p_1(\epsilon)+p_2(\epsilon) \,, \quad
p_1(\epsilon)=b\epsilon\,,  \quad b>0\,, \quad
p_2(\epsilon)=a+c\epsilon^d\,,\quad a>0\,, \quad c<0\,, \quad d<0\,,
\label{eq:polytrop}
\end{equation}
\end{adjustwidth}
throughout the whole QCD and EW eras, there are two independent contributions $p_1(\epsilon)$ and $p_2(\epsilon)$ to the overall pressure of the universe. While $p_1$ is always positive, the~second pressure $p_2$ is negative up to $\epsilon\lesssim (7-13)$~GeV$\cdot$fm$^{-3}$. Although~the corresponding value of the trace anomaly can not be directly deduced from the fitted EoS (\ref{eq:polytrop})
\begin{equation}
\Theta=\frac{\epsilon-3p}{T^4}= 
\frac{\epsilon(1-3b)-3a-3c\epsilon^d}{T^4} \,,
  \label{eq:tran3}
 \end{equation}
it is  positive for $\epsilon\gtrsim (3-4)$~GeV$\cdot$fm$^{-3}$, and for very large energy densities, it falls as $\Theta \sim g_{\rm eff}\epsilon^{-3/2}$, cf.~Equation~(\ref{eq:geff}).

Note that the sound velocities
\begin{equation}
 c_{s,1}^2(\epsilon)=\frac{dp_1}{d\epsilon}=b>0 \,, \qquad
 c_{s,2}^2(\epsilon)=\frac{dp_2}{d\epsilon}=cd \epsilon^{d-1} > 0 \,,
 \label{eq:polytropc}
 \end{equation}
are well defined, making it possible for each of two components to represent the EoS of some substance. Let us add that analytical expressions for the scale factor of the universe $a(t)$ and Hubble parameter $H(t)$ deduced from the EoS (\ref{eq:polytrop}) were recently discussed in Ref.~\cite{Tawfik:2021rvv}. 

While the first component of pressure (\ref{fig:eos}) corresponds roughly to the EoS of the massless gas of non-interacting particles, the second  one, in~agreement with the asymptotic freedom of QCD~\cite{Collins:1974ky}, dies out with increasing $\epsilon$. Interestingly, $p_2(\epsilon)$, up~to the constant additive term $a$, coincides with the generalized Chaplygin EoS used in Ref.~\cite{Bento:2002ps} to describe the evolution from a phase dominated by non-relativistic matter to a phase dominated by the Cosmological Constant (or DE)\endnote {One of the authors (M.\v{S}.) would like to thank Petr Jizba for pointing out this analogy.}. For~$d=-1$ in Equation~(\ref{eq:polytrop}), we obtain the ordinary Chaplygin gas~\cite{Chaplygin:1904, Kamenshchik:2001cp} with EoS $p=c/\epsilon$. 

Another possibility of how to read the term $p_2(\epsilon)$ in Equation~(\ref{eq:polytrop}) is, in~analogy with Refs.~\cite{Schneider:2001nf, Giacosa:2010vz} and Equation~(\ref{FZB1}), to~interpret it as the density-dependent bag function $\mathcal{B}(\epsilon)=-(a+c\epsilon^d)$, cf.~Equation~(\ref{eq:polytrop}). The~latter may account for the density-dependent character of the physical vacuum due to, e.g.,~the instanton liquid~\cite{Schafer:1996wv, Shuryak:1997vd}. Instantons~\cite{Gross:1980br}, classical solutions to the Euclidean equations of motion, are localized in all the four dimensions and correspond to tunneling events between degenerate classical vacua in Minkowski space. Since tunneling lowers the ground-state energy, the~instantons provide a simple understanding of the negative non-perturbative vacuum energy density. Yet, the~Euclidean-based instanton model is not the only solution representing the QCD vacuum. It remains, in~fact, questionable to what extent it represents the reality due to non-analyticity of the gluonic field operators and the associated color confinement property. Below,~in Section~\ref{sec:YMtheory}, we elaborate on a recently proposed alternative picture and new gluonic vacuum solutions, which are readily formulated in Minkowski and FLRW~spacetimes.

\subsection{Hydrodynamical Description of Dissipative Effects and the Early~Universe}
\label{Sec:Dissipat}

According to the currently accepted scenario, see, e.g.,~Refs.~\cite{Mukhanov:2005sc, Kolb:1990vq}, the~evolution of the early universe must include a number of dissipative processes in~order to explain the current large value of the entropy per baryon. Some of them, such as the decoupling of neutrinos during the radiation era~\cite{Weinberg:2008zzc} or different cooling rates of the fluid components in the expanding universe~\cite{Iorio:2014wma}, can result from the conventional physics; others, involving more exotic mechanisms, assume entropy production via string creation~\cite{Myung:1986jt} or the GUT phase transitions~\cite{Cheng:1991uu}. The~hydrodynamical description of dissipative effects is summarized in Appendix \ref{App:B}.

Let us now follow the evolution of the early universe in terms of the EoS including bulk viscosity $\zeta$ defined in Equation~(\ref{eq:ent2}). In~Ref.~\cite{Tawfik:2019qyd}, data from non-perturbative~\cite{Borsanyi:2016ksw} and perturbative~\cite{Laine:2006cp, Laine:2015kra} SM simulations were used to study behavior $\zeta$ over a wide range of temperatures $T$, entropy densities $s$ and energy densities $\varepsilon$. It was found that $\zeta/(T s)$ decreases exponentially with $T$ increasing. The~bulk viscosity dependence on the energy density  at zero baryon chemical potential $\mu_B=0$ is displayed on the right panel of Figure~\ref{fig:eos}. Looking first on the QCD contributions only, it is apparent that the non-monotonic dependence of $\zeta(\epsilon)$ can be divided into four regions. The~first, spanning $\epsilon \lesssim 100~$GeV/fm$^3$, corresponds to the hadron-QGP phase. The~second region, up~to $\sim$$5\times 10^7~$GeV/fm$^3$, contains both the QCD and the EW phases of matter. The~third one seems to form an asymmetric parabola with focus at the critical energy density of the universe, $\rho_c\simeq 10^{12}~$GeV/fm$^3$ \cite{Tawfik:2019qyd}. The~fourth region shows a rapid increase in $\epsilon$ emerging from a non-continuous~point.  

In the SM contributions, as shown in the~upper curve of Figure~\ref{fig:eos}, besides~gluons and ($2+1+1+1$) quarks, the~contributions of the gauge bosons: photons, $W^{\pm}$, and~$Z^0$, the~charged leptons: neutrino, electron, muon, and~tau, and~the Higgs bosons: scalar Higgs particle, were also taken into account. An~overall conclusion drawn in Ref.~\cite{Tawfik:2019qyd} is that over the entire range of energy densities, the SM contributions are very significant. It is worth mentioning that the characteristic structures observed with the QCD contributions only are almost removed when adding also the EW~contributions. 

It is worth noting that the bulk viscosity could be important even outside the realm of the Hot Big Bang. In~Ref.~\cite{Brevik:2017msy}, the theory of the inflationary epoch, covering cold and warm inflation, as~well as the models of late universe expansion in the presence of bulk viscosity, were analyzed. Assuming that the viscous effects during the inflationary epoch can be represented by a generalized and inhomogeneous EoS of the form:
\begin{equation}
 p=-\epsilon +A \epsilon^{\beta}+\zeta(H) =-\epsilon +A \epsilon^{\beta}+\bar{\zeta}\left(\frac{8\pi G \epsilon}{3}\right)^{\gamma/2} \,,
 \label{viscinf}
\end{equation}
where $A,\beta, \bar{\zeta}, ~$and$~\gamma$ are positive constants and $\zeta(H)=\bar{\zeta}H^{\gamma}$ is the Hubble parameter-dependent bulk viscosity, the authors have studied the behavior of various inflationary observables. Let us turn to important implications of non-perturbative QCD vacuum dynamics in cosmological~evolution.

\subsection{Theory of Hot Meson Plasma Interacting with the QCD~Vacuum}
\label{sec:hot-meson-plasma}

Shortly after the confinement phase transition, the~universe enters the state of hot meson plasma whose thermal evolution has been thoroughly explored in the framework of the Linear Sigma Model (L$\sigma$M) in Ref.~\cite{Prokhorov:2017nst}. This analysis exploits the non-perturbative method of a generating functional derived from the effective Lagrangian at finite temperatures (for further references on this method, see Refs.~\cite{Carter:1996rf,Carter:1998ti,Mocsy:2004ab,Bowman:2008kc}) and accounting for the quartic self-interactions of the $\sigma$-meson only. The~latter approximation corresponds to a realistic well-motivated configuration of the ``hadron gas'' interacting with the non-linear $\sigma$-field and reproduces some of the basic thermodynamical characteristics of the hot meson plasma observed also in other approaches, in~particular, in~L$\sigma$M-based scenarios~\cite{Mocsy:2004ab,Chen:2013oya} and Polyakov-loop extended Nambu–Jona-Lasinio (PNJL) models~\cite{Fukushima:2003fw,Buballa:2003qv,Blaschke:2014zsa}, but~also features additional properties such as a possibility for $\sigma\to \pi\pi$ decays in the plasma above the critical~temperature.

The effective L$\sigma$M chiral Lagrangian accounting for the lightest scalar and pseudoscalar degrees of freedom $\pi^\pm$, $\pi^0$, $K^\pm$, $K^0$, 
$\bar{K}^0$, $\eta$, $\eta'$, and~the $\sigma$-meson, in~the hadron gas approximation at $T=0$, reads~\cite{Prokhorov:2017nst},
\begin{align}
\mathcal{L}_{\rm eff}
= &
\frac{1}{2}\partial_{\mu}\sigma\partial^{\mu}\sigma
+ 2g^2 v_0^2 \sigma^2 - g^4 \sigma^4 + \nonumber
\\ &
\frac{1}{2}(
  \partial_{\mu}\pi_{\alpha}\partial^{\mu}\pi_{\alpha}
  + \partial_{\mu}\eta\partial^{\mu}\eta
  + \partial_{\mu}\eta'\partial^{\mu}\eta'
)
+ \partial_{\mu}\bar{K} \partial^{\mu}K - \label{L-gas} \nonumber
\\ &
\frac{1}{2}\Big[
  2\kappa g^2 (m_u+m_d)\sigma^2 \pi_{\alpha} \pi_{\alpha}
  +\frac{2}{3}\kappa g^2 (m_u+m_d+4m_s)\sigma^2 \eta^2 + \nonumber
  \\ & \quad
  \frac{4}{3}\kappa g^2 (m_u+m_d+m_s+\Lambda_{\rm an})\sigma^2 \eta'^2
\Big]
-
\kappa g^2 (m_u+m_d+2m_s)\sigma^2 \bar{K}K \,,
\end{align}
where $m_{u,d,s}$ are the constituent up, down and strange quark masses, respectively, $g$ is the $\sigma$ quartic coupling, and $\Lambda_{\rm an}\simeq 0.5$ GeV is the gluon anomaly term that provides an explicit breaking of \mbox{U(1)$_{\rm L}$ $\times$ U(1)$_{\rm R}$} symmetry. The~QCD order parameter of the hadron mater $v_0=265\pm 15\, \rm{MeV}$ represents the amplitude of the quark-gluon (quantum-topological) condensate, such that
\begin{eqnarray}
\epsilon_{\rm top}(T=0)=-\Big(\frac{b}{32}+\frac{(m_u+m_d+m_s)l_g}{4}\Big)
\langle 0 |\frac{\alpha_s}{\pi}G^{a}_{\mu\nu}G_{a}^{\mu\nu}|0\rangle\equiv -v_0^4 \,,
\label{eps-QCD-vac}
\end{eqnarray}
given in terms of the gluon correlation length $l_g \simeq (1.2$ GeV$)^{-1}$ \cite{Shifman:1978bx,Schafer:1996wv}, and~the coefficient of the one-loop $\beta$-function of three-flavor QCD, $b=9$. The~quark-gluon condensate contributes together with the perturbative hadronic vacuum $\epsilon^{\rm had}_{\rm vac}$ emerging due to regularized contributions from meson fluctuations to the net QCD ground-state energy density,
\begin{eqnarray}
\label{QCD-vac-eps}
\epsilon^{\rm QCD}_{\rm vac} &\equiv& \epsilon(T=0) = \epsilon_{\rm top}+\epsilon^{\rm had}_{\rm vac} = - v_0^4 - 
\frac{1}{128\pi^2}\big(m_{\sigma{\rm (vac)}}^4 + 3m_{\pi{\rm (vac)}}^4 \nonumber \\
&+& m_{\eta{\rm (vac)}}^4+4m_{K{\rm (vac)}}^4 + m_{\eta'{\rm (vac)}}^4\big)\simeq -7\times 10^9\,{\rm MeV}^4\,.
\end{eqnarray}
that satisfies the vacuum equation of state in the zero-temperature limit, $\epsilon(T=0)=-p(T=0)$. The~hadronic vacuum term is negative $\epsilon^{\rm had}_{\rm vac}<0$ and appears to have a relatively small magnitude, i.e.,~$\epsilon^{\rm had}_{\rm vac}/\epsilon^{\rm QCD}_{\rm vac}\approx 0.15$ \cite{Prokhorov:2017nst}. The~current $u,d,s$ quark masses break the global chiral \mbox{SU(3)$_{\rm L}$ $\times$ SU(3)$_{\rm R}$} symmetry explicitly, while it is also broken spontaneously by means of a $\sigma$-field expectation value,
\begin{eqnarray}
\sigma = \langle \sigma \rangle + \tilde{\sigma}\,, \qquad \langle \sigma \rangle \equiv \frac{v}{g} \,,
\label{sigma}
\end{eqnarray}

In order to generalize the effective Lagrangian approach to finite temperatures, following theoretical foundations laid out in, e.g.,~Refs.~\cite{Cornwall:1974vz,Norton:1974bm,Amelino-Camelia:1992qfe,Carter:1998ti}, one should resume the daisy and superdaisy contributions. This is effectively achieved by utilizing an approximation that the expectation values of different fields are independent of each other, while omitting odd-point correlation functions and factorizing the four-point correlation functions into a product of two-point ones, for~instance, \mbox{$\langle \eta^2 \tilde{\sigma}^2 \rangle = \langle \eta^2 \rangle \langle \tilde{\sigma}^2 \rangle$}, etc. Then, as~a result of the minimization procedure of the non-equilibrium vacuum potential, one obtains the equation of state for the $\sigma$-condensate,
\begin{eqnarray}
v^2&=&v_0^2-3g^2\langle\tilde{\sigma}^2\rangle - 
\frac12\kappa(m_u+m_d)\langle\pi_{\alpha} \pi_{\alpha}\rangle \nonumber \\
&-& \frac{1}{6}\kappa (m_u+m_d+4m_s)\langle\eta^2\rangle - 
\frac{1}{3}\kappa (m_u+m_d+m_s+\Lambda_{\rm an})\langle\eta'^2\rangle \nonumber \\
&-&\frac12\kappa (m_u+m_d+2m_s)\langle\bar{K}K\rangle \,, \label{eq-v}
\end{eqnarray}
as well as the equations of motion for the (pseudo)scalar field fluctuations about the evolving non-trivial ground state, for~example, $\partial_\mu\partial^\mu\tilde{\sigma}+m_\sigma^2\tilde{\sigma}=0$, etc. Those fluctuations after quantization correspond to physical mesons with masses $m_\sigma^2=8g^2v^2$, etc. that are, in~general, dependent on temperature. The~vacuum values of the pseudo-scalar pseudo-Goldstone meson masses are found in terms of the light quark condensates via the Gell-Mann--Oakes--Renner relation~\cite{Gell-Mann:1968hlm,Ioffe:1981kw,Reinders:1984sr}, while the $\sigma$-meson has been identified phenomenologically with the scalar $f_0(500)$ state with mass, \mbox{$m_{\sigma\rm{(vac)}}\simeq 400-500$~MeV}. 

The generating functional in the case of zeroth chemical potential is the free energy density of the considered meson plasma found in terms of the spatial part of the energy-momentum tensor as follows,
\begin{eqnarray}\nonumber
&& \mathcal{F}(T,v,m_{\sigma}^2,\mathcal{M}^2) \equiv \frac{1}{3}\langle\mathcal{T}_{i}^{i}\rangle\,, \label{F3}
\end{eqnarray}
whose minimization over the independent variables $m_{\sigma}^2$ and $\mathcal{M}^2\equiv v^2+g^2\langle\tilde{\sigma}^2\rangle$ provides the physical meson masses and the equation of state for the condensate $v^2$. Substituting the meson masses expressed in terms of the temperature $T$ and the order parameter $v$ into $\mathcal{F}$, one obtains the non-equilibrium Landau functional,
\begin{eqnarray}
\mathcal{F}_{\rm NE}(T,v)\equiv \mathcal{F}(T,v,m_{\sigma}(T,v),\mathcal{M}(T,v))\,,
\label{FTv}
\end{eqnarray}
that provides the critical temperature of the chiral phase transition, 
where the finite-condensate phase becomes unstable, by~means of
\begin{eqnarray}
\label{Tc}
\frac{d^2\mathcal{F}_{\rm NE}}{dv^2}\Big|_{T=T_c} = 0 \,,
\end{eqnarray}
while the stability condition of the low-symmetry phase reads $d^2\mathcal{F}_{\rm NE}/dv^2\geq 0$. Finally, resolving the order parameter as a function of temperature, i.e.,~$v=v(T)$, one finds the so-called equilibrium Landau functional as,
\begin{eqnarray}
\mathcal{F}_{\rm E}(T)\equiv \mathcal{F}_{\rm NE}(T,v(T))\,,
\label{FT}
\end{eqnarray}
that matches the usual free energy definition. A~variety of thermodynamic observables of the meson plasma are then derived in terms of $\mathcal{F}_{\rm E}(T)$ such as pressure, entropy density, energy density, heat capacity and the speed of sound squared,
\begin{adjustwidth}{-\extralength}{0cm}
\begin{eqnarray}
p(T)=-\mathcal{F}_{\rm E}(T) \,, \quad \sigma(T)=-\frac{d}{dT}\mathcal{F}_{\rm E}(T)\,, \quad \epsilon=\mathcal{F}_{\rm E}+T\sigma \,, \quad
c_V=T\frac{d\sigma}{dT}=\frac{d\epsilon}{dT}\,, \quad
u^2=\frac{\sigma}{c_V} \,,
\label{quantaties}
\end{eqnarray}
\end{adjustwidth}
respectively. While at $T=0$, the QCD vacuum equation of state, $\epsilon=-p$, is satisfied, both pressure and energy density grow with $T$ due to positive particle contributions. The~total energy density of the ``plasma + condensate'' system vanishes at $T_{\epsilon=0}\simeq 237$ MeV for $m_{\sigma\rm{(vac)}}\simeq 500$ MeV. 

Using the above formalism, in~Ref.~\cite{Prokhorov:2017nst}, the properties of different phases and transitions between them are explored in detail. For~instance, at~temperatures $T>T_c$, hadrons deconfine into quarks and gluons, while the condensates melts away, yielding a deconfined (or zero-condensate) phase of the QCD matter. Such a phase becomes metastable at temperature $T_0 < T_c$, when the $\sigma$-meson fluctuation becomes massless $m_\sigma(T_0)=0$, while $v(T_0)=0$ is still valid, and~$T_0$ is then found by resolving the extremum conditions on the generating functional. At~another $T_1$, two minima of $F_{\rm NE}(T,v)$ become equal such that the zero-condensate phase stabilizes and the corresponding temperature is found from the following equation, $F_{\rm NE}(T_1,0)=F_{\rm E}(T_1)$. The~first-order chiral phase transition to the zero-condensate phase then occurs at some temperature between $T_1$ and $T_c$, and~its strength increases with the $\sigma$-meson mass in the vacuum. Note, as~expected from the first-order nature of this transition, the entropy density of the meson plasma grows with $T$ and remains finite at all values of $T$, while the heat capacity appears to have a singularity, and the speed of sound squared vanishes at the critical temperature $T=T_c$.

\begin{figure}[H]
{\includegraphics[width=0.6\textwidth]{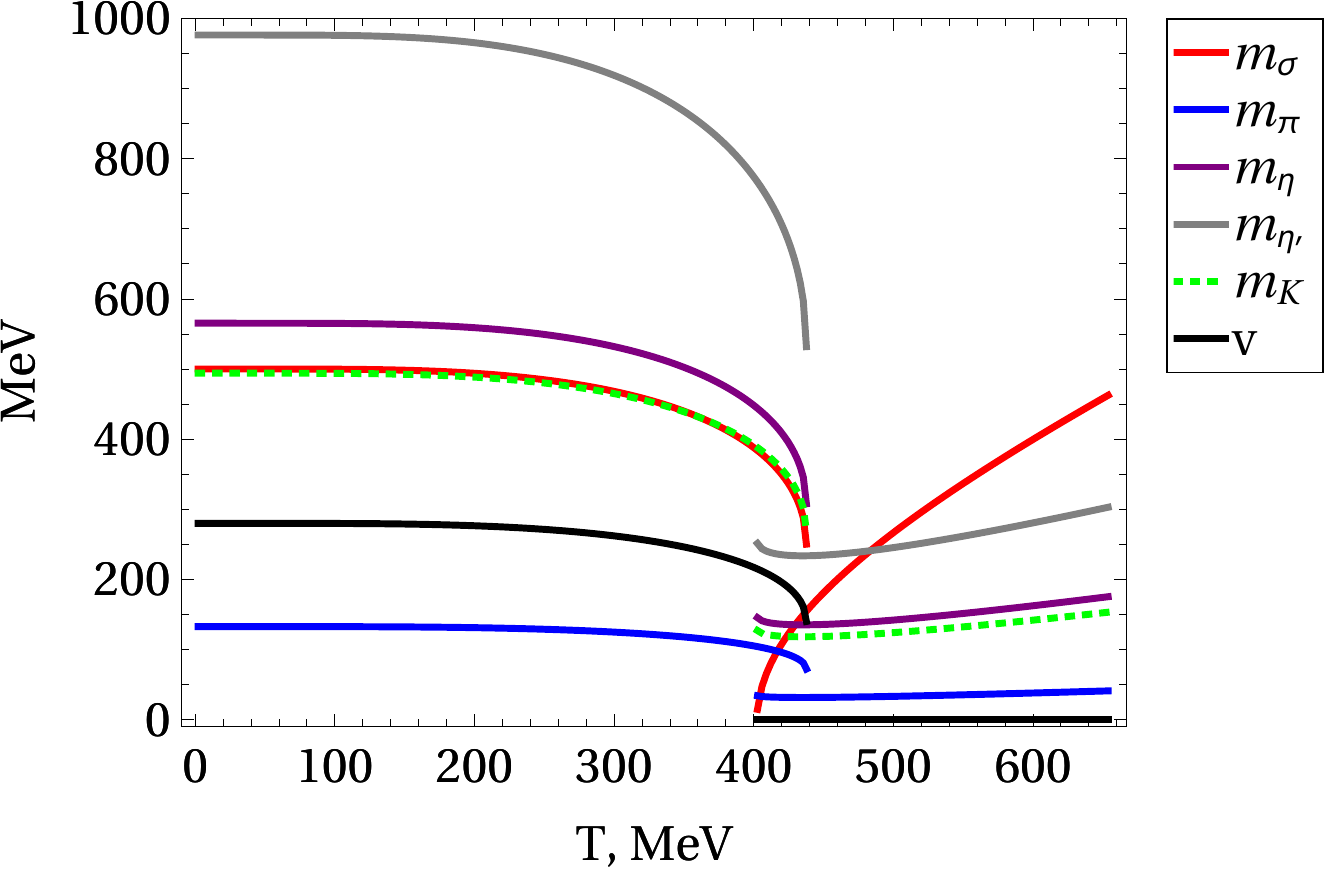}}
\caption{The condensate and meson masses as function of $T$. Here, $m_{\sigma{\rm (vac)}}=500\,{\rm MeV}$, $g^2=0.4$, $T_c = 438\,{\rm MeV}$, $T_0 = 402\,{\rm MeV}$, $T_1 = 430\,{\rm MeV}$.}
\label{fig:meson-masses}
\end{figure}

\begin{figure}[H]
\begin{minipage}{0.325\textwidth}
 \centerline{\includegraphics[width=1\textwidth]{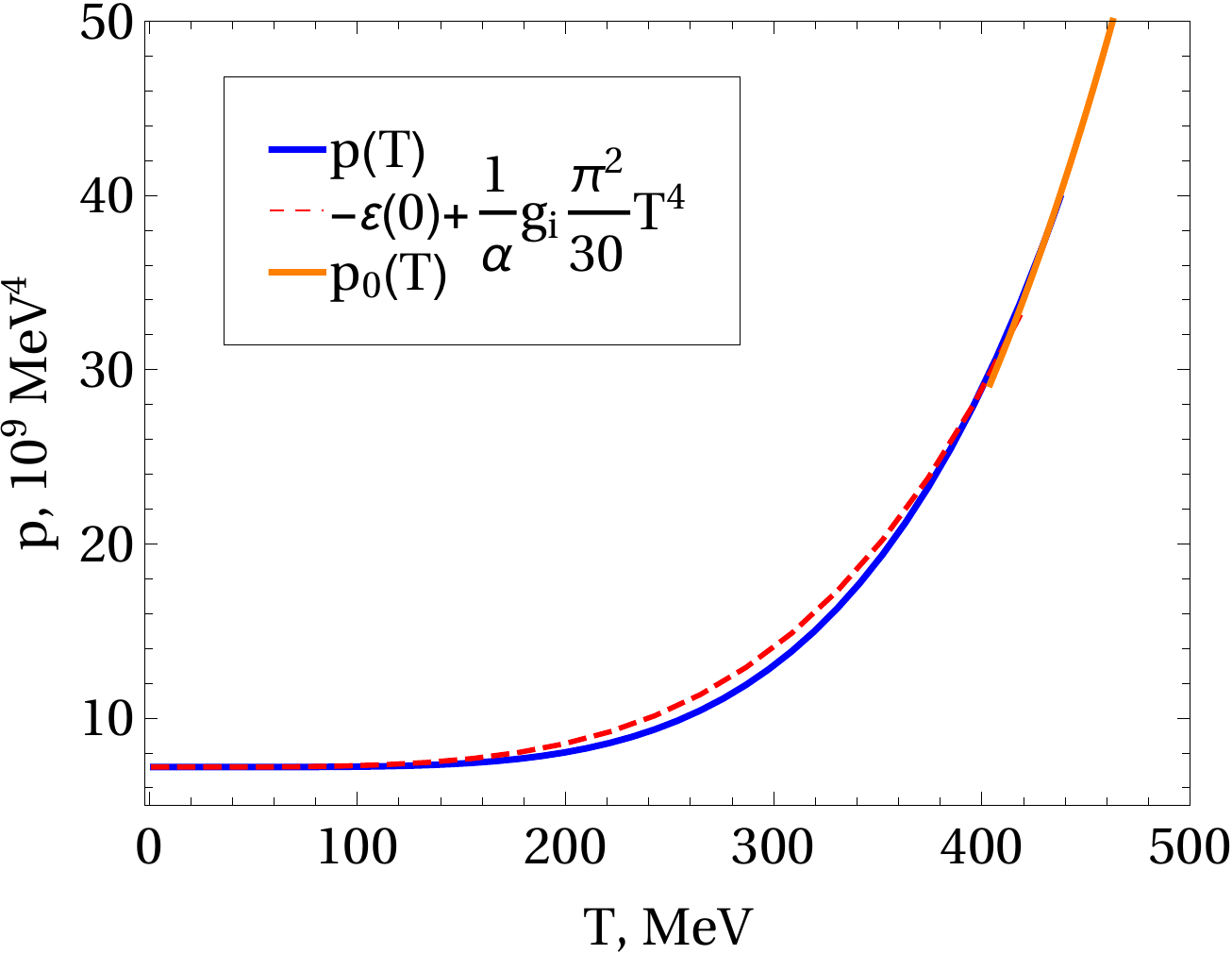}}
\end{minipage}
\begin{minipage}{0.325\textwidth}
 \centerline{\includegraphics[width=1\textwidth]{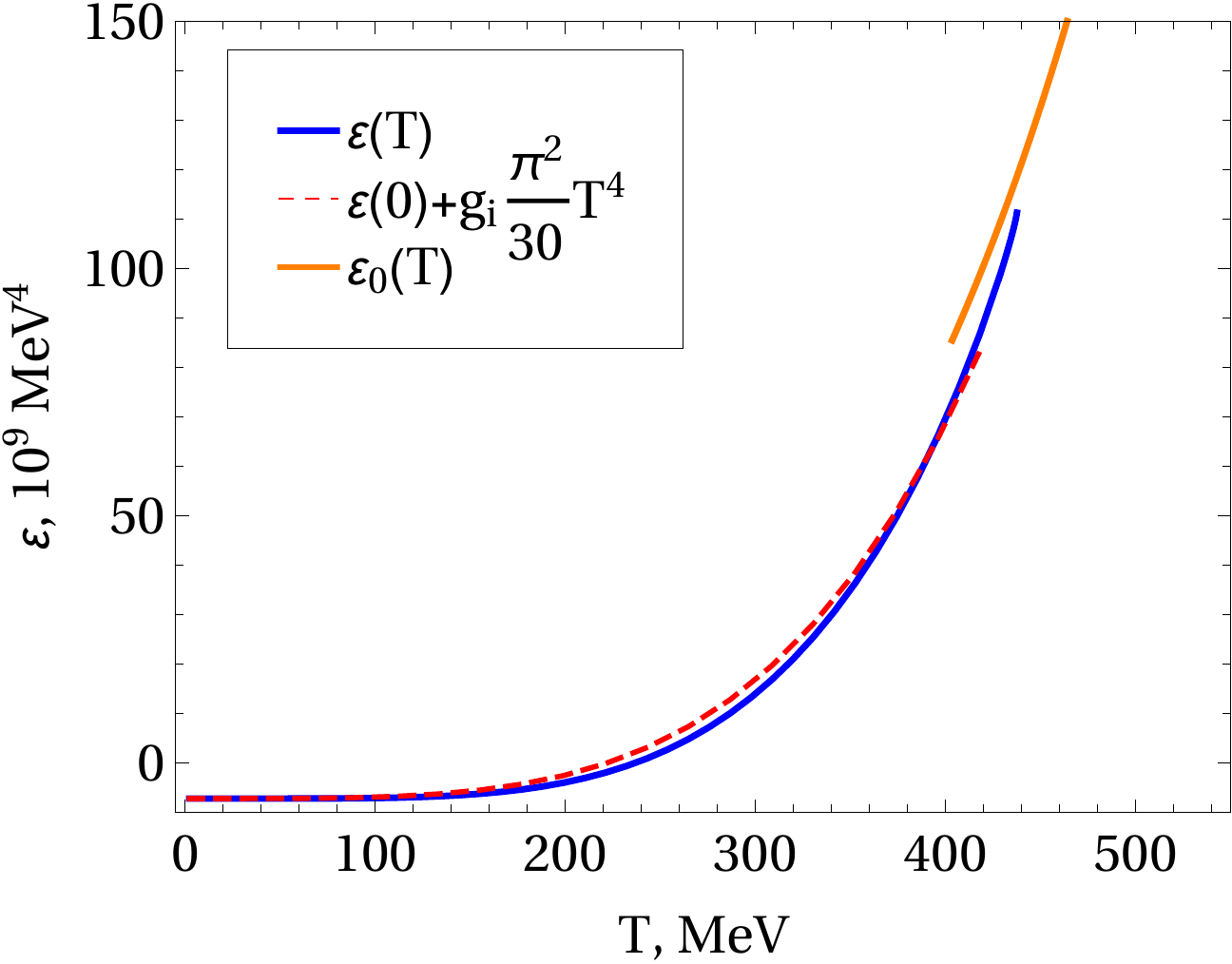}}
\end{minipage}
\begin{minipage}{0.325\textwidth}
 \centerline{\includegraphics[width=1\textwidth]{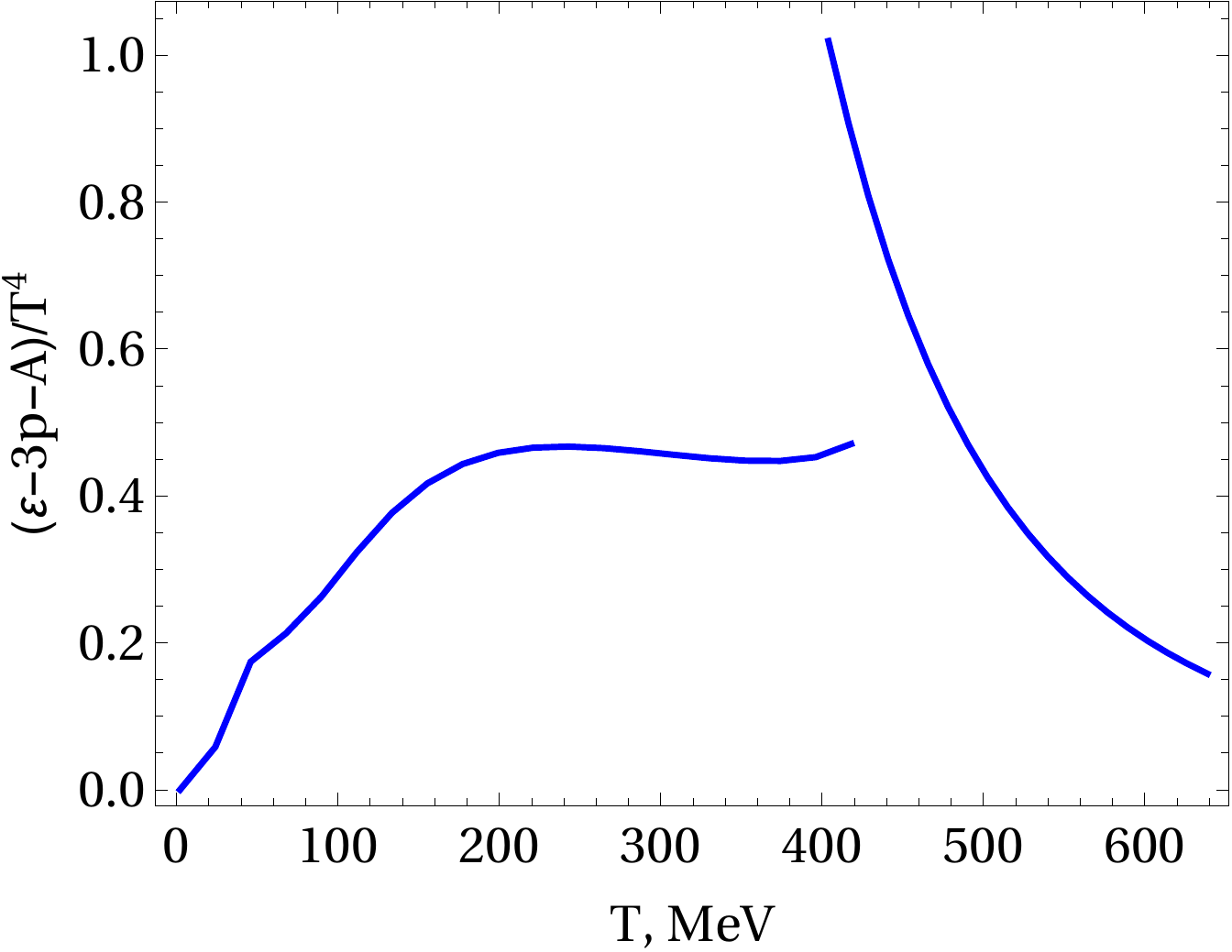}}
\end{minipage}
\caption{(\textbf{Left}) Pressure $p(T)$ as a function of temperature for the finite-condensate $v(T)\not = 0$ phase compared to that in the zero-condensate $v(T) = 0$ phase, $p_0(T)$ (solid lines), and~to the approximated result (dashed line).
(\textbf{Middle}) The same but for the energy density.
(\textbf{Right}) The normalized EoS $(\epsilon(T)-3p(T)-A)/T^4$ as a function of temperature, where $A\equiv \epsilon(T=0)-3p(T=0)$ is the net vacuum contribution.}
\label{fig:p-eps-EoS}
\end{figure}

The thermal evolution of the meson mass spectrum and the condensate of the L$\sigma$M is shown in Figure~\ref{fig:meson-masses}. Interestingly enough, both the condensate and the masses of all the mesons decrease with temperature for $T < T_c = 438\,{\rm MeV}$. The~critical temperature becomes reduced if the fermions (quarks and baryons) are introduced~\cite{Mocsy:2004ab}. As~a result of the ``hadron gas'' approximation, the~$\sigma$-mass rapidly falls to zero at $T_0=402$ MeV and then grows much faster than the masses of other mesons (such that $m_\sigma>2m_\pi$ almost for any values of $T$) in contrast with the corresponding predictions of other existing approaches such as PNJL. As~a result, at~low temperatures, the hadronic plasma is dominated by pions. There is also a significant phase co-existence domain of size $(T_c-T_0)/T_c\simeq 0.1$. Pressure $p(T)$, energy density $\epsilon(T)$ and the EoS $(\epsilon(T)-3p(T)-A)/T^4$, where $A=\epsilon(T=0)-3p(T=0)$ is the net vacuum contribution, are shown in Figure~\ref{fig:p-eps-EoS} from left to right, respectively. Due to the positive ``hadron gas'' contribution, both $p(T)$ and $\epsilon(T)$ rise with temperature. Their profiles can be approximately reconstructed as a sum of the negatively-definite constant QCD vacuum term $\epsilon(T=0)$ and the contribution of the relativistic hadron plasma $\propto T^4$ (dashed lines), with~the numerical coefficient $\alpha\simeq 3.5$ and the effective number of DoFs, $g_i\simeq 9$.
\unskip

\subsection{Cosmological Constant and Vacuum~Catastrophe}
\label{sec:QCD-vac-CC}

The tight observational constraint on the DE EoS
\begin{eqnarray}
w_{\rm DE}=-1.03 \pm 0.03 \,, \label{w-DE}
\end{eqnarray}
comes through a combination of the cosmological data from various sources such as the Type Ia supernovae, the~baryon acoustic oscillations, the~CMB anisotropies, and the~weak gravitational lensing, etc.; for details on the confidence level and datasets, see Ref.~\cite{Planck:2018vyg}. These constraints are consistent with the standard cosmological model known as the $\Lambda$CDM ($\Lambda$ term plus DM in the form of CDM, as~two dominant components of the universe). Specifically, the~DE is considered to be in the form of Cosmological Constant or~$\Lambda$-term density,
\begin{equation}
    \epsilon_{\rm DE} = \epsilon_{\Lambda} \,, \qquad \epsilon_{\Lambda}\equiv \frac{\Lambda}{\kappa}\,, \qquad \kappa = 8\pi G \,,
\end{equation}
expressed in terms of $\Lambda$-term in conventional normalization, $\Lambda$; see Equation~(\ref{eq:Eeq}). The~latter satisfies the EoS $w_{\rm DE}=-1$ exactly. For~a detailed recent review on achievements and challenges of the $\Lambda$CDM, see, e.g.,~Ref.~\cite{Bull:2015stt}.

Classically, an~arbitrary $\Lambda$-term density $\epsilon_0$ can be readily added to the right-hand side of the Einstein equations of GR that determine the macroscopic evolution of the universe,
\begin{eqnarray}
R_{\mu\nu} - \frac{1}{2} g_{\mu\nu}R=\kappa(\epsilon_0 g_{\mu\nu} + T_{\mu\nu})\,, \qquad
T_{\mu\nu} = - \frac{2}{\sqrt{-g}}\frac{\delta S_m}{\delta g^{\mu\nu}} \,, \qquad 
S_m=S_m[\phi,\psi,A_{\mu},g_{\mu\nu}] \,,
\end{eqnarray}
in terms of action of matter fields $S_m$ and~their energy-momentum tensor $T_{\mu\nu}$. In~quantum theory, a~non-trivial contribution to the ground-state energy density emerges as an average of the energy-momentum tensor over the Heisenberg vacuum state~\cite{Sakharov:1966aja,Sakharov:1967pk}
\begin{eqnarray}
\langle 0 |T_{\mu\nu}|0\rangle = \epsilon_{\rm vac} g_{\mu\nu}\,, \qquad \epsilon_{\rm vac} \not=0 \,.
\end{eqnarray}

The latter is proportional to the trace of the energy-momentum tensor; hence, it represents an effect of conformal symmetry breaking in a given fundamental QFT through either the formation of a Bose-Einstein condensate in a massless theory or through nonzero mass-dimensional terms in the original Lagrangian. It is generically ill-defined and should be renormalized, with~the classical (``bare'') $\epsilon_0$ being treated as a counter-term in the initial Lagrangian, such that the divergences are cancelled between the two yielding a finite, but~renormalization scale $\mu$ dependent, vacuum energy density, $\epsilon_{\Lambda}(\mu)$. This is the physical vacuum energy density that emerges in cosmological measurements performed at some fixed scale $\mu=\mu_{IR}$ in the present universe, such that
\begin{eqnarray} \label{vac-eos}
\epsilon_{\Lambda}(\mu_{IR})\equiv \epsilon_0+\epsilon_{\rm vac}\,.
\end{eqnarray}

A macroscopic Cosmological Constant effect causing the universe to expand with acceleration (de-Sitter phase) is usually identified with energy density of the quantum vacuum that acquires contributions from all the incident vacuum subsystems existing in the SM and beyond. These would correspond to all quantum fields existing at energy scales ranging from the quantum gravity (Planck) scale, $M_{\rm PL}\sim 10^{19}$ GeV, down to the QCD confinement scale, $M_{\rm QCD} \sim 1$ GeV --- the maximal and minimal energy scales of particle physics, respectively. The~current vacuum state of the universe is considered to be produced in the aftermath of the latest QCD phase transition associated with hadronization of the cosmological~plasma.

In the framework of SM, besides~the zero-point energy contributions to the ground state coming from each elementary particle, there are two major vacuum condensates whose characteristics are well established in particle physics --- the weakly coupled classical Higgs condensate responsible for spontaneous EW symmetry breaking giving masses to the SM vector bosons and fermions and~the strongly-coupled quantum-topological quark-gluon condensate in~QCD. 

One of the important aspects of the cosmological QCD transition epoch concerns the formation of the negatively-definite (CM) contribution to the ground-state of the universe that has received little attention in the literature so far. For~illustration of this effect, let us consider the conformal anomaly term in the trace of the effective QCD energy-momentum tensor~\cite{Crewther:1972kn,Chanowitz:1972da,Collins:1976yq},
\begin{equation}
\label{traceA}
T_{\mu}^{\mu,{\rm QCD}}=\frac{\beta(g_s^2)}{2}F_{\mu\nu}^aF^{\mu\nu}_a+\sum_{q=u,d,s}m_q\bar{q}q\,,
\end{equation}
where $m_q$ are the light (sea) quark masses $q=u,d,s$, $g_s$ and $\beta$ are the QCD coupling constant and the $\beta$-function, respectively, and~$F_{\mu\nu}^a$ is the gluon field stress tensor. Taking the vacuum average, we obtain
\begin{adjustwidth}{-\extralength}{0cm}
\begin{eqnarray}
\langle
0|T_{\mu}^{\mu,{\rm QCD}}|0\rangle &=&-\frac{9}{32}\langle0|:\frac{\alpha_{\rm S}}{\pi}F^a_{\mu\nu}(x)
F_a^{\mu\nu}(x):|0\rangle + \frac14 \Big[\langle0 | :m_u\bar uu: | 0\rangle + \langle0 | :m_d\bar dd: |0\rangle \nonumber \\
&+&
\langle0|:m_s\bar ss:|0\rangle\Big] \simeq -(5\pm 1)\times 10^{-3}\; \text{GeV}^4\,, \qquad \alpha_{\rm S}=\frac{g_s^2}{4\pi} \,,
 \label{Lambda-QCD}
\end{eqnarray}
\end{adjustwidth}
representing the maximal value of the averaged quantum-topological QCD contribution to the physical vacuum energy density, whose spacetime dynamics is not fully understood and yet to be established. This contribution, also known as the quark-gluon condensate, is predicted by the theory of QCD instantons~\cite{Shifman:1978bx,Schafer:1996wv} and plays an important role in the chiral symmetry breaking and in dynamics of color confinement as well as in generation of mass of light mesons in hadron physics as suggested by the Gell-Mann--Oakes--Renner relation~\cite{Gell-Mann:1968hlm,Ioffe:1981kw,Reinders:1984sr}.

The so-called ``Vacuum Catastrophe'' reflects the fundamental problem of consistent matching between the macroscopic observable $\epsilon_{\Lambda}$ value, close to the critical energy density of the universe $\rho_{c}$,
\begin{eqnarray}
\epsilon_{\Lambda}\simeq 0.7 \rho_{c} \simeq 2.5 \times 10^{-47}\, 
{\rm GeV}^4 > 0\,, \qquad \rho_{c}\equiv \frac{3H_0^2}{\kappa} \,,
\label{Lcosm}
\end{eqnarray}
and the characteristic sizes of microscopic QFT predictions for the energy scale of each separate vacuum condensate~\cite{Martin:2012bt,Sola:2013gha}. Indeed, considering the topological QCD vacuum energy density $\epsilon_{\rm vac}^{\rm QCD}$ alone at macroscopic time and space separations typical for cosmological measurements, see Equation~(\ref{Lambda-QCD}), its value is off by over forty orders of magnitude and has a wrong sign compared the observable $\epsilon_{\Lambda}$. 

Indeed, such a large and negative contribution to the vacuum density creates a big problem for existence of the spatially flat universe, as the right-hand side of the corresponding Friedmann equation must be positive at all times (see, e.g.,~Refs.~\cite{Bull:2015stt,Pasechnik:2013sga,Pasechnik:2016sbh,Pasechnik:2016twe}). The~presence of a negative cosmological constant of the QCD scale would necessarily trigger a fast collapse of the universe at the time scale of a microsecond, as the other components of the cosmological plasma energy-density decay as $\propto 1/a^n$ (with $n=4$ for relativistic and $n=3$ for non-relativistic media). This would prevent the universe from traversing the QCD horizon scale such that no macroscopic evolution would be possible. For~a recent review on the status of this problem and the existing approaches, see, e.g.,~Ref.~\cite{Pasechnik:2016sbh} and references~therein. 

A consistent resolution of the so-called ``old'' Cosmological Constant problem (why is $\epsilon_{\Lambda}$ small and positive?) and the ``new'' Cosmological Constant problem (why is $\epsilon_{\Lambda}$ non-zeroth and exists at all?) \cite{Bull:2015stt} may require a dynamical mechanism for compensation of different short-distance vacuum configurations in the infrared limit of the corresponding QFT. Such a vacuum self-alignment in the non-perturbative regime would be desired in order to avoid a major fine tuning of different parameters of the fundamental theory~\cite{Polchinski:2006gy}, and it may be considered as a new physical phenomenon~\cite{Copeland:2006wr,Pasechnik:2013sga,Pasechnik:2013poa,Pasechnik:2016sbh}.

As has been pointed out in Ref.~\cite{Hofmann-book}, in~the thermal SU(2)/SU(3) YM theories, the~confining phase at low temperatures is expected to be void of energy density and pressure. Along these lines, one introduces a natural hypothesis about a heterogeneous structure of the non-perturbative QCD vacuum in the infrared limit of QCD~\cite{Pasechnik:2016sbh}. Such a structure is characterized by the presence of at least two distinct vacuum subsystems which contribute with opposite signs to the net QCD vacuum energy density and mutually eliminate each other on average at large space and time separations, $\Delta x\sim \Delta t \gg 1/\Lambda_{\rm QCD}$. Then, it is reasonable to assume that a phase transition in the QCD vacuum has occurred in the course of cosmological evolution. Such a transition has led to a dramatic drop in the net vacuum energy density due to an (almost) exact cancellation between the different vacuum subsystems in the IR limit of~QCD.

Generically, an~ultimate goal would be to develop a universal framework that consistently describes quantum vacua (condensates) dynamics in real cosmological time at both macroscopic (IR limit) and microscopic (UV limit) separations and~then identify the phase transitions between those. This remains one of the major unsolved problems of fundamental physics~\cite{Weinberg:1988cp,Wilczek:1983as} (see also Refs.~\cite{Sola:2013gha,Pasechnik:2016sbh}).

\begin{figure}[H]
\begin{minipage}{0.45\textwidth}
 \centerline{\includegraphics[width=1\textwidth]{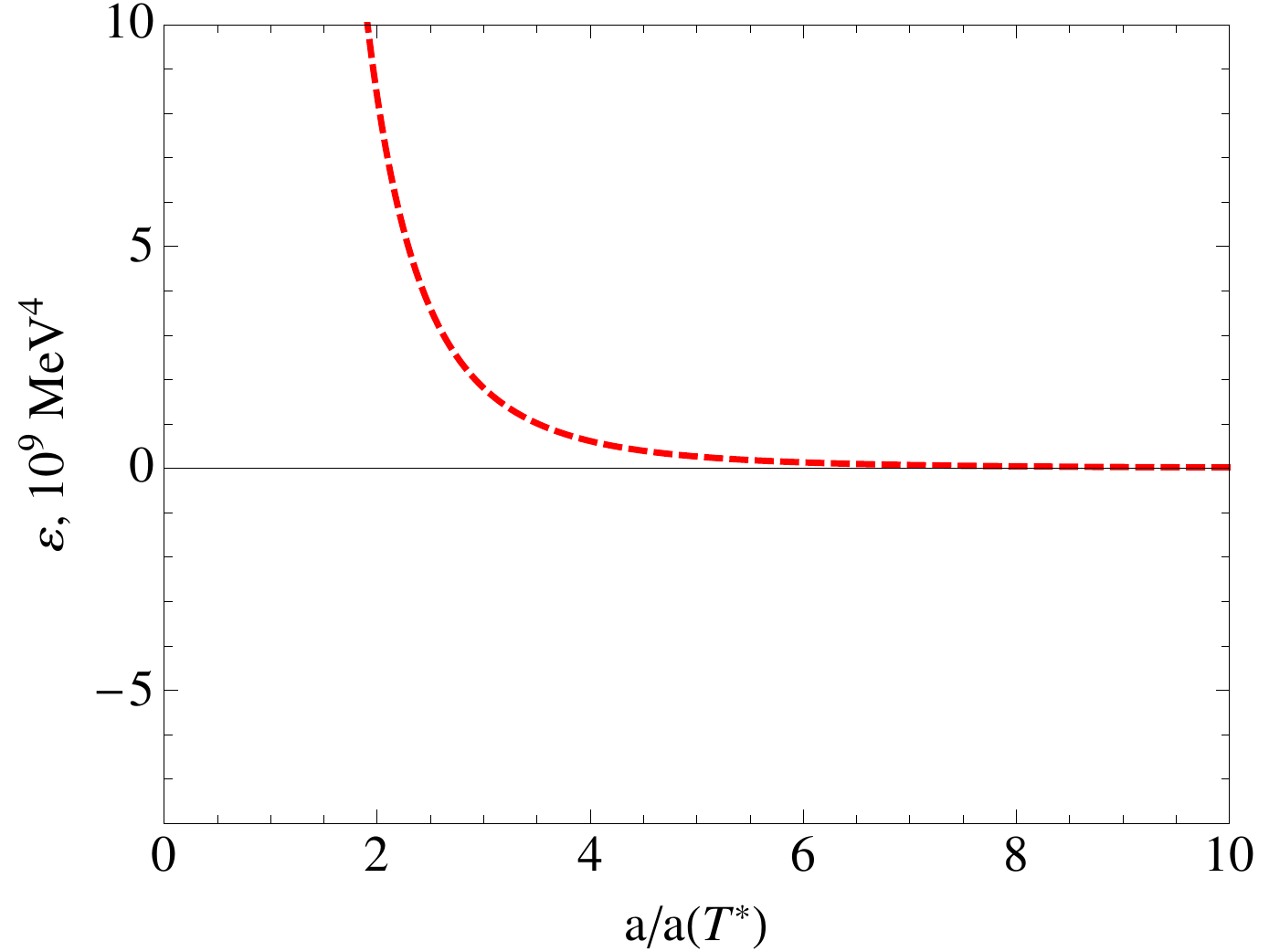}}
\end{minipage}
\begin{minipage}{0.45\textwidth}
 \centerline{\includegraphics[width=1\textwidth]{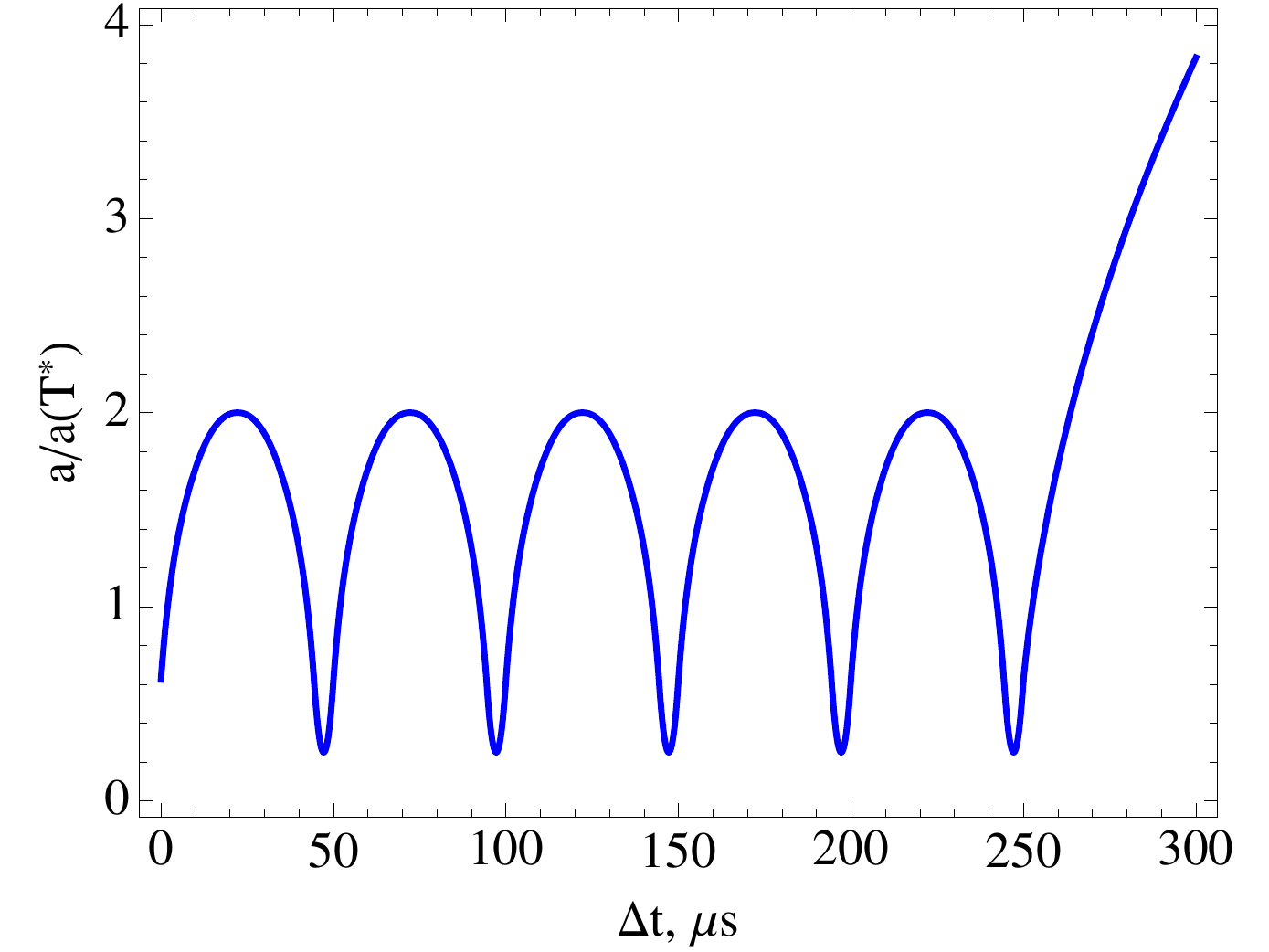}}
\end{minipage}
\caption{(\textbf{Left}) The QCD vacuum energy density as a function of the (normalized) scale factor $a$ in the meson plasma model (solid line) and the same quantity but with an extra positive $\Lambda$-term (dashed line) that exactly compensates the negative (topological) QCD term at large time-scales. (\textbf{Right}) A scenario of the universe oscillating during the QCD phase transition epoch with stochastic generation of a positively-definite QCD-scale $\Lambda$-term ``compensator''.}
\label{fig:oscU}
\end{figure}

Since the effect of the negative QCD condensate term must be eliminated somehow beyond the Fermi scale of QCD, Ref.~\cite{Prokhorov:2017nst} explores the simplest scenario for cosmological evolution by invoking an additional positively-definite cosmological constant in the framework of the effective meson plasma model, and~also using the Bag-like model of the QCD crossover transition~\cite{Ferroni:2008ej} for~comparison. The~basic working assumption adopted in this work was that a positive cosmological constant has been formed (stochastically) at the QCD scale together with the negative (topological) term, which means that the QCD vacua effects are dynamically eliminated at distances beyond the typical hadron scale. In~Figure~\ref{fig:oscU} (left), the~net QCD vacuum energy density $\epsilon$ is shown as a function of the normalized scale factor in both scenarios with and without an additional positive $\Lambda$-term ``compensator''. In~this scenario, as~the QCD vacuum evolves with temperature and the universe eventually collapses, a~``backward'' QCD transition from the meson plasma to QGP may occur above the critical QCD temperature. Provided that there exists a mechanism for a bounce from the singularity (for possible scenarios for such a bounce, see, e.g.,~Refs.~\cite{Novello:2008ra,Mukhanov:1991zn,Szydlowski:2005qb,Dabrowski:1995ae} and references therein), a~possible series of such sequential ``direct'' and ``backward'' QCD transitions implies that the universe may, in~principle, oscillate around the QCD epoch for some time. Eventually, the~negative QCD vacuum effect is eliminated, the universe enters the phase of unbound expansion, and the standard cosmological evolution takes off~\cite{Prokhorov:2017nst} (see Figure~\ref{fig:oscU}, right).

\section{Dynamics of Ground State in YM~Theories}
\label{sec:YMtheory}

Let us discuss the properties of quantized YM theories and their major implications in cosmology while focusing primarily on QCD-like strongly-coupled dynamics and its connections to confinement and to the nearly vanishing value of the cosmological~constant.

\subsection{YM Ground State as a Time~Crystal}
\label{sec:time-crystal}

While Gross, Wilczek and Politzer proved the asymptotic freedom of non-Abelian gauge theories at large momentum transfers~\cite{Gross:1973id, Gross:1973ju, Politzer:1973fx}, Savvidy showed that the perturbative QCD vacuum at zero field strength is unstable~\cite{Savvidy:1977as} and thereby demonstrated the existence of the vacuum condensate (see also Batalin, Matinyan and Savvidy~\cite{Batalin:1976uv}). Nielsen and Olesen worked out an argument for why the explanation of vacuum condensates in terms of a homogeneous field filling the vacuum is problematic: such a field mode is unstable, at~least on a static Minkowski background~\cite{Nielsen:1978rm}. The~ground state of QCD is a non-perturbative quantum-topological state of the YM theory that is of primary importance for the understanding of color confinement dynamics~\cite{Olesen:1980nz} as well as of hadronic and effective quark masses. For~a thorough discussion on the QCD vacuum and its implications, see, e.g.,~Ref.~\cite{Shuryak:1983ni} and references~therein.

Despite the possible theoretical frameworks that are available for the description of the quantum ground state in YM theories, an~interesting case of the formation of a metastable condensate in the Savvidy approach, with~gravitational back-reaction providing a stationary stabilization, was studied in Ref.~\cite{Addazi:2018ctp}. Intriguingly, the~authors showed that the relaxation process induced by the QCD phase transition provides a novel mechanism for the production of GWs in the early universe. Such production is enabled through the SSB of time translation invariance that is reminiscent of what happens in the time-crystals that were theoretically predicted by Wilczek in Refs.~\cite{Wilczek:2012jt, Wilczek:2013uca} and observed by the Monroe group~\cite{MonG} (for a detailed review, see Ref.~\cite{Sacha:2017fqe}). Within~the setting of early cosmology, as discussed in Ref.~\cite{Addazi:2018ctp} that included the perturbative all-order effective action, it was derived that the energy density of the quark-gluon mean-field decays monotonically in time, while the pressure density undergoes violent oscillations at the characteristic QCD scale. This mechanism entails the generation of a primordial multi-peaked GW signal that eventually is shifted into the radio frequencies' domain. If~detected, such a signal would represent an unprecedented echo of the QCD phase transition and it is, in~principle, observable through forthcoming measurements at the FAST and SKA~telescopes.

The scenario depicted in Ref.~\cite{Addazi:2018ctp} requires the emergence of a quark-gluon condensate characterized, as~mentioned above, by~violent oscillations of the pressure density. Such oscillations would be periodic in multiples of the inverse characteristic scale \mbox{$\Lambda_{\rm QCD}\simeq 0.1$~GeV.} A very similar result that is consistent with this analysis was derived in Ref.~\cite{Attems:2018gou} in which the authors deployed holography with the aim of analyzing relativistic collisions in a one-parameter family of strongly coupled gauge theories that were undergoing thermal phase transitions. An~oscillating behavior of the pressure density was discovered also in this latter work, and~again, the~period of the oscillations was found to be a multiple of \mbox{$\Lambda^{-1}_{\rm QCD}$}. It was concluded that out-of-equilibrium physics smoothes out the details of the~transition.

Most parts of analyses developed that involve lattice QCD tacitly assume an analytical continuation of the results obtained on the Euclidean space to the Minkowski spacetime. The~underlying argument, as~has been commonly advocated, relies on the idea that locally, on~any FLRW spacetime, the~dynamics of QCD can be studied on a ``frozen'' Minkowski-like background and, thus, results may be analytically continued to the Euclidean space. Relying on this assumption, a~notable theorem due to Maiani and Testa~\cite{Maiani:1990ca} showed that the scattering of asymptotic states can be counter-Wick rotated only in the infinite volume limit (with the scale being the threshold amplitude) and for time scales much smaller than the level spacing due to momentum~discretization.

Specifically, the~authors of~\cite{Maiani:1990ca} started out with the Osterwalder-Schrader theorem that ensures that the Euclidean correlation functions can be analytically continued back to the Minkowski spacetime. However, this theorem heavily relies on the so-called reflexion positivity condition~\cite{Glimm:1987ng}, and this condition is not fulfilled in the aforementioned cases when the FLRW dynamics are studied at time scales that exceed the Hubble time by one order of magnitude. Indeed, for~the cases discussed, for~instance in Refs.~\cite{Attems:2018gou,Addazi:2018ctp}, non-perturbative effects that are originating from violent oscillations of the pressure density definitely spoil the time reflexion positivity condition. This insight suggests that the Maiani-Testa theorem is inapplicable in the context of the current~discussion.

In fact, recent analyses developed according to different frameworks surprisingly confirm a quite different picture than the one on which the lattice QCD analyses are based: for instance, the~determined period of oscillations in~\cite{Attems:2018gou, Addazi:2018ctp} that exceeds the Hubble time by one order of magnitude (in Planck units), as~was mentioned above. Thus, effects of the spacetime curvature cannot be neglected due to the influence of non-trivial non-perturbative effects that are recovered beyond the characteristic-time of QCD, $\Lambda^{-1}_{\rm QCD}$. This provides an argument that should influence the confidence of the application of lattice QCD methods in cosmology; specifically when such methods aim to determine the order of the phase~transition.

Finally, we mention, as~a further approach to the problem of determining the order of the QCD phase transition in the early universe, a~method provided by conformal field theories prescribed by the foundation of the modern understanding of quantum field theory and particle physics (for a comprehensive overview of the basic concepts, see Ref.~\cite{Poland:2018epd}). Having been hitherto deepened so as to unveil the universal properties of scale invariant critical points, these frameworks were applied to describe continuous phase transitions in fluids and magnets as~well as in many other materials. Substantial efforts were devoted to the study of non-perturbative strongly coupled conformal field theories, especially concerning their symmetries and theoretical constraints. Such work has opened the pathway to the development of the so-called conformal bootstrap~\cite{Ferrara:1973yt,Polyakov:1974gs}, which has proven to be a successful framework in two dimensions and which finally was extended within the last decade so as to account for higher dimensionality including the physically relevant cases of three and four dimensions (for a detailed pedagogical review on the conformal bootstrap approach in $d$ dimensions, see, e.g.,~Refs.~\cite{Rychkov:2016iqz,Simmons-Duffin:2016gjk}). Notably, significant progress in analytical methods has been achieved, shedding light on the possibilities of how to cast the bootstrap equations and, in~parallel, more powerful numerical techniques were developed in attempts to find their solution~\cite{Rattazzi:2008pe}. This has brought about ground-breaking results including the determination of critical exponents and correlation function coefficients in the Ising \mbox{O($N$)} models in three dimensions~\cite{Kos:2016ysd}.

\subsection{Effective Action~Approach}
\label{sec:effectiveActionTheory}

In this section, the~effective action approach is outlined in order to provide a background to the discussion of the contribution to the vacuum energy density through the trace of the energy-momentum tensor (EMT). In~1977, Matinyan and Savvidy~\cite{Matinyan:1976mp} and Savvidy~\cite{Savvidy:1977as} investigated the asymptotic behavior of the effective Lagrangian density in gauge theories building on earlier work by Heisenberg and Euler and by Schwinger (see references therein). The~behavior was studied using RG methods in order to relate the effective picture of strong fields to the short-range properties of gauge theories. The~quantum corrections to the classical action as found by Schwinger were discussed both in these two publications and further summarized in a recent review~\cite{Savvidy:2019grj}.

The investigation begins with an examination of corrections to the classical action as suggested by Schwinger, i.e.,~corrections that allow for an expansion of the effective YM action in the gauge fields $\bar{A}_\mu^a$ (also known as connections):
\begin{align}
  \label{eq:YMeffectiveAction}
  \Gamma[A]
  & =
  \int \! \mathrm d x \, \mathcal L_\text{eff}
  \nonumber
  \\ & =
  \sum_n \int \! \mathrm d x_1 \cdots \mathrm d x_n \,
    \Gamma_{\phantom{(n) \, }\mu_1 \cdots \mu_n}^{(n) \, a_1 \cdots a_n} \,
    {\bar A}_{\mu_1}^{a_1}(x_1) \cdots {\bar A}_{\mu_n}^{a_n}(x_n)
  \nonumber
  \\ & =
  S_\text{cl} + W^{(1)} + W^{(2)} + \ldots \,.
\end{align}
Here, the~effective Lagrangian density $\mathcal{L}_\text{eff}$ undergoes a perturbative expansion so that the \mbox{$n$-loop} corrections provide a deviation from the classical action $S_\text{cl}$. The~charged vector connection \mbox{${\bar A}_\mu^a(x) \equiv \langle0\lvert A_\mu^a(x) \rvert0\rangle$} is the vacuum expectation value of the field operator and \mbox{$\Gamma^{(n)}$} is the one-particle irreducible (1PI) vertex function. To~each order, \mbox{$W^{(n)}$} provides the $n$-loop correction to the classical~action.

The effective Lagrangian at all-loop order in an \mbox{SU($N$)} YM theory can be defined in terms of an order parameter $\mathcal J$ and a running coupling \mbox{$\bar g(\mathcal J)$.} It should be noted that the latter is different from the bare coupling $g_\text{YM}$ of the classical theory~\cite{Pasechnik:2016twe, Addazi:2018fyo}. In~a non-stationary cosmological background characterized by the FLRW metric, the~conventional effective (quantum) YM Lagrangian can be written, through a rescaling of the fields according to \eq{rescaleNotation}, as~\begin{equation}
  \label{eq:YMeffectiveLagrangian}
  {\mathcal L}_\text{eff} = \frac{\mathcal J}{4 {\bar g}^2},
  \quad
  {\bar g}^2 = {\bar g}^2(\mathcal J),
  \quad
  \mathcal J = -\frac{{\mathcal F}_{\mu\nu}^a {\mathcal F}^{a\,\mu\nu}}{\sqrt{-g}},
\end{equation}
where ${\mathcal A}_\mu^a$ are the rescaled \mbox{SU($N$)} connections, ${\mathcal F}_{\mu\nu}^a$ are the rescaled field-strength components, and~the equality entering the covariant field-strength ${\mathcal F}_{\mu\nu}^a$ in a curved background
  \mbox{$
  \nabla_\mu {\mathcal A}_\nu^a - \nabla_\nu {\mathcal A}_\mu^a
  =
  \partial_\mu {\mathcal A}_\nu^a - \partial_\nu {\mathcal A}_\mu^a
  $}
has been accounted for. Furthermore, \mbox{$g \equiv \det(g_{\mu\nu})$}, where \mbox{$g_{\mu\nu} = a(\eta)^2 \, \text{diag}(1, -1, -1, -1)$} is the FLRW metric as a function of the conformal time $\eta$. Now, $\mathcal J$ simplifies to
\beql{explicitOrderParameter}
  \mathcal J
  =
  \frac{2}{\sqrt{-g}} \sum_a
    \big( \vec{E}_a \cdot \vec{E}_a - \vec{B}_a \cdot \vec{B}_a \big)
  \equiv
  \frac{2}{\sqrt{-g}} \big( \vec{E}^2 - \vec{B}^2 \big) \,,
\eeq
which is an expression that emphasizes the dependence on the components of the CE and CM fields: $\vec{E}_a$, $\vec{B}_a$. The~running of the coupling $\bar g$ as a function of the order parameter $\mathcal J$ in \eq{YMeffectiveLagrangian} fully determines the dynamics of the effective YM theory and is given by the solution of the RG evolution equation~\cite{Savvidy:2019grj, Addazi:2018fyo}
\beql{RGequation}
  2\mathcal J \frac{\mathrm d {\bar g}^2}{\mathrm d \mathcal J}
  =
  {\bar g}^2 \beta({\bar g}^2) \,,
\eeq
where $\beta$ is the standard beta-function of the YM theory~\cite{Callan:1970yg, Symanzik:1970rt}.

As an aside, it should be noted that apart from $\mathcal J$, a~second, independent, invariant may be constructed using the dual field strength~\cite{Matinyan:1976mp}:
\beql{secondInvariant}
  \mathcal G
  =
  -\frac{{\mathcal F}_{\mu\nu}^a {\mathcal F}^{*a \, \mu\nu}}{\sqrt{-g}}
  =
  \frac{4}{\sqrt{-g}} \vec{E} \cdot \vec{B} \,,
  \qquad
  {\mathcal F}^{*a \, \mu\nu} = \frac{1}{2}\epsilon^{\mu\nu\rho\sigma}{\mathcal F}_{\rho\sigma}^a \,.
\eeq

This quantity is usually disregarded in the effective YM approach, since it vanishes for all fields that have orthogonal electric and magnetic components but it may as well, in~principle, be incorporated into the effective~Lagrangian.

In order to study the vacuum dynamics on cosmological scales, the~spatially averaged quantity \mbox{$\langle \mathcal J \rangle$} should be considered, and two cases are distinguished in which: (i)~\mbox{$\langle \mathcal J \rangle$} is positive, meaning that the averaged CE components \mbox{$\langle \vec{E}^2 \rangle$} dominate over the averaged CM terms \mbox{$\langle \vec{B}^2 \rangle$}; (ii)~vice~versa, that is the case of a CM-dominated state with \mbox{$\langle \mathcal J \rangle < 0$} that corresponds to a CM condensate. For~the purpose of studying the basic features of the cosmological evolution of the CM and CE condensates in pure gluodynamics, it is sufficient to consider the effective \mbox{SU(2)} YM theory, since \mbox{SU(2)} subgroups can always be picked out of the \mbox{SU($N$)} YM theory, and such a subgroup is the part that accounts for the cosmological application~\cite{Addazi:2018fyo}. The~explicit brackets \mbox{$\langle\ldots\rangle$} will be dropped for the remainder of the~discussion.

Applying the variational principle to the effective YM action as in the classical field theory, one straightforwardly obtains the effective YM equations of motion as described in Appendix~\ref{App:C}. Similarly, the~EMT of the effective YM theory can be found as
\beql{YMEMT}
  T_{\phantom{\nu}\mu}^\nu
  =
  \frac{1}{{\bar g}^2} \Big[ \frac{\beta({\bar g}^2)}{2} - 1 \Big]
  \left(
    \frac{{\mathcal F}_{\mu\lambda}^a {\mathcal F}^{a\,\nu\lambda}}{\sqrt{-g}}
    +
    \frac{1}{4}\delta_{\phantom{\nu}\mu}^\nu \mathcal J
  \right)
  -
  \delta_{\phantom{\nu}\mu}^\nu \frac{\beta({\bar g}^2)}{8{\bar g}^2} \mathcal J \,,
\eeq
which is particularly useful for our discussion of cosmological evolution of the quantum YM system in what~follows.

\subsection{Mirror Symmetry of the Ground-State~Solutions}
\label{sec:mirrorSymmetry}

The one-loop ground-state solution behaves differently depending on the sign of $\mathcal J$ and of the running coupling \mbox{$\bar{g}_1$}. In~the case of the CM solution, it is again noted that \mbox{$\mathcal J < 0$} and, hence, \mbox{$\mathcal F > 0$.} In addition, one refers to a positive \mbox{$\bar{g}^2 > 0$} in this case so that an absolute value in the one-loop effective Lagrangian, see \eq{SavvidyOneLoopLagrangian}, may be removed. Considering the CM branch to one-loop order, which corresponds to a choice of the initial condition in the RG equation such that \mbox{${\bar g}_1^2(\mu_0^4) > 0$}, the~RG solution as derived in \eq{oneLoopCoupling} can be written on the compact form
\beql{compactOneLoopCoupling}
  {\bar g}_1^2(\mathcal J) = \frac{96\pi^2}{bN\ln(-\mathcal J/\lambda^4)}\,.
\eeq
Here, note that \mbox{${\bar g}_1^2(\mathcal J) > 0$} when \mbox{$-\mathcal J > \lambda^4$}, with
\beql{lambdaDef}
  \lambda^4
  \equiv
  \mu_0^4 \exp\bigg[ -\frac{96\pi^2}{bN{\bar g}_1^2(\mu_0^4)} \bigg]\,,
\eeq
and this gives yet another frequently used representation for the CM \mbox{SU($N$)} Lagrangian~\cite{Addazi:2018fyo}
\beql{compactOneLoopLagrangian}
  {\mathcal L}_\text{eff, CM}^{(1)}
  =
  \frac{bN}{384\pi^2}\mathcal J \ln\bigg( \frac{-\mathcal J}{\lambda^2} \bigg)\,.
\eeq

The minimum of the effective Lagrangian at \mbox{$\mathcal J_*$} is taken to be the physical scale of the quantum YM theory, which in the case of a CM vacuum reads
\beql{physicalScale}
  -\mathcal J_* = \mu_0^4 \,,
\eeq
\textls[-15]{which is the well-known phenomenon of dimensional transmutation. Readily from \eq{compactOneLoopLagrangian}, the~minimal value of the effective CM Lagrangian reads}
\beql{CMminimum}
    {\mathcal L}_\text{eff, CM}^{(1)}(\mathcal J_*)
    =
    \frac{\mathcal J_*}{4{\bar g}_1^2(\mathcal J_*)} < 0 \,,
    \qquad
    \text{with }
    {\bar g}_1^2(\mathcal J_*) > 0 \,,
\eeq
and this is a negative value. In~the standard notation, the~CM minimum corresponds to
\beql{CMSavvidyMinimum}
  2 g_\text{YM}^2 {\mathcal F}^* = e\mu^4 \equiv \Lambda_\text{QCD}^4
  \quad \to \quad
  -\mathcal J_* \equiv 2\Lambda_\text{QCD}^4 \,,
  \qquad
  {\mathcal L}_\text{eff, CM}^{(1)}(\mathcal J_*)
  =
  \frac{-\Lambda_\text{QCD}^4}{2 g_\text{YM}^2} \,,
\eeq
expressed in terms of the conventional scale, $\Lambda_\text{QCD}$.

The exact ground-state solution for positive \mbox{${\mathcal J}_* > 0$} may be found immediately by inspection of the all-loop YM equation of motion; see \eq{operatorForm}. This is the CE condensate characterized by
\beql{CEparameters}
  \beta({\bar g}_*^2) = 2\,,
  \quad
  {\bar g}_*^2 = {\bar g}^2({\mathcal J}_*) \,,
  \quad
  {\mathcal J}_* > 0 \,,
\eeq
from which it follows that the equation of motion is trivially satisfied. It should be pointed out that such a special solution is universal for any \mbox{SU($N$)} symmetry, i.e.,~it is independent of $N$.
  
The contribution to the vacuum by the CE condensate comes from the trace of the EMT (see e.g.,~Ref.~\cite{Agasian:2016hcc}). The~trace at the minimum as given by Equation~\eqref{eq:YMEMT} is
\beql{CEVeV}
  T_{\phantom{\nu}\mu}^\mu
  =
  -\frac{\beta({\bar g}_*^2)}{2{\bar g}_*^2} {\mathcal J}_*
  =
  -\frac{1}{{\bar g}_*^2} {\mathcal J}_* \,.
\eeq

The CE condensate hence contributes positively to the vacuum energy density, and this observation is intriguing when remembering that the CM condensate of Savvidy theory comes as a negative-definite contribution~\cite{Savvidy:2019grj}.

A striking and very interesting property of the YM effective Lagrangian will be discussed here, namely a mirror symmetry. It is apparent that the Lagrangian of \eq{YMeffectiveLagrangian} is \mbox{$\mathbb{Z}_2$}-symmetric under simultaneous sign changes of $\mathcal J$ and ${\bar g}^2$. The~invariance of the Lagrangian under this $\mathbb{Z}_2$ symmetry results in that the two condensates (CE and CM) are associated with two, apart from the overall sign, equal minima of the Lagrangian. Since the running of the coupling is a non-linear function of $\mathcal J$ in general, this symmetry may only be realized close to the ground state given in \eq{CEparameters}~\cite{Addazi:2018fyo}. Therefore, in~the vicinity of the ground state, the~action is symmetric under the simultaneous transformation
\beql{Z2Symmetry}
  \mathbb{Z}_2: \qquad
  {\mathcal J}_* \longleftrightarrow -{\mathcal J}_* \,,
  \qquad
  {\bar g}_*^2 \longleftrightarrow -{\bar g}_*^2 \,.
\eeq

As implicated by the form of the $\beta$-function, see \eq{oneLoopBeta}, the~imposed symmetry forces an additional change of sign in precisely $\beta$: \mbox{$\beta({\bar g}_*^2)\longleftrightarrow -\beta(-{\bar g}_*^2)$}. There are important consequences of this symmetry of the action. Mainly, the~conventional CE condensate effectively becomes mapped onto the CM gluon condensate with \mbox{${\mathcal J}_* < 0$} and \mbox{${\bar g}_*^2 > 0$}~\cite{Savvidy:2019grj}.
  
Taking the order parameter at the ground state to be the physical scale of the YM theory, one has, for~the two condensates \mbox{$\mu_0^4 = \lvert {\mathcal J}_* \rvert$}. Then, for~the CM branch, with~\mbox{${\bar g}_*^2 > 0$}, the~running coupling for the one-loop solution may be rewritten as
\beql{compactOneLoopCouplingCM}
  {\bar g}_1^2(\mathcal J)
  =
  \frac{96\pi^2}{bN\ln(\lvert\mathcal J\rvert/\lambda^4)} \,,
  \qquad
  \lambda^4
  =
  \lvert{\mathcal J}_*\rvert
  \exp\bigg[ \! -\frac{96\pi^2}{bN{\bar g}_1^2({\mathcal J}_*)} \bigg] \,,
\eeq
c.f.~\eq{compactOneLoopCoupling} and \eq{oneLoopCoupling}. Insertion of this expression back into the effective Lagrangian results in 
\beql{compactOneLoopLagrangian2}
  {\mathcal L}_\text{eff, CM}^{(1)}
  =
  \frac{bN}{384\pi^2}\mathcal J
  \ln\bigg( \frac{\lvert\mathcal J\rvert}{\lambda^2} \bigg) \,,
\eeq
c.f.~\eq{compactOneLoopLagrangian}.

Due to the $\mathbb{Z}_2$ mirror symmetry, the~minima of the effective Lagrangian on the two symmetry branches where ${\bar g}_*^2$ is either positive or negative come with the same value but with different scales: for the two condensates; the~scale is modified by the exponential in \eq{compactOneLoopCouplingCM}, so that
\beql{twoScaleParameter}
  \lambda_\pm
  =
  \lvert{\mathcal J}_*\rvert
  \exp\bigg[ \! \mp\frac{96\pi^2}{bN\lvert{\bar g}_1^2({\mathcal J}_*)\rvert} \bigg] \,.
\eeq

The upper sign stands for the CM condensate while the lower sign is the CE branch with \mbox{${\bar g}_1^2({\mathcal J}_*) < 0$}.
  
As is clear from \eq{twoScaleParameter} and from the discussion in Ref.~\cite{Addazi:2018fyo}, the~minimum for which \mbox{${\mathcal J}_* > 0$} appears in the non-perturbative region defined by \mbox{$0 < {\mathcal J}_* < \lambda^4$}. Therefore, this minimum corresponds to the CE condensate found in \eq{CEparameters}. The~mirror minimum is then swiftly found by applying the $\mathbb{Z}_2$-symmetry transformation. Note that the physical scale for the mirror minimum associated with the CM condensate is exponentially suppressed relative to the minimum point \mbox{$\lvert\mathcal{J}_*\rvert$} with the consequence that the CM condensate, with~\mbox{${\mathcal J}_* < 0$}, appears in the perturbative region where \mbox{$\lvert{\mathcal J}_*\rvert > \lambda^4$}.

Finally, we may return to the contribution of the CM condensate to the vacuum energy density. Applying the mirror symmetry to the CE minimum gives \mbox{$\beta \to -2$}, and the equations of motion, as~presented in \eq{operatorForm}, are no longer trivially satisfied. However, as~pointed out in~Ref.~\cite{Addazi:2018fyo}, the~dynamical equation in the vicinity of the CM condensate becomes
\beql{CMeom}
  \hat{\mathcal D}_\nu^{ab}
  \bigg[ \frac{{\mathcal F}^{b\,\mu\nu}}{{\bar g}^2\sqrt{-g}} \bigg] = 0 \,.
\eeq
This expression bears a close resemblance to the classical YM equations of motion that are valid in the vicinity of the ground state. The~EMT for the CM minimum involves extra terms in comparison to \eq{CEVeV}:
\beql{CMEMT}
  T_{\phantom{\nu}\mu}^\nu \big( \beta = -2 \big)
  =
  \frac{-2}{{\bar g}_*^2}
  \bigg(
    \frac{{\mathcal F}_{\mu\lambda}^a {\mathcal F}^{a\,\nu\lambda}}{\sqrt{-g}}
    +
    \frac{1}{4}\delta_{\phantom{\nu}\mu}^\nu {\mathcal J}_*
  \bigg)
  +
  \delta_{\phantom{\nu}\mu}^\nu \frac{1}{4{\bar g}_*^2}{\mathcal J}_* \,.
\eeq

However, in~the trace of this expression, the~terms within the parenthesis cancel out, yielding
\beql{CMEMTtrace}
  T_{\phantom{\mu}\mu}^\mu = +\frac{1}{{\bar g}_*^2} {\mathcal J}_* \,.
\eeq

The conclusion of Ref.~\cite{Addazi:2018fyo} is therefore that the contribution of the two condensates to the vacuum energy density cancel each other with equal magnitude and opposite signs as long as the averaging over macroscopic volumes that contains many CE and CM vacuum ``pockets'' of typical microscopic length-scales of $\sim \lambda_{\pm}^{-1}\sim \Lambda_{\rm QCD}^{-1}$ is considered:
\beql{vacuumContributions}
  \epsilon_\text{vac}
  =
  \frac{1}{4}\langle T_{\phantom{\mu}\mu}^\mu\rangle
  =
  \mp{\mathcal L}_\text{eff}({\mathcal J}_*) \,.
\eeq

Provided that the energy scales of ``electric gluon'' and ``magnetic gluon'' condensation are not the same, as~was elaborated above, the~condensates are formed at different spacetime separations. However, they evolve toward the same absolute value of the energy density, but~with opposite signs, due to their cosmological attractor nature (see ~\sec{cosmologicalAttractors} below). This effectively causes the cancellation of ``electric gluon'' and ``magnetic gluon'' contributions to the QCD ground state at sufficiently large separations, i.e.,~in the deep infrared limit of the theory, although~at the expense of a loss of homogeneity at typical length (Fermi) scales of QCD $\sim \Lambda_{\rm QCD}^{-1}$. As~the QCD vacuum appears to be locally inhomogeneous at these length-scales, gravity is expected to react to its electric and magnetic pockets in opposite ways such that the local metric fluctuations become averaged out beyond the length-scale of QCD to a small net effect compatible to that of the global observed cosmological $\Lambda$-term~\cite{Pasechnik:2013poa}.

In the framework of the YM effective action approach, one may reach an intriguing conclusion that the exact compensation of the CE and CM gluon condensate components in the QCD vacuum (averaged over spacetime volumes above the QCD Fermi scale) is the necessary and sufficient condition for confinement in QCD. Indeed, as~will be discussed below, the~CE and CM vacuum pockets are always separated by non-analytic domain walls effectively blocking the gluon field from propagating over length-scales beyond the Fermi scale of each such pocket. The~domain walls that separate different CM and CE pockets of the QCD vacuum prevent the color fields from propagating over macroscopic distances and, thus, effectively confine them within such pockets. An~exact cancellation of their contributions to the net vacuum energy-density emerges in experimental observations as a complete disappearance of the gluon DoFs in the IR limit of the theory, i.e.,~beyond the QCD Fermi scale. In~this picture of confinement, it would be natural to consider such pockets (with quarks and gluons being locked inside of them) as hadronic vacuum excitations. This is fully compatible with the classical limit of the YM theory where the conformal anomaly is absent and where only the radiation-like medium (hadron gas) remains at large separations. It remains to be seen exactly how the domain-wall picture of confinement readily formulated in Minkowski spacetime relates to a more standard center-vortex mechanism of confinement~\cite{Aharonov:1978jd,Vinciarelli:1978kp,Ambjorn:1980ms,DelDebbio:1995gc,DelDebbio:1996lih,Faber:1997rp,Engelhardt:1998wu} strongly supported by lattice simulations in Euclidean spacetime (for a recent review of the current status of this research field, see, e.g.,~Ref.~\cite{Greensite:2011zz,Pasechnik:2021ncb}).

Note, with~respect to the exact CM/CE cancellation and, hence, color confinement, the~restoration of a discrete (mirror) symmetry between ``electric gluon'' and ``magnetic gluon'' contributions at the level of the ground state is an intrinsic property of the pure YM theory and the RG flow equations. This generic property is intricately connected to the fact that the QCD-induced component of the cosmological constant term vanishes for averages over macroscopic volumes of physical spacetime. A~residual effect of the CE/CM cancellation, emerging due to an effective dynamical breaking of the mirror symmetry by gravitational interactions in the QCD vacuum, rigorously matches (in both the sign and an order of magnitude) the observed value of the cosmological constant~\cite{Pasechnik:2013poa}.

Thus, the~color confinement phenomenon and the tiny value of the cosmological constant are the direct and closely connected consequences of the mirror symmetry of the QCD vacuum in the infrared regime. An~important implication of the domain walls in the QCD vacuum is that no analyticity of the scattering amplitudes can be assumed in such a case, causing potential problems with the standard imaginary time and Euclidean formulations such as lattice QCD that is relying on the analyticity properties and vacuum triviality of the~theory.

\subsection{YM Cosmological~Attractors}
\label{sec:cosmologicalAttractors}

The temporal evolution of the condensate(s) discussed in the subsections above may be described on cosmological scales where short-distance fluctuations are averaged (integrated) out. A~simple background can be obtained by splitting the full gauge field into a background field component ${\bar A}_\mu$ and a fluctuating field $a_\mu$: \mbox{${\mathcal A}_\mu = {\bar A}_\mu + a_\mu$}. This scheme was developed in general by Wetterich in, e.g.,~Ref.~\cite{Wetterich:1992yh, Reuter:1997gx}, first for scalar fields in \mbox{$SO(N)$} and later for gauge fields, and it was further studied by, e.g.,~Gies~\cite{Gies:2002af} and Eichhorn et~al.~\cite{Eichhorn:2010zc}.

The (up to a rescaling) unique \mbox{SU(2)} pure YM theory will be parameterized in terms of a scalar time-dependent but spatially homogeneous field component (the background) on large scales due to the local isomorphism of the isotropic \mbox{SU(2)} gauge group and the \mbox{SO(3)} group of spatial 3-rotations~\cite{Khvedelidze:1998hm, Khvedelidze:2002hv, Cervero:1978db, Henneaux:1982yr, Hosotani:1984wj}. One may therefore obtain a unique decomposition of the gauge field into this spatially homogeneous isotropic part (describing the YM condensate) and a non-isotropic/inhomogeneous component (accounting for the YM waves), the~latter being the fluctuations, according to
\beql{background}
  {\mathcal A}_{\phantom{a}k}^a = U(t) \delta_{\phantom{a}k}^a + \tilde{A}_{\phantom{a}k}^a(t, \vec{x}) \,.
\eeq

Here, the~decomposition has been performed using the gauge condition \mbox{$\mathcal{A}_{\phantom{a}0}^a = 0$}~\cite{Addazi:2018fyo}. It should be stressed that the fluctuations that parameterize the inhomogeneous YM wave modes interpreted as gluons in the field-theoretical framework average out over large distances by definition in the sense that
\mbox{
    $\langle \tilde{A}_{\phantom{a}k}^a(t, \vec{x}) \rangle
    \equiv
    \int \! \mathrm d^3 \vec{x} \,
      \tilde{A}_{\phantom{a}k}^a(t, \vec{x})
    = 0$
  }.
Further, the~homogeneous YM condensate itself can be considered positively definite, \mbox{$U(t) > 0$}, and it contributes to the ground state of the theory. The~parameterization of the gauge field in \mbox{SU(2)} as a spatially homogeneous isotropic condensate and wave modes may be generalized to \mbox{SU(3)} for an application to~QCD.

\begin{figure}[H]
{\includegraphics[width=0.6\textwidth]{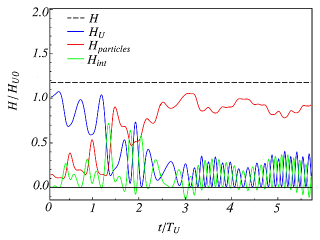}}
   \caption{The time dependence of the Hamiltonian corresponding to the YM condensate, $H_{\rm U}$, YM wave modes (or particles), $H_{\rm particles}$, as~well as the interaction term between them, $H_{\rm int}$ of the YM ``condensate + waves'' system (with total energy $H$) in the quasi-classical approximation of small wave amplitudes. Adapted from Ref.~\cite{Prokhorov:2013xba}.}
 \label{fig:YMC-int}
\end{figure}

A quasi-classical theory of \mbox{SU(2)} YM quantum-wave excitations of the classical homogeneous condensate (i.e., without accounting for the vacuum polarization effects) has been thoroughly discussed in Ref.~\citep{Prokhorov:2013xba}. The~formalism enables a proper extension to an arbitrary gauge and symmetry group with at least one \mbox{SU(2)} subgroup. Among~the key results: an~excitation of longitudinal wave (plasma) modes as well as an energy swap between the evolving homogeneous condensate and waves have been established in the linear and next-to-linear approximations. As~is shown in Figure~\ref{fig:YMC-int}, the~condensate tends to loose its energy, leading to the growth of YM wave amplitudes denoted as ``particles''. This represents a possible mechanism of particle production due to the dynamical vacuum decay which can be particularly relevant for cosmology and also in QGP production in heavy ion collisions. The~effect has further been observed in the maximally supersymmetric \mbox{${\cal N}=4$} YM theory and in the more complicated two-condensate \mbox{SU(4)} gauge theory. As~the next step, it would be important to perform an analogical study of quantum-wave dynamics in the effective action approach, i.e.,~in the case of a quantum YM vacuum, in~order to study the impact of vacuum polarization phenomena on the energy balance in the ``condensate + waves'' system and hence on the growth of the wave~modes.

Now, the~dynamical behavior of the homogeneous YM condensate \mbox{$U(t)$} introduced in \eq{background} will be discussed. The~Einstein equations for the pure YM theory in a non-trivial spacetime are obtained through the principle of variation starting from the effective action \citep{Pasechnik:2013sga,Pasechnik:2016twe} and read as follows:
\begin{adjustwidth}{-\extralength}{0cm}
\begin{align}
  \label{eq:EMeqs1}
  & \frac{1}{\kappa} \left( R_{\phantom{\nu}\mu}^\nu - \frac{1}{2}\delta_{\phantom{\nu}\mu}^\nu R \right)
  = \\ & \qquad
  \bar{\epsilon} \delta_{\phantom{\nu}\mu}^\nu
  +
  \frac{b}{32\pi^2} \frac{1}{\sqrt{-g}} \left[
    \bigg(
      -\mathcal{F}_{\mu\lambda}^a \mathcal{F}^{a \, \nu\lambda}
      +
      \frac{1}{4} \delta_{\phantom{\nu}\mu}^\nu \mathcal{F}_{\sigma\lambda}^a \mathcal{F}^{a \, \sigma\lambda}
    \bigg)
    \ln\frac{e\lvert \mathcal{F}_{\alpha\beta}^a \mathcal{F}^{a \, \alpha\beta} \rvert}{\sqrt{-g} \lambda^4}
    -
    \frac{1}{4} \delta_{\phantom{\nu}\mu}^\nu \mathcal{F}_{\sigma\lambda}^a \mathcal{F}^{a \, \sigma\lambda}
  \right] \,, \nonumber
  \\
  \label{eq:EMeqs2}
  & \left( \frac{\delta^{ab}}{\sqrt{-g}}\partial_\nu \sqrt{-g} - f^{abc} \mathcal{A}_\nu^c \right)
  \left(
    \frac{\mathcal{F}^{b \, \mu\nu}}{\sqrt{-g}}
    \ln\frac{e\lvert \mathcal{F}_{\alpha\beta}^a \mathcal{F}^{a \, \alpha\beta} \rvert}{\sqrt{-g} \lambda^4}
  \right)
  =
  0 \,.
\end{align}
\end{adjustwidth}

Here, \mbox{$\lambda = \xi\Lambda_\text{QCD}$}, with~\mbox{$\Lambda_\text{QCD} \sim 0.1$ GeV} being the QCD scale and where $\xi$ has been introduced for scaling purposes, $e$ is the base of the natural logarithm and \mbox{$\kappa$} is the gravitational constant. Finally, $\bar{\epsilon}$ describes the ground-state energy density s.t.~\mbox{$\bar{\epsilon} = \epsilon_\text{top}^\text{QCD} + \epsilon_\text{CC}$} in terms of the quantum-topological contribution from QCD and the contribution from the Cosmological Constant. For~confined QCD, \mbox{$\epsilon_\text{top}^\text{QCD} \sim -5 \times 10^9$ MeV$^4$} for an \mbox{SU(3)} color symmetric theory, and this value may be extracted from the evaluation of non-perturbative quantum-topological fluctuations of the quark and gluon fields. The~contribution from the Cosmological Constant as obtained from astrophysical measurements is \mbox{$\epsilon_\text{CC} \sim 3 \times 10^{-35}$ MeV$^4$}, which is a~comparatively minuscule and positive value. The~fact that \mbox{$\epsilon_\text{top}^\text{QCD}$} contributes to the ground-state energy of the universe with a large negative value is a severe problem for all existing cosmological paradigms, or~more accurately, for~the particle physics theories from which it arises. This is because the large negative contribution must be compensated for, in~any first-order approximation, to~a remarkable precision, resulting in the observed cosmological constant value to an accuracy of a few tens of decimal~digits.
  
The conformal dynamics of the ground state (the condensate) \mbox{$U = U(\eta)$} and of the scale factor \mbox{$a = a(\eta)$} are described by the following equations of motion as derived from Equations~\eqref{eq:EMeqs1} and \eqref{eq:EMeqs2}:
\begin{align}
  \label{eq:firstEoM}
  \frac{6}{\kappa} \frac{a''}{a^3} & = 4\bar{\epsilon} + T_{\phantom{\mu}\mu}^{\mu, U} \,,
  \qquad
  T_{\phantom{\mu}\mu}^{\mu, U}
  =
  \frac{3b}{16\pi^2a^4}\left[ (U')^2 - \frac{1}{4} U^4 \right] \,,
  \\ \label{eq:Umotion}
  \frac{\partial}{\partial\eta}
  \bigg(
    & U' \ln\frac{6e\lvert (U')^2 - \tfrac{1}{4} U^4 \rvert}{a^4 \lambda^4}
  \bigg)
  +
  \frac{1}{2} U^3 \ln\frac{6e\lvert (U')^2 - \tfrac{1}{4} U^4 \rvert}{a^4 \lambda^4}
  = 0 \,.
\end{align}

It should be noted that a particular exact solution to \eq{Umotion} can be obtained if the logarithm evaluates to zero at all times: that is, if \mbox{$\lvert Q \rvert = 1$} for
\beql{defQ}
  Q \equiv
  6e \left[ (U')^2 - \frac{1}{4} U^4 \right]
  a^{-4} (\xi\Lambda_\text{QCD})^{-4} \,.
\eeq

This may be solved for the two special cases \mbox{$Q = \pm 1$}, and the solutions are shown in Figure \ref{fig:homogeneousBgd}. The~homogeneous background \mbox{$U = U(t)$} displays quasi-periodic singularities in physical time in both cases. It should be stressed that the exact compensation of the CE and CM gluon condensate contributions to the QCD ground-state energy density, as~discussed earlier, is realized in particular if the two components \mbox{$Q = \pm 1$} co-exist in the universe. The~cancellation happens over macroscopic distances as the average of the background vanishes in the large-time limit, and importantly, this occurs without any fine tuning. Crucially, the~cancellation will arise due to the time-attractor nature of the contributions coming from the two minima; a property that is demonstrated in the following.
\begin{figure}[H]
  \begin{minipage}{.48\textwidth}
    \resizebox{\textwidth}{!}{\input{fig/homogenuousBgd/UposQ}}
  \end{minipage}
  \begin{minipage}{.48\textwidth}
    \resizebox{\textwidth}{!}{\input{fig/homogenuousBgd/UnegQ}}
  \end{minipage}
  \caption{The homogeneous QCD condensate amplitude oscillations. The~homogeneous component \mbox{$U(t)$} displays quasi-periodic singularities in the physical time \mbox{$t = \int \! a \, \mathrm d\eta$,} plotted here in units of the characteristic time scale \mbox{$\Lambda_\text{QCD}^{-1}$.} To the left, the~CE vacuum solution of \eq{defQ} with \mbox{$Q = 1$} is shown, and to the right, the~CM ditto with \mbox{$Q = -1$} is displayed. \mbox{$\xi = 4.0$} has been used along with initial conditions \mbox{$U = 0$} and \mbox{$U' = 0$}, respectively. These results are compatible with those of Ref.~\cite{Addazi:2018fyo} up to the scaling of the figure on the right-hand side.
  \label{fig:homogeneousBgd}}
\end{figure}

The conformal integral of \eq{firstEoM} is
\begin{align}
\label{eq:integratedFirstEoM}
  \frac{3}{\kappa} \frac{(a')^2}{a^4}
  & =
  \bar{\epsilon} + T_{\phantom{0}0}^{0, U} \,,
  \\
  T_{\phantom{0}0}^{0, U}
  & =
  \frac{3b}{64\pi^2 a^4}
  \left\{
    \left[ (U')^2 + \frac{1}{4}U^4 \right]
    \ln\frac{6e \lvert (U')^2 - \tfrac{1}{4}U^4 \rvert}{a^4 \lambda^4} + (U')^2 - \frac{1}{4}U^4
  \right\} \,,
  \nonumber
\end{align}
\textls[-25]{which may be verified by differentiating this expression with respect to ~$\eta$ and making use of \eq{Umotion}.}
  
Equations~\eqref{eq:firstEoM} and \eqref{eq:integratedFirstEoM} can now be combined in order to find a solution for the scale factor $a$, the~trace of the EMT \mbox{$T_{\phantom{\mu}\mu}^{\mu}$} and the total energy density \mbox{$T_{\phantom{0}0}^{0}$}. This is possible since the latter equation incorporates the constraint of \eq{Umotion}. Solving this set of equations provides the benefit of obtaining solutions for observable quantities that must necessarily be smooth functions in time. Hence, the~quasi-periodic singularities of \mbox{$U(t)$} may be avoided. Since \mbox{$t = \int \! \mathrm{d}\eta \, a(\eta)$}, Equations~\eqref{eq:firstEoM}--\eqref{eq:integratedFirstEoM} may be recast in terms of the physical time as
\beql{physTime}
  \frac{6}{\kappa}
  \left[
    \frac{\ddot{a}}{a} + \frac{\dot{a}^2}{a^2}
  \right]
  =
  4\bar{\epsilon} + T_{\phantom{\mu}\mu}^{\mu, U}
  \equiv
  T_{\phantom{\mu}\mu}^{\mu}(t) \,,
  \qquad
  \frac{3}{\kappa} \frac{\dot{a}^2}{a^2}
  =
  \bar{\epsilon} + T_{\phantom{0}0}^{0, U}
  \equiv
  T_{\phantom{0}0}^{0}(t) \,,
\eeq
where the energy density of the gluon condensate and the trace in the one-loop effective YM theory read
\begin{align}
  \label{eq:EMTtrace}
  T_{\phantom{\mu}\mu}^{\mu, U}
  & =
  \frac{3b}{16\pi^2a^4}
  \left[ a^2\dot{U}^2 - \frac{1}{4}U^4 \right] \,,
  \\
  \label{eq:EnergyDens}
  T_{\phantom{0}0}^{0, U}
  & =
  \frac{3b}{64\pi^2a^4}
  \left\{
    \left[ a^2\dot{U}^2 + \frac{1}{4}U^4 \right] \ln\frac{6e \lvert a^2\dot{U}^2 - \tfrac{1}{4}U^4 \rvert}{a^4 (\xi\Lambda_\text{QCD})^4}
    +
    a^2\dot{U}^2 - \frac{1}{4}U^4
  \right\} \,.
\end{align}

In order to eliminate the explicit dependence on \mbox{$U(t)$} and, hence, the~obstructing singularities in the two equations,~\cite{Addazi:2018fyo} introduced a universal analytic function \mbox{$g(t)$} that parameterizes the relation between the trace of the EMT and the total energy density. The~defining equation of this function is
\begin{align}
  \label{eq:gDef}
  T_{\phantom{\mu}\mu}^{\mu, U} - C
  & =
  \big( g(t) + 1 \big)
  \left[
    T_{\phantom{0}0}^{0, U} - \frac{C}{4}
  \right],
  \\ \nonumber
  C & \equiv -4 \epsilon_\text{top}^\text{QCD}
  =
  \frac{3b}{16\pi^2}
  \frac{(\xi \Lambda_\text{QCD})^4}{6e} \,.
\end{align}

Using $g(t)$, \eq{physTime} can be written entirely in terms of continuous functions:
\begin{align}
  \label{eq:contEq1}
  \frac{6}{\kappa}
  \left[
    \frac{\ddot{a}}{a} + \frac{\dot{a}^2}{a^2}
  \right]
  & =
  4\epsilon_\text{CC}
  +
  \big( g(t) + 1 \big)
  \left[
    T_{\phantom{0}0}^{0, U} - \frac{C}{4}
  \right] \,,
  \\
  \label{eq:contEq2}
  \frac{3}{\kappa} \frac{\dot{a}^2}{a^2}
  & =
  \epsilon_\text{CC} - \frac{C}{4}
  +
  T_{\phantom{0}0}^{0, U} \,.
\end{align}

Note here that \mbox{$T_{\phantom{0}0}^{0, U} = T_{\phantom{0}0}^{0, U}(U,\dot{U},a)$.} After excluding \mbox{$T_{\phantom{0}0}^{0, U}$} above, the~resulting equation for the scale factor that is left to be solved is
\beql{scalFac}
  6\frac{\ddot{a}}{a}
  +
  3\big( 1 - g(t) \big) \frac{\dot{a}^2}{a^2}
  +
  \kappa\epsilon_\text{CC} \big( g(t) - 3 \big)
  =
  0 \,.
\eeq

  The general solution is\endnote{By means of the following ansatz:
  \mbox{$a(t) = a^* \exp\big[ f(t) \big]$},
  the equation for the scale factor can be rewritten as
  \begin{equation*}
    \ddot{f} - \tilde{h}(t) (\dot{f})^2 + A\tilde{h}(t)
    = 0,
  \end{equation*}
  with
  \mbox{$\tilde{h}(t) = \tfrac{1}{2}\big( g(t) - 3 \big)$}
  and
  \mbox{$A = \frac{\kappa\epsilon_\text{CC}}{3}$.}
  The introduction of \mbox{$m(t) \equiv \dot{f}$},  results in a first-order equation that may be solved. It is explicitly
  \begin{equation*}
    \dot{m} - \tilde{h}(t) m^2(t) + A\tilde{h}(t)
    = 0 \,.
  \end{equation*}
  The scale factor is therefore found in terms of the integral of the solution for \mbox{$m(t)$} as
  \begin{equation*}
    a(t)
    =
    a^* \exp\left[
      \int_{t_0}^t \! \mathrm{d}t \, m(t)
    \right] \,.
  \end{equation*}} 

\begin{adjustwidth}{-\extralength}{0cm}
\begin{align}
\label{eq:scaleFac}
  a(t)
  & = \\
  & a^*
  \exp\left[
    \sqrt{\frac{\kappa \epsilon_\text{CC}}{3}}
    \int_{t_0}^t \!
      \frac{
        1 + \sqrt{\tfrac{\epsilon_\text{CC}}{\epsilon_0}}
        +
        \left( 1 - \sqrt{\tfrac{\epsilon_\text{CC}}{\epsilon_0}} \right)
        \exp\left\{
          \sqrt{\tfrac{\kappa \epsilon_\text{CC}}{3}}
          \left(
            -3(t' - t_0)
            +
            \int_{t_0}^{t'} \! g(\tau) \, \mathrm d\tau
          \right)
        \right\}
      }{
        1 + \sqrt{\tfrac{\epsilon_\text{CC}}{\epsilon_0}}
        -
        \left( 1 - \sqrt{\tfrac{\epsilon_\text{CC}}{\epsilon_0}} \right)
        \exp\left\{
          \sqrt{\tfrac{\kappa \epsilon_\text{CC}}{3}}
          \left(
            -3(t' - t_0)
            +
            \int_{t_0}^{t'} \! g(\tau) \, \mathrm d\tau
          \right)
        \right\}
      }
      \, \mathrm d t'
  \right] \,, \nonumber
\end{align}
\end{adjustwidth}
in terms of the initial values of the scale factor \mbox{$a^* \equiv a(t=t_0)$} and the total energy density \mbox{$\epsilon_0 \equiv T_{\phantom{0}0}^{0}(t=t_0)$,} respectively. Note that \mbox{$\epsilon_\text{CC} \ll \epsilon_0$.}

The total energy density, \mbox{$T_{\phantom{0}0}^{0}(t)$,} and the trace of the EMT, \mbox{$T_{\phantom{\mu}\mu}^{\mu}(t)$,} both explicitly defined in \eq{physTime}, can be found by insertion of the solution above into \eq{contEq2} together with a manipulation of \eq{gDef}. The~result is
\begin{adjustwidth}{-\extralength}{0cm}
\begin{align}
  \label{eq:eDens}
  \frac{T_{\phantom{0}0}^{0}(t)}{\epsilon_\text{CC}}
  & =
  \left[
    \frac{
      1 + \sqrt{\tfrac{\epsilon_\text{CC}}{\epsilon_0}}
      +
      \left( 1 - \sqrt{\tfrac{\epsilon_\text{CC}}{\epsilon_0}} \right)
      \exp\left\{
        \sqrt{\tfrac{\kappa \epsilon_\text{CC}}{3}}
        \left(
          -3(t - t_0)
          +
          \int_{t_0}^{t} \! g(\tau) \, \mathrm d\tau
        \right)
      \right\}
    }{
      1 + \sqrt{\tfrac{\epsilon_\text{CC}}{\epsilon_0}}
      -
      \left( 1 - \sqrt{\tfrac{\epsilon_\text{CC}}{\epsilon_0}} \right)
      \exp\left\{
        \sqrt{\tfrac{\kappa \epsilon_\text{CC}}{3}}
        \left(
          -3(t - t_0)
          +
          \int_{t_0}^{t} \! g(\tau) \, \mathrm d\tau
        \right)
      \right\}
    }
  \right]^2 \,,
  \\
  \label{eq:EMT}
  \frac{T_{\phantom{\mu}\mu}^{\mu}(t)}{\epsilon_\text{CC}}
  & =
  4
  +
  \frac{
    4\big( g(t) + 1 \big)
    \left( 1 -\tfrac{\epsilon_\text{CC}}{\epsilon_0} \right)
    \exp\left\{
      \sqrt{\tfrac{\kappa \epsilon_\text{CC}}{3}}
      \left(
        -3(t - t_0)
        +
        \int_{t_0}^{t} \! g(\tau) \, \mathrm d\tau
      \right)
    \right\}
  }{
    \left[
      1 + \sqrt{\tfrac{\epsilon_\text{CC}}{\epsilon_0}}
      -
      \left(
        1 - \sqrt{\tfrac{\epsilon_\text{CC}}{\epsilon_0}}
      \right)
      \exp\left\{
        \sqrt{\tfrac{\kappa \epsilon_\text{CC}}{3}}
        \left(
          -3(t - t_0)
          +
          \int_{t_0}^{t} \! g(\tau) \, \mathrm d\tau
        \right)
      \right\}
    \right]^2
  } \,.
\end{align}
\end{adjustwidth}

It shall be pointed out here that the above solutions for the scale factor, the~energy density, and~the trace of the EMT do not rely on any approximations but are the general solutions that can be obtained from \eq{physTime}. These cosmological observables may therefore be studied on the full range from $t_0$ to $t$, provided that \mbox{$g(t)$} is known.
  
For practical analyses, the~auxiliary function $g$ may be studied in the vicinity of the exact, large-time cancellation point where \mbox{$Q(t) \sim 1$.} This is done by introducing an expansion of the YM energy density around the asymptotic value of the exact solution, where \mbox{$T_{\phantom{0}0}^{0, U*} = C/4$}, such that
\beql{expEden}
  T_{\phantom{0}0}^{0, U}(t) \simeq C/4 + \delta\epsilon(t), \qquad \delta\epsilon \ll C \,.
\eeq

Depending on the relation between the expansion parameter \mbox{$\delta\epsilon$} and the remaining scale $\epsilon_\text{CC}$, the~time derivatives $\dot{a}$ and \mbox{$\dot{T}_{\phantom{0}0}^{0, U}(t)$} take two different asymptotic forms. Firstly, in~the case of large \mbox{$\delta\epsilon(t) \gg \epsilon_\text{CC}$}, these are
\begin{align}
  \label{eq:greatAss}
  \dot{a}
  & \simeq
  \sqrt{\frac{\kappa}{3}} a \sqrt{\delta\epsilon},
  \\ \nonumber
  \dot{T}_{\phantom{0}0}^{0, U}
  & \simeq
  \sqrt{\frac{\kappa}{3}}
  \big( g(t) - 3 \big)
  \big( \delta\epsilon \big)^{3/2} \,,
\end{align}
when keeping only the leading terms in \mbox{$\delta\epsilon(t) \ll C$.} Secondly, in~the opposite case when \mbox{$\delta\epsilon(t) \ll \epsilon_\text{CC}$,} the same quantities instead become
\begin{align}
  \label{eq:smallAss}
  \dot{a}
  & \simeq
  \sqrt{\frac{\kappa\epsilon_\text{CC}}{3}}
  a
  \left(
    1 + \sum_{n=1}^\infty \! {\tfrac{1}{2} \choose n} \left( \frac{\delta\epsilon}{\epsilon_\text{CC}} \right)^n
  \right) \,,
  \\ \nonumber
  \dot{T}_{\phantom{0}0}^{0, U}
  & \simeq
  \sqrt{\frac{\kappa\epsilon_\text{CC}}{3}}
  \big( g(t) - 3 \big)
  \delta\epsilon
  \left(
    1 + \sum_{n=1}^\infty \! {\tfrac{1}{2} \choose n} \left( \frac{\delta\epsilon}{\epsilon_\text{CC}} \right)^n
  \right) \,.
\end{align}

The difference between Equations~\eqref{eq:greatAss} and \eqref{eq:smallAss} introduces only a very small correction to the period of \mbox{$g(t)$}, and this correction can safely be neglected, as will be shown later in this~section.

It is now possible to study $g$ in the vicinity of the asymptote (\eq{expEden}) and in the two limits of \mbox{$\delta\epsilon(t)$} relative to $\epsilon_\text{CC}$. First, solve Equations~\eqref{eq:EMTtrace} and \eqref{eq:EnergyDens} for $U$, $\dot{U}$ in terms of $T_{\phantom{0}0}^{0, U}$, $T_{\phantom{\mu}\mu}^{\mu, U}$. Then, insert the expansion of the energy density from Equation~\eqref{eq:expEden} in the resulting expressions as well as in Equation~\eqref{eq:gDef}. Explicit expressions for $U$, $\dot{U}$ in terms of the EMT components and the scale factor allows for the formulation of
\beql{uDiff}
  \partial_t U(t) - \dot{U} \equiv 0 \,.
\eeq

Explicit computation of the first term results in a differential equation for \mbox{$g(t)$} when the expansions in Equation~\eqref{eq:greatAss} are inserted. The~final form of such a differential equation is
\beql{gEq}
  {\dot{g}}^4
  -
  \frac{8 \big( \xi\Lambda_\text{QCD} \big)^4}{3e}
  \big( 1 - g^2 \big)^3
  =
  0 \,.
\eeq

Its implicit analytic solution can be found for the inverse function \mbox{$t(g)$} over half a period of oscillation of \mbox{$g(t)$} as
\beql{gInvEq}
  t(g)
  =
  -\frac{(6e)^{1/4}}{2\xi\Lambda_\text{QCD}}
  \left[ _2F_1\left( \frac{1}{2}, \frac{3}{4}, \frac{3}{2}; g^2 \right) g - k \right],
  \qquad
  0 < t(g) < T_g/2 \,.
\eeq

The constant \mbox{$k \approx 2.622$} is defined through the above equation, and the initial condition \mbox{$g(t_0) = 1$} is adopted for simplicity. These conditions determine \mbox{$g = g(t)$} as a periodic quasi-harmonic function with unit amplitude. The~period of oscillation may be found from Equation~\eqref{eq:gEq} as
\beql{gPeriod}
  T_g
  =
  \frac{2 (6e)^{1/4}}{\xi\Lambda_\text{QCD}}
  \int_0^1 \! \frac{\mathrm{d}g}{(1 - g^2)^{3/4}}
  =
  \frac{2 k (6e)^{1/4}}{\xi\Lambda_\text{QCD}} \,.
\eeq

Using instead the expansions in Equation~\eqref{eq:smallAss}, the~above calculation can be repeated for the case of \mbox{$\delta\epsilon(t) \ll \epsilon_\text{CC}$.} The analogue of Equation~\eqref{eq:gEq}, now with an additional term proportional to \mbox{$\alpha \ll 1$}, is
\begin{align}
  \label{eq:gEq2}
  & \frac{\mathrm{d} g}{\mathrm{d} \tau} \pm 2(1 - g^2)^{3/4} - \alpha(1 - g^2) = 0,
  \qquad
  \tau = t\frac{\xi\Lambda_\text{QCD}}{(6e)^{3/4}} \,,
  \\
  & T_g' = T_g \left[ 1 + \frac{\pi}{4k^2} \alpha^2 \right] \,, \nonumber
  \\
  & \alpha
  =
  \frac{2(6e)^{1/4}}{\xi\Lambda_\text{QCD}}
  \sqrt{\frac{\kappa\epsilon_\text{CC}}{3}}
  \approx
  4.8 \times 10^{-24}, \qquad \text{for } \Lambda_\text{QCD} \sim 210~\text{MeV} \,.
  \nonumber
\end{align}
Hence, the~additional term may safely be neglected.

Note that the function \mbox{$g(t)$} satisfies the following integral constraint
\beql{gConstr}
  \int_0^t \! \mathrm{d}\tau \, g(\tau)
  =
  \pm \frac{(6e)^{1/4}}{\xi\Lambda_\text{QCD}} (1 - g^2)^{1/4} \,,
  \qquad
  \frac{nT_g}{2} < t < \frac{(n+1)T_g}{2} \,,
\eeq
where the upper sign corresponds to even $n$ and the lower corresponds to odd $n$. It should be noted that this constraint can be used in Equations~\eqref{eq:scaleFac}, \eqref{eq:eDens} and \eqref{eq:EMT} in order to explicitly express the general solutions for \mbox{$a(t)$}, \mbox{$T_{\phantom{0}0}^{0}(t)$} and \mbox{$T_{\phantom{\mu}\mu}^{\mu}(t)$} in terms of $g$. A~very good analytic approximation to the exact \mbox{$g(t)$} may be constructed when keeping only the first two non-vanishing terms of the harmonic Fourier expansion s.t.
\beql{FourierApprox}
  g(t)
  \approx
  A \cos\left( \frac{2\pi t}{T_g} \right)
  +
  (1 + A) \cos\left( \frac{6\pi t}{T_g} \right) \,.
\eeq

The amplitude $A$ is found through
\beql{A}
  A = \frac{2}{k}
  \int_0^1 \! \mathrm{d}g \,
    \frac{g}{(1 - g^2)^{3/4}}
    \cos\left( \frac{\pi}{2k} \int_g^1 \frac{\mathrm{d}x}{(1 - x^2)^{3/4}} \right)
  \approx
  1.14 \,.
\eeq

\begin{figure}[H]

  \resizebox{0.6\textwidth}{!}{\input{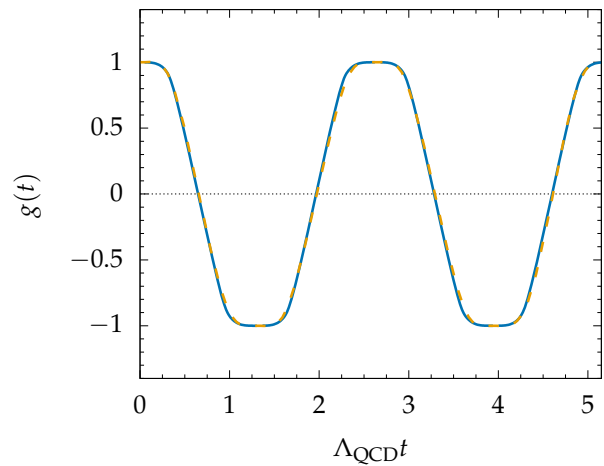}}

  \caption{The time dependence of the quasi-harmonic universal function \mbox{$g = g(t)$}. The~exact solution in Equation~\eqref{eq:gInvEq} (solid line) has been extrapolated from the solution over a single period \mbox{$T_g/2$}. A~harmonic approximation in Eq.~\eqref{eq:FourierApprox} (dashed line) captures the behavior of \mbox{$g(t)$} well.
  \label{fig:g}}
\end{figure}

A comparison between this approximation and the exact solution \mbox{$g(t)$} is provided in Figure~\ref{fig:g} from which it is clear that the approximation is indeed capturing the universal function to a very good~accuracy.
For a particular approximation to \mbox{$g(t)$} discussed above, the physical observables have been plotted in Figure~\ref{fig:condensateSolutions} for illustration.
\begin{figure}[H]
  \begin{minipage}{.33\textwidth}
    \resizebox{\linewidth}{!}{\input{fig/T00/T00}}
  \end{minipage}%
  \begin{minipage}{.33\textwidth}
    \resizebox{\linewidth}{!}{\input{fig/Tmumu/Tmumu}}
  \end{minipage}%
  \begin{minipage}{.33\textwidth}
    \resizebox{\linewidth}{!}{\input{fig/a/a}}
  \end{minipage}
  \caption{Solutions for the total energy density
  \mbox{$T_{\phantom{0}0}^0(t)$} (left), the~trace of the total QCD EMT
  \mbox{$T_{\phantom{\mu}\mu}^\mu(t)$} (middle) and the scale factor \mbox{$a(t)$} (right). The~asymptotic values for which \mbox{$Q \to 1$} are indicated by horizontal lines in the left and middle panels, respectively.
  The initial conditions have been chosen as
  \mbox{$U_0 = 0$,}
  \mbox{$\dot{U}_0 = (\xi \Lambda_\text{QCD})^2/\sqrt{3e}$,}
  \mbox{$Q_0 > 1$,}
  \mbox{$\xi = 4.0$}
  \mbox{$\Lambda_\text{QCD} = 332$ MeV} and
  \mbox{$\kappa = 10^{-7}$ MeV$^{-2}$.}
  The energy density and the trace are plotted in dimensionless units, rescaled by $\Lambda_\text{QCD}^4$ and for illustrative purposes, $\epsilon_\text{CC}$ was set to $\sim0.5\,\%$ of $\bar{\epsilon}$. These results are compatible with the qualitative picture in Ref.~\cite{Addazi:2018fyo}.
  \label{fig:condensateSolutions}}
\end{figure}

\subsection{\texorpdfstring{SU($N$)}{TEXT} and the Functional RG Approach}
\label{sec:SUNFRG}

So far, we have addressed gluodynamics resorting to an all-order effective perturbative approach. Nonetheless, we can extend our investigation so as to include non-perturbative all-order analyses, resorting to the Functional Renormalization Group (FRG) approach~\mbox{\cite{Wetterich:1992yh, Morris:1993qb, Papenbrock:1994kf, Fischer:2002hna, Fischer:2006vf,Fischer:2009tn, Fischer:2008uz, Ellwanger:1995qf,Ellwanger:1996wy, Bergerhoff:1997cv, Pawlowski:2003hq}}. This latter is a Wilsonian momentum-shell-wise integration method for the path integral, which was developed to delve into the dynamics of interacting quantum field theories and statistical systems in a non-perturbative way when couplings cannot be dealt with using perturbative techniques. A~regulator function $R_k$ is taken into account so as to suppress quantum fluctuations at momenta lower than some physical scale, i.e.,~at an IR cut-off scale $k$. This scale is in principle different to the one defined in the subsections above, namely $\mu_0$. The~former denotes the scale above which all quantum fluctuations are integrated out, while the latter captures instead the one-loop renormalization scale and may further be extended to all-loop accuracy. The~regulator function has been discussed by, e.g.,~Gies in Ref.~\cite{Gies:2002af}. A~scale-dependent effective action that flows with $k$ is then recovered, i.e.,~$\Gamma_k$, which encodes quantum fluctuations effects at momenta larger than the IR cut-off $k$. Varying $k$ then allows to smoothly interpolate among the microscopic/short-scale action and the full quantum effective action $\Gamma_{k \rightarrow 0}$.

This elucidates why this procedure looks to be tailored ad~hoc for cosmological applications, at~the IR scale. The~Wetterich equation for a non-zero background that fixes the running dependence on the cut-off scale in the FRG approach reads~\cite{Wetterich:1992yh, Morris:1993qb}
\begin{equation}
  \label{FRGE}
  \partial_t \Gamma_k
  =
  \frac{1}{2} {\rm STr} \big(\Gamma^{(2)}_k + R_k \big)^{-1} \, \partial_t R_k \,,
\end{equation}
\textls[-15]{where $\Gamma^{(2)}_k$, a~matrix in the field space, denotes the second functional variation of the effective (running) action with respect to the field content of the theory\endnote{For the case of the simple background considered in Section~\ref{sec:cosmologicalAttractors},
  \mbox{${\mathcal A}_\mu = {\bar A}_\mu + a_\mu$} \,,
  then
  \mbox{$\Gamma_k^{(n,m)}\big[\bar A, a\big]
  =
  \tfrac{\delta^n}{(\delta \bar A)^n}
  \tfrac{\delta^m}{(\delta a)^m}
  \Gamma_k\big[\bar A, a\big]$}.
  }.
Above, STr is the super-trace, including a summation over all field components and discrete indices, as~well as all the eigenvalues of the Laplacian in the kinetic term. The~FRG equation retains a dependence on the full (field-dependent) non-perturbative regularized propagator~\cite{Papenbrock:1994kf}, namely, $(\Gamma^{(2)}_k +R_k)^{-1} $.}

The FRG approach has been then adapted to YM \mbox{SU($N$)} theories~\cite{Reuter:1997gx,Gies:2002af,Eichhorn:2010zc}. Specifically, in~Ref.~\cite{Eichhorn:2010zc}, a numerical extrapolation among low and high energy scales for the full propagators was deployed in order to derive the gluon condensate. Refining these results in~Ref.~\cite{Dona:2015xia}, the~FRG approach was extended, within~a cosmological setting, to~the \mbox{SU(2)} case.

In Ref.~\cite{Dona:2015xia}, resorting to approximations that are necessary to solve the FRG equation, which is otherwise too complicated to provide analytic results, the~authors considered replacing $\Gamma_k$ in Equation~\eqref{FRGE} with the bare action $S$, allowing for integration on both sides of the FRG equation, namely,
\begin{equation}
  \Gamma_k = - \int L_{\rm eff}
  =
  \int \! \mathrm d k \,
    \frac{1}{2} {\rm STr} (S^{(2)}+ R_k)^{-1} \partial_t R_k
  =
  \frac{1}{2} {\rm STr} \, \log (S^{(2)}+ R_k) + {\rm const.}\,.
\end{equation}

Setting the bare action to the standard expression \mbox{$S=\frac{1}{4}\int dx \, F_{\mu \nu}^a F^{a\,\mu \nu}$}, which here corresponds to the UV limit of the effective theory, the~integration constant can be fixed by requiring that $\Gamma_k$ vanishes for a vanishing field~strength.  

The inversion of the regularized propagator then requires the use of a harmonic gauge fixing, entering the action $S_{\rm gf}$ and depending on the $\alpha$-parameter, with~associated Faddeev-Popov ghosts, in~$S_{\rm gh}$, namely,
\begin{equation}
  S_{\rm gf}
  =
  \frac{1}{2 \alpha} \int \! \mathrm dx \,
    \bar{D}^\mu \bar{a}^a_\nu \bar{D}^\nu a_\mu^a \,,
  \qquad\qquad
  S_{\rm gh}
  =
  \int \! \mathrm dx \,
    \bar{D}_\mu \bar{c}_\nu {D}^\mu c^\nu \,,
\end{equation}
where the background methods have been deployed, and~barred quantities are calculated with respect to background fields. Then, in~the Landau gauge where \mbox{$\alpha \rightarrow{0}$}, the~super-trace recasts along the transverse sector as
\begin{equation}
  \frac{1}{2} {\rm STr} \, \log (S^{(2)}+ R_k)
  =
  \frac{1}{2} {\rm Tr}_{\rm trans} \log \big[\bar{D}^{\mu \nu}_{T}+R_k(\bar{D}^{\mu \nu}_{T})\big]
  -
  \frac{1}{2} {\rm Tr}_{\rm gh} \log \big[\bar{D}^{\mu \nu}_{\rm gh}+R_k(\bar{D}^{\mu \nu}_{\rm gh})\big]\,,  
\end{equation}
where the differential operators can be expressed in terms of the \mbox{SU($N$)} structure constant $f^{abc}$ and the YM coupling constant $g_\text{YM}$ as
\begin{equation}
  \bar{D}^{\mu\nu}_{\rm T}
  =
  \bar{\Box} \, \delta^{cb} \delta^{\mu\nu}
  +
  g_\text{YM}\, \bar{F}^{a\, \mu \nu} f^{abc} \,,
  \qquad\qquad
  \bar{D}^{\mu \nu}_{\rm gh} = \eta^{\mu \nu} \bar{\Box} \,.
\end{equation}

In order to disentangle the emergence of a condensate as a solution to the FRG equation, one restricts the consideration to the case of \mbox{SU(2)}. Although~this could seem to be limiting in the wider theoretical perspective, nonetheless, the restriction to \mbox{SU(2)} shall be considered as a selection of a subgroup of \mbox{SU($N$)}.

Bearing the discussion above in mind, one may select a straightforward expression for the regulator function in terms of a cut-off scale, $R_k(\mathcal{D}) = k^2$. The~super-trace \mbox{STr} may then be evaluated using the Schwinger formula
\begin{eqnarray}
\ln A \equiv \int_0^{\infty}\frac{ds}{s} \Big[e^{-sA}-e^{-s}\Big] \,, \qquad A>0 \,,
\end{eqnarray}
and opting for a convenient choice of the background, the~self-dual one, as~investigated in Ref.~\cite{Eichhorn:2010zc}. These choices allow for the evaluation of the super-trace as a sum over the eigenvalues of the kinematic terms. In~general, the~eigenvalues of $\bar{\mathcal D}_\text{T}$ are not known for an arbitrary background. However, those were found in the case of a self-dual background considered in Ref.~\cite{Eichhorn:2010zc}, and~a general discussion on the trace technology used in this approach can be found, e.g.,~in Refs.~\cite{Reuter:1997gx,Gies:2002af}. Applying the results of Ref.~\cite{Eichhorn:2010zc} in the case of a self-dual background dominated by the color-magnetic field $B$, Ref.~\cite{Dona:2015xia} obtained an expression for the regularized effective action in the following form:
\begin{adjustwidth}{-\extralength}{0cm}
\begin{equation} 
\label{Leffe}
  L_{\rm eff} =
  \frac{g_\text{YM}^2 \theta }{4\pi^2}
  \int_0^\infty \! \frac{\mathrm d s}{s} \,
    \Big[e^{- A s} - e^{-s} \Big]
    \left(\frac{1}{4 \sinh^2 (s)} + 1 - \frac{1}{4s^2} \right)\,,
    \quad
    A = \sqrt{\frac{k^4}{g_\text{YM}^2 \theta }} \,, \quad \theta \equiv B^2 > 0 \,.
\end{equation}
\end{adjustwidth}

This expression can be recast in terms of the order parameter in the CM domain, \mbox{$\mathcal J = 2g_\text{YM}^2 \theta > 0$}, which was introduced earlier and then analytically continued to the CE branch $\mathcal J < 0$ as follows~\cite{Addazi:2018fyo},
\begin{adjustwidth}{-\extralength}{0cm}
\beql{allLoopEffLagrangian}
{\mathcal L}_\text{eff} =
\frac{2\mathcal J}{16\pi^2}
\int_0^\infty \frac{\mathrm d s}{s} \,
\big[e^{-s A[\mathcal J]} - e^{-s}\big]
\bigg[\frac{1}{4\sinh^2 s} - \frac{1}{4s^2} + 1\bigg] \,,
\qquad
A = \sqrt{\frac{\lambda^4}{\mathcal J \tanh(\mathcal J/\lambda^4\varepsilon)}} \,,
\eeq
\end{adjustwidth}
for \mbox{$\varepsilon \ll 1$}. Indeed, due to
\beql{epsilonLimit}
  \mathcal J \tanh\bigg( \frac{\mathcal J}{\varepsilon \lambda^4} \bigg) \Big\lvert_{\varepsilon\to 0}
  \to \lvert\mathcal J\rvert \,,
\eeq
the transition between Equation~(\ref{Leffe}) and \eq{allLoopEffLagrangian} becomes apparent. Performing the explicit expansion of the rightmost parenthesis above, the~form matches that of the one-loop effective Lagrangian in Equation~\eqref{eq:compactOneLoopLagrangian} for \mbox{SU(2)}. 

The results of Appendix~\ref{sec:one-loop}, valid at one-loop order, may now be compared to the all-loop order results. In~Figure~\ref{fig:Leff-gBarInv-betaQ} (left), we show a direct comparison of the one-loop with the all-loop effective Lagrangian for $\mathcal J > 0$ (CE branch only). A~unique non-trivial minimum is found in the non-perturbative region, $0<\mathcal{J}^*<\lambda^4$, and~it is therefore identified with the CE condensate~\cite{Addazi:2018fyo} whose values for one-loop and all-loop cases differ at a permille level and thus express a remarkable consistency of the one-loop approximation. The~all-loop running coupling may be straightforwardly extracted from the all-loop effective Lagrangian as
\beql{allLoopCoupling}
  \big({\bar g}^2\big)^{-1}
  =
  \frac{2}{4\pi^2}
  \int_0^\infty \frac{\mathrm d s}{s} \,
    \big[e^{-s A[\mathcal J]} - e^{-s}\big]
    \bigg[\frac{1}{4\sinh^2 s} - \frac{1}{4s^2} + 1\bigg] \,,
\eeq
and this is shown in Figure~\ref{fig:Leff-gBarInv-betaQ} (middle) for the one-loop and all-loop cases. To~find an expression for the \mbox{$\beta$-function}, one may study the RG equation given by \eq{RGequation}. Using
\beql{InverseDerivative}
  \frac{\mathrm d {\bar g}^2}{\mathrm d \mathcal J}
  =
  -\big({\bar g}^2\big)^2
  \frac{\mathrm d}{\mathrm d \mathcal J} \Big(\frac{1}{{\bar g}^2}\Big) \,.
\eeq

We may express $\beta$ as
\begin{align}
  \frac{\beta}{{\bar g}^2}
  & =
  -2\mathcal J
  \frac{\mathrm d}{\mathrm d \mathcal J} \Big(\frac{1}{{\bar g}^2}\Big)
  \nonumber
  \\ & =
  -2\mathcal J \frac{2}{4\pi^2}
  \bigg( \!\!-\!\frac{\mathrm d A}{\mathrm d \mathcal J} \bigg)
  \int_0^\infty \! \mathrm d s \,
    e^{-s A[\mathcal J]}
    \bigg[\frac{1}{4\sinh^2 s} - \frac{1}{4s^2} + 1\bigg] \,,
\end{align}
where
\beql{Aderivative}
  \frac{\mathrm d A}{\mathrm d \mathcal J}
  =
  -\frac{1}{2\mathcal J}
  \sqrt{\frac{\lambda^4}{\mathcal J \tanh(\mathcal J/\lambda^4\varepsilon)}}
  \,
  \left( 1 + \frac{\mathcal J}{\lambda^2\varepsilon}\frac{1 - \tanh^2 (\mathcal J/\lambda^2\varepsilon)}{\tanh (\mathcal J/\lambda^2\varepsilon)} \right) \,.
\eeq

The $\beta$ over the running coupling at one-loop order compared to the corresponding all-loop quantity for \mbox{SU(2)} is displayed in Figure~\ref{fig:Leff-gBarInv-betaQ} (right). It is clear from the figures that the one-loop approximation accurately captures the main features of the pure YM theory as the differences from the corresponding all-loop quantities are negligible.
\begin{figure}[H]
\begin{minipage}{0.323\textwidth}
  \centering
  \resizebox{1\textwidth}{!}{
    \input{fig/effLagrangian/effLagrangian}
  }
\end{minipage}
~
\begin{minipage}{0.323\textwidth}
  \centering
  \resizebox{1\textwidth}{!}{
    \input{fig/gBarInverse/gBarInverse}
  }
\end{minipage}
~
\begin{minipage}{0.323\textwidth}
  \centering
  \resizebox{1\textwidth}{!}{
    \input{fig/betaQuota/betaQuota}
  }
\end{minipage}
\caption{(\textbf{Left}) All- and one-loop effective Lagrangian of \mbox{SU(2)} as dependent on \mbox{$\mathcal J / \lambda^4$}. Plotted here for the CE branch with \mbox{${\mathcal J} > 0$}, it is clear that the one-loop result captures the main features of pure YM theory. In~the non-perturbative regime shown (\mbox{$\mathcal J < \lambda^4$}), the~two minima in the vicinity of \mbox{${\mathcal J}_* > 0$} coincide. Here, \mbox{$\epsilon = 0.01$} has been used. (\textbf{Middle}) The inverse running coupling \mbox{${\bar g}^{-2}$} of \mbox{SU(2)}.  The~(dashed) one-loop result captures well the overall behavior of the running coupling in comparison to the all-loop coupling (solid). Here, \mbox{$\epsilon = 0.01$} has been used. (\textbf{Right}) $\beta$ over the running coupling at one-loop order compared to the all-loop quantity for \mbox{SU(2)}. The~ratio of the $\beta$-function to the coupling coincides with the all-loop result at one-loop level for small $\mathcal J$, while a discrepancy is seen away from zero. Here, \mbox{$\epsilon = 10^{-5}$} has been~used.}
  \label{fig:Leff-gBarInv-betaQ}
\end{figure}

As outlined in Ref.~\cite{Dona:2015xia}, when considering the formation of the non-perturbative CE condensate as the YM system rolls down toward the minimum of the effective Lagrangian depicted in Figure~\ref{fig:Leff-gBarInv-betaQ} (left), the~equation that dictates the evolution of this system in cosmological time is found from the continuity equation~\cite{Dona:2015xia} (see also Ref.~\cite{Savvidy:2021ahq} for a more recent discussion)
\beql{continuityEquation}
  \dot{\rho}_\text{YM} + 3 \frac{\dot a}{a}\big( \rho_\text{YM} + p_\text{YM} \big) = 0 \,.
\eeq

Accounting for only homogeneous CE YM fields (i.e., for $\theta = E^2$) in the EMT, for~simplicity, the~expressions of the energy and pressure densities of the YM system $\rho_\text{YM}$, $p_\text{YM}$ can be written as functionals of the effective action~\cite{Dona:2015xia}:
\begin{align}
\label{YMCenergyDensity}
\rho_\text{YM} = - {\mathcal L}_\text{eff}(\theta) + 2 \theta {\mathcal L}_\text{eff}'(\theta) \,, 
\qquad
p_\text{YM} = {\mathcal L}_\text{eff}(\theta) - \frac{2}{3} \theta {\mathcal L}_\text{eff}'(\theta) \,,
\end{align}
where prime denotes the functional variation with respect to $\theta$. In~this case, \eq{continuityEquation} transforms to
\begin{equation}
\label{CC}
\dot{\theta} \, ( {\mathcal L}_{\rm eff}'+ 2 \theta  {\mathcal L}_{\rm eff}'') + 4 \frac{\dot{a}}{a} \theta {\mathcal L}_{\rm eff}' = 0\,,
\end{equation}
in comoving coordinates. Equation~\eqref{CC} may be integrated, given that ${\mathcal L}_\text{eff}(\theta)$ is sufficiently well behaved, such that
\begin{equation}
\label{CCR}
\sqrt{\theta} {\mathcal L}_{\rm eff}'(\theta) = {\cal C} \, a^{-2} \,,
\end{equation}
where ${\cal C}$ denotes a coefficient of proportionality that is fixed by the initial conditions. \mbox{Equation~\eqref{CCR}} may be solved by inversion, providing the dynamics of a \mbox{SU(2)} YM condensate in cosmological time. Note that Equation~(\ref{YMCenergyDensity}) is only valid for the dominant electric-field configurations. An~analogical analysis of magnetic-field quantum configurations and their cosmological evolution in the effective Lagrangian approach at the one-loop level has been performed in Ref.~\citep{Savvidy:2021ahq}. Note, generic YM configurations contain both magnetic and electric components. We refer the reader to Section~\ref{sec:cosmologicalAttractors} above for a thorough discussion of those more generic configurations and their cosmological (real-time) dynamics.

\section{Cosmological Implications of Gauge-Fields Driven~Inflation}
\label{Sect:YMinflation}

Gauge-fields driven inflation has been considered in several models, starting form the paper by Ford~\cite{Ford:1989me}, in~which a hypercharge cosmological inflationary scenario was envisaged. The~theory that was taken into account included, besides~the Einstein-Hilbert action, the~action for a massive hypercharge field, individuated by the Lagrangian
\begin{equation}
L=\frac{1}{4} F_{\mu \nu} F^{\mu \nu} + V(A^\alpha A_\alpha)\,.
\end{equation}

Anisotropies that are naturally present in the Maxwell tensor impose to consider a Bianchi type-I metric of the form
\begin{equation}
ds^2 = dt^2 - a^2(t) (dx^2 + dy^2) - b^2(t) dz^2  \,.  
\end{equation}

Provided the expression for the energy-momentum tensor of the massive vector field
\begin{equation}
T_{\mu \nu}= F_{\mu \beta} F_{\nu}^{\ \beta} - \frac{1}{4} g_{\mu \nu} F_{\alpha \beta }  F^{\alpha \beta } - g_{\mu \nu} V + 2 V' A_\mu A_\nu\,,
\end{equation}
with prime denoting the functional derivative in $A^\alpha A_\alpha$, the~Einstein equations were recast as
\begin{equation} \label{FriFor}
2\, \frac{\dot{a} \dot{b}}{ab} + \frac{\dot{a}^2}{a^2}=8 \pi \epsilon\,, \qquad \qquad 2 \frac{\ddot{a}}{a} + \frac{\dot{a}^2}{a^2}= - 8 \pi p_z\,,
\end{equation}
with $\epsilon$ as the energy density and~$p_z$ as the pressure density component along the $z$-axis. The~conservation law, derived assuming that the pressure density components along the $x$ and $y$ axes equal $p_x=p_y$, encodes
\begin{equation} \label{ConFor}
\dot{\epsilon} +\left(2 \frac{\dot{a}}{a} + \frac{\dot{b}}{b} \right) \epsilon + 2\frac{\dot{a}}{a} p_x + \frac{\dot{b}}{b} p_z=0\,,
\end{equation}
and finally, the~only non-vanishing component of the vector field fulfills the equation
\begin{equation} \label{GauFor}
\ddot{A}_z + \left[ 2 \frac{\dot{a}}{a} - \frac{\dot{b}}{b} \right] \dot{A}_z + 2 V' A_z=0\,.
\end{equation}

The solutions to this system of equations, Equations~\eqref{FriFor}--\eqref{GauFor}, immediately outlined the impossibility to suppress anisotropies over the dynamical evolution of the background and the vector field. Thus, already with the seminal attempt of~Ref.~\cite{Ford:1989me}, it was evident that the over-production of anisotropies at the cosmological level would have provided a main issue, eventually ruling out this class of~models.

A possible way to overcome the abundance of anisotropies, severely constrained by the CMB observations by WMAP and Planck, was imagined by Golovnev, Mukhanov and Vanchurin~\cite{Golovnev:2008cf}, who considered a stochastic distribution of $N\simeq 10^{12}$ vector fields, randomly spanning the space directions. Each vector field was assumed to be massive and~regulated by the action
\begin{equation}
S=\int d^4x \sqrt{-g} \left( -\frac{R}{16 \pi} -\frac{1}{4} F_{\mu \nu} F^{\mu \nu} + \frac{1}{2} (m^2 +\frac{R}{6})A_\mu A^\mu \right) \,,    
\end{equation}
where $m$ denotes the mass of the hypercharge vector field considered, such that \linebreak $F_{\mu \nu}=\nabla_{\mu} A_\nu - \nabla_{\nu} A_\mu=\partial_{\mu} A_\nu - \partial_{\nu} A_\mu$. 

The equations of motion for the gauge field, which in a covariant fashion recast
\begin{equation}
\frac{1}{\sqrt{-g}} \frac{\partial}{\partial x^\mu} \left(\sqrt{-g} F^{\mu \nu} \right) + \left( m^2 +\frac{R}{6} \right) A^\nu=0\,, 
\end{equation}
split into a temporal ``0'' component, implying $A_0=0$, and~into space components, which finally provide the equation fro the ``field strength'' $B_i\equiv A_i/a$, i.e.,
\begin{equation}
\ddot{B}_i+ 3 H \do{B}_i +m^2 B_i =0\,,   
\end{equation}
having introduced comoving coordinates such that $ds^2=dt^2-a^2(t)d\vec{x}^2$, and~these were denoted with dot derivative with respect to the cosmological time, so that $H=\dot{a}/a$. For~a homogeneous vector field, it was then noticed that the energy-momentum tensor can be expressed by
\begin{eqnarray}
T^0_0&=&\frac{1}{2} \left(\dot{B}^2_k + m^2 B_k^2 \right) \,, \nonumber\\ 
 T^i_j&=&\left[  -\frac{5}{6} \left( 
 \dot{B}_k^2 - m^2 B_k^2 \right) -\frac{2}{3} H \dot{B}_k B_k - \frac{1}{3}(\dot{H}+ 3H^2 )B_k^2 \right]\delta^i_j+
 \dot{B}^i \dot{B}_j + H(\dot{B}^i B_j  \\ 
 &+& \dot{B}_j B^i) + ( \dot{H} + 3H^2 - m^2 )B^i B_j\,, \nonumber
\end{eqnarray}
where the summation over the index $k$ is meant. Summing up now the contributions of a triplet of mutually orthogonal fields $B^{(a)}_i$ with the same magnitude $|B|$, the~total energy momentum tensor is averaged to the quantity
\begin{equation}
T^0_0= \epsilon= \frac{3}{2} \left( 
 \dot{B}_k^2 - m^2 B_k^2 \right)\,, \qquad \qquad T^i_j = - p \delta^i_j = -\frac{3}{2}  \left( 
 \dot{B}_k^2 - m^2 B_k^2 \right) \delta^i_j\,,
\end{equation} 
where $B_k$ satisfies
\begin{equation}
\ddot{B}_i+ 3 H \dot{B}_i + m^2 B_i=0 \,.
\end{equation} 

These latter relations for the energy-momentum tensor can be proved by the fact that for $B^{(a)}_i$, it holds:
\begin{equation}
\sum_i B^{(a)}_i B_{(b)\ i} = |B|^2 \delta^{(a)}_{(b)} \rightarrow \sum_a B^{(a)}_j B^{(a) \ i} = |B|^2 \delta^{i}_{j}\,.
\end{equation} 

Summing up over a large amount of $N$ triple of fields, the~energy-momentum tensor finally acquires the expression
\begin{equation}
T^0_0= \epsilon \simeq \frac{N}{2} \left( \dot{B}_k^2 + m^2 B_k^2 \right)\, \qquad \qquad T_{i j} \propto \sum_{a=1}^N  B^{(a)}_i B^{(a)}_j \simeq \frac{N}{3} B^2 \delta^i_j + O(1) \sqrt{N} B^2\,.
\end{equation} 

Within this scenario, anisotropies are then proven to fall off as $1/\sqrt{N}$, hence justifying consistency with the experimental~data.

Finally, accounting for an inflationary slow-roll phase, with~$\dot{B}_i\simeq 0$, one finally finds for the first Friedmann equation,
\begin{equation}
H^2=\frac{8 \pi}{3} \epsilon \simeq  \frac{4 \pi}{3} N m^2 B^2 \,.
\end{equation}

A different perspective was suggested by Maleknejad and Sheikh-Jabbari, who imagined in Refs.~\cite{Maleknejad:2012fw,Maleknejad:2011jw,Maleknejad:2011sq} that it could become relevant at the level of cosmic perturbations, while at the same time driving the inflationary background evolution of the universe. Isotropy could be guaranteed here by aligning the internal indices (of the adjoint representation) of a \mbox{SU(2)} subgroup of \mbox{SU($N$)} to the space directions, namely, for~the space component of the connection $A$
\begin{equation}
 A^a_i =\psi(t) e^a_i = \psi(t) a(t) \delta^a_i\,, \qquad \qquad   e^a =  a(t) \delta^a_i\,,
\end{equation}
where $\psi(t)$ is a scalar field, $e^a_i$ denotes the triads, which recombine into the space metric $h_{ij}=e^a_i e^a_j$ and are expressed here in the comoving coordinates. For~compactness of notation, the~authors reshuffled $\phi(t)=\psi(t) a(t)$.

Specifically, the~authors considered a YM action provided with the square of a Pontriagin term, i.e.,
\begin{equation}
S=\int d^4 x \sqrt{-g} \left[ -\frac{R}{2} -\frac{1}{4} F^a_{\mu \nu} F_a^{\mu \nu} + \frac{\kappa}{384} \left(\varepsilon^{\mu \nu \lambda \sigma} F^a_{\mu \nu} F^a_{\lambda \sigma} \right)^2 \right]\,,
\end{equation}
having set $8 \pi G=1$, denoted with $\varepsilon^{\mu \nu \lambda \sigma} $ the totally antisymmetric tensor, and~chosen the real parameter $\kappa$ to be~positive. 

Labeling with a subscript ``YM'' the contributions arising from the YM terms, and~with $\kappa$ the ones related to the Pontriagin terms, it was found that
\begin{equation}
\epsilon_{\rm YM}= \frac{3}{2} \left(\frac{\dot{\phi}^2}{a^2}+ \frac{g^2 \phi^4}{a^4} \right) \,, \qquad \qquad \epsilon_\kappa= \frac{3}{2} \frac{\kappa g^2 \phi^4 \dot{\phi}^2 }{a^6}\,,
\end{equation}
where $g$ is the YM coupling constant entering the definition of the field strength of the \mbox{SU(2)} connection
\begin{equation}
F^a_{\mu \nu}= \partial_\mu A^a_\nu -  \partial_\nu A^a_\mu - g \epsilon^{a b c } \, A^b_\mu A^c_\nu \,,    
\end{equation}
and where $\epsilon=\epsilon_{\rm YM}+ \epsilon_{\kappa}$ and $p=\frac{1}{3} \epsilon_{\rm YM} -\epsilon_{\kappa}$. Having provided these definitions, it was possible to show the emergence of a slow-roll phase of inflation, which was dominated by the $\kappa$-term contribution $\epsilon_\kappa$ over the YM contribution $\epsilon_{\rm YM}$, namely $\epsilon_\kappa \gg \epsilon_{\rm YM}$. Furthermore, within~the scenario introduced in Refs.~\cite{Maleknejad:2012fw,Maleknejad:2011jw,Maleknejad:2011sq}, it was conceivable that sizable effects could be turned on, so as to source primordial magnetic fields at the cosmic perturbation level, with~specific possible observational features on the CMB and primordial cosmic magnetic fields~outlined.

A novel framework with respect to the ones so far discussed was elaborated by Alexander, Marcian\`o and Spergel in Ref.~\cite{Alexander:2011hz}, in~which the authors considered the interaction a model of inflation with as a new ingredient the interaction of an Abelian gauge field with a fermionic charge. This scenario then dramatically differs from only adopting gauge fields in order to generate a realistic inflationary epoch. As~a by-product of this approach, researchers considered the possibility of generating the a net-lepton asymmetry. The~Sakharov conditions are realized in the model presented in Ref.~\cite{Alexander:2011hz} during the inflationary epoch, due to a dynamical inter-change of the gauge field fluctuations into the lepton asymmetry of the~universe.

The action the authors moved from involved, as~well as a U(1) hypercharge field $A_\mu$, also a massive scalar field $\theta$, interacting with $A_\mu$ through a Chern-Simons term, i.e.,
\begin{eqnarray}
S&=&S_D + \int d^4 x \sqrt{-g}\,  \left[ \frac{M_{\rm PL}^2 R}{8\pi} - \frac{1}{2} \partial_\mu \theta \partial^\mu \theta - \frac{1}{4} F_{\alpha \beta}  F^{\alpha \beta} + \frac{\theta}{4 M_*} F_{\alpha \beta}  \tilde{F}^{\alpha \beta}
\right]\,,
\nonumber\\
S_D&=&\int d^4 x \sqrt{-g} \, \left( \imath \bar{\psi} \gamma^\mu \nabla_\mu \psi+ cc. + M \bar{\psi} \psi + q \bar{\psi} \gamma^\mu \psi A_\mu \right)\,, 
\end{eqnarray}
where $M_{\rm PL}$ denotes the Planck mass, $M_*$ is the mass-scale of the pseudo-scalar decay constant, regulating with $theta$ the CP-violating Chern-Simons such as term, $F_{\mu \nu}=\partial_{[\mu} A_{\nu]}$ is the field strength of the U(1) connection $A$, $\tilde{F}^{\alpha\beta}= \varepsilon^{\alpha\beta \lambda \sigma} \tilde{F}_{\lambda \sigma}$ is the gravitational Hodge-dual, and~$\gamma^\mu=e^\mu_I \gamma^I$, with~$\gamma^I$ Dirac matrices, $e^\mu_I$ inverse tetrad and internal indices $I=0, \dots 3$. 

The model, which is then based on the interaction between a homogeneous and isotropic configuration of a U(1) gauge field and a fermionic charge density $\mathcal{J}_0$, relies on the regulated fermionic charge density as generated from a Bunch-Davies vacuum state, using the procedure outlined by Koksma and Prokopec in~Ref.~\cite{Koksma:2009tc}. In~conformal coordinates, this was found to redshift as $1/a(\eta)$. Then, within~the scenario of Ref.~\cite{Alexander:2011hz}, the~time-like component of the hypercharge gauge field was found to be sourced by the fermionic charge, consistently with a growth in the gauge field proportional to the scale factor, namely,
\begin{equation}
A_0(\eta) \sim a(\eta)\,.     
\end{equation}

This motivates results with inflation dominated by the energy density stored within the interaction among the gauge field and the fermionic charge, namely $A_0 \mathcal{J}_0$, which is approximately constant over the inflationary epoch. The~appealing feature of Ref.~\cite{Alexander:2011hz} stands in the possibility to obtain an epoch of cosmic inflation involving the physical  description fields already existing in nature, specifically the time-like U(1) gauge field interacting with a fermionic charge density. Nonetheless, the~role of the scalar field cannot be underestimated, retaining a certain relevance in producing baryogenesis, and~providing a graceful exit from inflation. Indeed, the~mechanism that accounted for the graceful exit is strictly interconnected to the one advocated for reproducing the baryogenesis, with~the right baryon asymmetry index. The~Chern-Simons term, through the coupling to the pseudo-scalar field, converts gauge field fluctuations into lepton number, while the rapid oscillation of the pseudo-scalar field near its minimum allows achieving thermalization of the gauge field and thus to end inflation. The~relevance of the coupling between scalar modes, there interpreted as axions, was further investigated in Ref.~\cite{Domcke:2019mnd}.

An improvement of the scheme first addressed in Ref.~\cite{Alexander:2011hz} was provided in Ref.~\cite{Alexander:2014uza}, where the authors analyzed the consistency of the model via the St\"uckelberg mechanism~\cite{Stueckelberg:1938zz}. This provided an incorporation of the longitudinal scalar DoFs into the hypercharge field, which is now massive. This could be thought again as a further step-forward, toward the realization of an inflationary mechanism relying on the YM dynamics, the~hypercharge sector being eventually recognized as an Abelian subgroup of the \mbox{SU($N$)} gauge~sector.

The action of the theory was then considered to be
\begin{eqnarray}
S= \int d^4 x \sqrt{-g} \left[ \frac{M_{\rm PL}^2 R}{8 \pi} -\frac{1}{4} G_{\alpha \beta} G^{\alpha \beta} - \frac{1}{2} m^2 C_\mu C^\mu + C_\mu \mathcal{J}^\mu + L_{\rm D}   \right] \nonumber\\
L_{\rm D}= - \imath \bar{\psi} \gamma^\mu \nabla_\mu \psi + c.c. + M \bar{\psi} \psi \,,
\end{eqnarray}
having introduced the massive St\"uckelberg field $C_\mu = A_\mu -\frac{1}{m} \partial_\mu \theta$ and its field strength $G_{\mu \nu} = \partial_{[\mu} C_{\nu]}=\partial_{[\mu} A_{\nu]}=F_{\mu \nu}$, and~the fermionic vector current $\mathcal{J}^\mu= q \bar{\psi} \gamma^\mu \psi$. Gauge invariance is ensured in this framework by the transformations
\begin{equation}
A_\mu \rightarrow  A_\mu ' =  A_\mu  +\partial_\mu \Lambda \,, \qquad \qquad \theta \rightarrow  \theta ' =  \theta  + m \Lambda\,.
\end{equation}

Within this framework, the~authors could prove the existence and the stability of dynamical attractor solutions for the cosmological inflation epoch, which is again driven by the coupling among the fermions and a (massive) gauge field. Numerical analyses then showed that stability is attained for a large basin of the initial conditions, making this inflationary scenario almost independent on these latter: inflation arises without fine tuning and~without the need of postulating any effective potential or any non-standard~coupling. 

An alternative scenario featuring a coupling between an axion-like field and an \mbox{SU($2$)} gauge field is known as the chromo-natural inflation~\cite{Adshead:2012kp}. The~rotationally invariant homogeneous condensate of the gauge field satisfies an attractor solution that enables it to drive cosmic inflation for the axion decay constant having a natural value at a sub-Planckian scale. Interestingly enough, this scenario features a possibility for termination of inflation as soon as the axion potential vanishes, simultaneously providing a small tensor-to-scalar perturbation~ratio.

An inflationary scenario, taking into account non-trivial topological features deployed in~Refs.~\cite{Barvinsky:2006uh, Barvinsky:2006tu, Barvinsky:2007vb}, was extended in~Ref.~\cite{Barvinsky:2017lfl} so as to account, over~the universe expansion, for~a sector of a strongly coupled QCD-like gauge theory. The~idea at the base of investigations in~Refs.~\cite{Barvinsky:2006uh, Barvinsky:2006tu, Barvinsky:2007vb} is to perform a periodic (between the $\Sigma$ and $\Sigma'$ surfaces) path integral over Euclidean geometries,
\begin{equation}
e^{-\Gamma}=\int_{ {g,\phi}|_\Sigma={g,\phi}_{\Sigma'} } D[g, \phi] e^{-S_E{g,\phi}}\,,
\end{equation}
with $g$ and $\phi$, respectively, representing the metric and matter field and $S_E$ denoting the Euclidean action, so as to extract the `density matrix' of the universe,
\begin{equation}
\rho[\varphi, \varphi']= e^{\Gamma} \int_{ {g,\phi}|_{\Sigma,\Sigma'} } D[g, \phi] e^{-S_E[g,\phi]}
\,,
\end{equation}
which describes a microcanonical ensemble, $\varphi$ denoting field configurations that encode both gravitational and matter~variables.  

\textls[-15]{The uniform distribution over the Euclidean spaces actually corresponds, over~Lorentzian spaces, to~a distribution that is peaked about complex saddle points of the path integral. The~latter can be then represented by cosmological instantons, entailing a bounded range values for the cosmological~constant.}

On the other hand, inflationary cosmologies can be engendered by the very same instantonic solutions~\cite{Barvinsky:2006uh, Barvinsky:2006tu, Barvinsky:2007vb}. The~low energy of the accelerated expansion can be then attained at its late stage, resorting to the dynamical evolution of extra dimensions specifically postulated in string theory framework~\cite{Barvinsky:2007vb}. This results in a bounded range for the very early (inflationary) cosmological constant, which provides a constraint on the available landscape of the string vacua. Finally, the~same mechanism can be advocated to give rise to a possible DE candidate, accounting for the quasi-equilibrium decay of the microcanonical state of the universe. Within~this scenario, Barvinsky and Zhitnitsky promoted a new picture for the emergence of an inflationary spacetime~\cite{Barvinsky:2017lfl}, resorting to considerations developed in Refs.~\cite{Zhitnitsky:2013pna, Zhitnitsky:2014aja, Zhitnitsky:2015dia} on the generation, in~a strongly coupled QCD-like gauge theory, of~the vacuum energy from non-trivial topological~features. 

The limits of the usual semi-classical expansion were overcome by the dominant contribution of the numerous conformal modes. Integration over the modes then provides the quantum effective action of the conformal field theory $\Gamma_{\rm CFT}[g_{\mu \nu}]$, which can be calculated with methods similar to those ones implemented in determining the conformal anomaly. Starting from the FLRW background, accounting for a periodic factor $a(\tau)$ --- this is due to the fact that functions of the Euclidean time are supported to the circle $\mathbb{S}^1$ --- and finally using a local conformal transformation to the static Einstein universe and the very same well-known trace anomaly, one finds
\begin{equation}
  g_{\mu \nu}
  =
  \frac{\delta \Gamma_{\rm CFT} }{ \delta g_{\mu \nu}}
  =
  \frac{1}{4 (4\pi)^2} g^{1/2}\,
  \left(
    \alpha \Box R + \beta E \gamma C^2_{\mu\nu\rho\sigma}        
\right)  \,,    
\end{equation}
where we introduced the Gauss-Bonnet term $E=R^2_{\mu \nu \alpha \gamma }- 4 R^2_{\mu \nu }+R^2$ and the Weyl tensor $C_{\mu\nu\rho \sigma}$.

Considering a spacetime with topology $\mathbb{S}^3 \times \mathbb{S}^1$, and~moving from the expression of the energy of the gauge field holonomy, winding across the compactified coordinate of the length $\mathcal{T}$, Barvinsky and Zhitnitsky found that
\begin{equation}
\rho=\rho_{\rm vac} [\mathbb{S}^3 \times \mathbb{S}^1 ] - \rho_{\rm vac} [\mathbb{R}^4]=\frac{ \bar{c}_{\mathcal{T}} \Lambda^3_{\overline{QCD}}  }{\mathcal{T}}\,,   
\end{equation}
with $\Lambda_{\overline{QCD}}$ being the scale of the QCD-like gauge theory, $\bar{c}_{\mathcal{T}}$ being a dimensionless constant of order one, and~similarly, the~full period of the proper Euclidean time on these periodic $m$-fold garland instantons is given by the analogous integral,
\begin{equation}
\mathcal{T}=\oint_{\mathbb{S}_1} d\tau \,, 
\end{equation}
which in an FLRW metric background reduces to the $2m$-multiple result
\begin{equation}
\mathcal{T}= 2 m \int_{a_-}^{a^+} \frac{da}{a} \,, 
\end{equation}
where the integral is between the two neighboring turning points of $a(\tau)$ such that $\dot{a}(\tau_\pm)=0$.

\section{Summary}
\label{Sect:summary}

In this review, we have made a brief outlook of the current status of confined and de-confined QCD dynamics in the early universe as well as the key methodology for studies QCD in the strongly coupled regime relevant for cosmological evolution. The~covered research areas are broadly inter-disciplinary, and~our discussion may not be fully exhaustive. Still, we have identified a few quite unexpected and intriguing connections between currently pursued research in particle physics and possible dynamics of the early universe. Such fundamental questions as the gauge-fields-driven inflation, cyclic universe, particle production mechanisms, non-perturbative real-time dynamics of the QCD ground state, a~rather challenging problem of dynamical generation of cosmological DE and DM, the~structure of the QCD vacuum, the~QCD phase transitions and the role of QCD matter in late-time universe evolution are among the key points of this review. Such a wide breadth of topics, with~deep roots into QCD or, more generically, quantum YM field theories, exhibits enormous and critical significance of microscopic dynamics of particle physics and confined field theories for understanding of the macroscopic cosmic evolution. The~picture is far from its final shape though, and~many more pillars of such connections and possible interplay are yet to be established. We believe that our review can be useful for both young researchers and for more senior experts specialized in both particle physics and cosmology research~areas.

\vspace{6pt} 

\authorcontributions{Conceptualization, A.A., A.M., T.L., R.P. and M.\v{S}.; validation, R.P., T.L. and M.\v{S}.; formal analysis, R.P., T.L. and M.\v{S}.; investigation, R.P., T.L. and M.\v{S}.; resources, R.P. and M.\v{S}.; writing --- original draft preparation, A.M., T.L., R.P. and M.\v{S}.; writing --- review and editing, A.A., A.M., T.L., R.P. and M.\v{S}.; supervision, R.P.; project administration, R.P.; funding acquisition, R.P. and M.\v{S}. All authors have read and agreed to the published version of the manuscript.}
\funding{Work of A.A.~is supported by the Talent Scientific Research Program of College of Physics, Sichuan University, Grant No.1082204112427 \& the Fostering Program in Disciplines Possessing Novel Features for Natural Science of Sichuan University,  Grant No. 2020SCUNL209 and 1000 Talent program of Sichuan province 2021. A.M.~wishes to acknowledge support by the Shanghai Municipality, through the grant No.~KBH1512299, by~Fudan University, through the grant No.~JJH1512105, the~Natural Science Foundation of China, through the grant No.~11875113, and~by the Department of Physics at Fudan University, through the grant No.~IDH1512092/001. R.P.~is supported in part by the Swedish Research Council grant, contract number 2016-05996, as~well as by the European Research Council (ERC) under the European Union's Horizon 2020 research and innovation programme (grant agreement No 668679). M.\v{S}.~is partially supported by the grants LTT17018 and LTT18002 of the Ministry of Education of the Czech~Republic.}
\institutionalreview{Not applicable.}
\informedconsent{Not applicable.}
\dataavailability{Not applicable.} 
\conflictsofinterest{The authors declare no conflict of interest. The funders had no role in the design of the study; in the collection, analyses, or interpretation of data; in the writing of the manuscript, or in the decision to publish the results.} 

\appendixtitles{yes}
\appendixstart
\appendix
\section[\appendixname~\thesection]{Elements of Relativistic~Hydrodynamics}\label{App:A}

Equations of relativistic hydrodynamics are based on conservation of the energy-momentum and the current 
\begin{equation}
\partial_\mu T^{\mu\nu} = 0\,, \quad \partial_\mu j_i^\mu = 0\, ,
\label{eq:conservationT}
\end{equation}
where $j_i^\mu, i=B, Q, L, \ldots$ denotes the conserved currents corresponding to baryon number $B$, electric charge $Q$, lepton number $L$, etc. Both $T^{\mu\nu}$ and $j_i^\mu$ can be decomposed into time-like and space-like components using natural projection operators, the~local flow four-velocity $u^\mu$,  and~the second-rank tensor perpendicular to it $\Delta^{\mu\nu} = g^{\mu\nu} - u^\mu u^\nu$\mbox{ \cite{Yagi:2005yb, Hirano:2008hy, Kovtun:2012rj, Romatschke:2017ejr}}:
\begin{eqnarray}
T^{\mu\nu} &=& \epsilon u^\mu u^\nu - p\Delta^{\mu\nu} + W^\mu u^\nu + W^\nu u^\mu + \pi^{\mu\nu}\,,\label{eq:decompositionT}\\
j_i^\mu    &=& n_i u^\mu + V_i^\mu\,,
\label{eq:decompositionN}
\end{eqnarray}
where $\epsilon=u_\mu T^{\mu\nu}u_\nu$ is the energy density, $p\equiv p_s+\Pi  = -\frac13\Delta_{\mu\nu}T^{\mu\nu}$ is the total (hydrostatic $p_s$ + bulk $\Pi$) pressure, $W^\mu=\Delta^{\mu}_{ \alpha}T^{\alpha\beta}u_\beta$ is the energy (or heat) current, $n_i=u_\mu j^{\mu}_i$ is the charge density, $V_i^\mu = \Delta^{\mu\ }_{\ \nu} j_i^\nu$ is the charge current, and~$\pi^{\mu\nu} = \left< T^{\mu\nu}\right> $ is the shear stress tensor. The~angular brackets in the definition of the shear stress tensor $\pi^{\mu\nu}$ stand for the following operation:
\begin{equation}
\left<A^{\mu\nu}\right> = \left[\frac12(\Delta^{\mu\ }_{\ \alpha}\Delta^{\nu\ }_{\ \beta}+\Delta^{\mu\ }_{\ \beta}\Delta^{\nu\ }_{\ \alpha})-\frac13\Delta^{\mu\nu}\Delta_{\alpha\beta}\right]A^{\alpha\beta}\,.
\label{eq:bracket}
\end{equation}

To further simplify our discussion, we restrict ourselves in the following to only the one conserved charge, the~baryon number $B$, and~denote the corresponding baryon current as $j^\mu \equiv j_B^\mu$. The~various terms appearing in the decompositions (\ref{eq:decompositionT}) and (\ref{eq:decompositionN}) can then be grouped 
into ideal and dissipative parts as follows
\begin{eqnarray}
T^{\mu\nu} &=& T_{id}^{\mu\nu} + T_{dis}^{\mu\nu} = \left[\epsilon u^\mu u^\nu - p_s\Delta^{\mu\nu}\right]_{id} +\left[-\Pi\Delta^{\mu\nu}+W^\mu u^\nu+W^\nu u^\mu + \pi^{\mu\nu}\right]_{dis}
\label{Tmn}\\
j^\mu &=& j_{id}^\mu + N_{dis}^\mu \; =  \left[nu^\mu \right]_{id} + \left[V^\mu \right]_{dis} \,.
\label{jm}
\end{eqnarray}

Neglecting the dissipative parts, the~energy-momentum conservation and the current conservation (\ref{eq:conservationT}) define ideal hydrodynamics. In~this case (and for a single conserved charge), a~solution of the hydrodynamical Equation~(\ref{eq:conservationT}) for a given initial condition describes the spacetime evolution of the six variables --- three state variables $\epsilon(x)$, $p(x)$, $n(x)$, and~three space components of the flow velocity $u^\mu$. However, since (\ref{eq:conservationT}) constitutes only five independent equations, the~sixth equation relating $p$ and $\epsilon$, the~EoS $p(\epsilon)$, has to be added by hand in order to solve~them.

Two definitions of flow can be found in the literature, see e.g.,~Refs.~\cite{Yagi:2005yb, Hirano:2008hy}; one related to the flow of conserved charge (Eckart):
\begin{equation}
u_E^\mu=\frac{j^\mu}{\sqrt{j_\nu j^\nu}}\,,
\end{equation}
the other related to the flow of energy (Landau): 
\begin{equation}
\label{eq:landau}
u_L^\mu=\frac{T^{\mu\ }_{\ \,\nu}u_L^{\nu}}{\sqrt{u_L^\alpha T_{\alpha\ }^{\ \beta}T_{\beta\gamma}u_L^{\gamma}}}=\frac{1}{e}T^{\mu\ }_{\ \,\nu}u_L^{\nu}\,.
\end{equation}

Let us note that $W^\mu = 0$ ($V^\mu = 0$) in the Landau (Eckart) frame. In~the case of vanishing dissipative currents, both definitions represent a common flow. 
The Landau definition is more suitable when describing the evolution of matter at zero chemical potential, such as in the case of the mid-rapidity particle production in ultra-relativistic HIC at the LHC and at the top RHIC energy, or~in the early universe. In~this case, all momentum density is due to the flow of energy density, $u_{\mu}T_{id}^{\mu\nu}=\epsilon u^{\nu}$ and  $u_{\mu}T_{dis}^{\mu\nu} = 0$, i.e.,~the heat conduction effects can be~neglected.

\section{Hydrodynamical Description of Dissipative~Effects}
\label{App:B}

In its modern formulation, relativistic fluid dynamics provides an effective description of a system that is in local thermal equilibrium, and~it can be derived from the underlying kinetic description through Taylor expansion of the entropy four-current $S^\mu$ = $s u^\mu$ in gradients of the local thermodynamic variables~\cite{Romatschke:2017ejr}. In~zeroth order in gradients, one obtains ideal fluid dynamics 
\begin{equation}
\partial_{\mu}S^{\mu}=\partial_{\mu}(s u^\mu)=u^{\mu}\partial_{\mu}s+ s\partial_{\mu}u^\mu=0\,, \label{eq:ent1}   
\end{equation}
and the evolution of the scale factor of the universe is driven solely by  the entropy conservation  $s(t) a^3(t)={\rm const}$. The~higher orders describe effects due to irreversible thermodynamic processes such as the frictional energy dissipation between the fluid elements that are in relative motion or~their heat exchange with its surroundings on its way to approach thermal equilibrium with the whole~fluid. 

When solving the hydrodynamic equations with the dissipative terms, it is customary to introduce the following two phenomenological definitions (so-called \textit{constitutive equations}) for the shear stress tensor $\pi^{\mu\nu}$ and the bulk pressure $\Pi$  appearing in \mbox{Equation}~(\ref{Tmn})~\cite{Hirano:2008hy},
\begin{eqnarray}
\pi^{\mu\nu} = 2\eta\left<\nabla^{\mu}u^\nu\right>\,, \quad
\Pi = -\zeta \partial_\mu u^\mu = -\zeta \nabla_\mu u^\mu\,, 
\label{eq:bp}
\end{eqnarray}
where the angular brackets $\langle \ldots \rangle$ are defined in Equation~(\ref{eq:bracket}) and $\nabla^{\mu}=\Delta^{\mu\nu}\partial_{\nu}$. Neglecting the charge current $V_{\mu}$ in Equation~(\ref{jm}), the first-order expansion of $S^\mu$ is completely determined by the \textit{shear viscosity} $\eta$ and \textit{bulk viscosity} $\zeta$ coefficients~\cite{Hirano:2008hy}:
\begin{equation}
T \partial_{\mu} S^{\mu}= 
\pi_{\mu \nu}\langle  \nabla^{\mu} u^{\nu} \rangle -\Pi \partial_{\mu}u^{\mu} =
\frac{\pi_{\mu \nu}\pi^{\mu \nu}}{\eta} + \frac{\Pi^2}{\zeta} =
2\eta \langle  \nabla^{\mu} u^{\nu} \rangle ^2 + \zeta (-\partial_{\mu}u^{\mu})^2 \geq 0\,.
\label{eq:ent2}
\end{equation}

A well-known example of the flow involving both coefficients $\eta$ and $\zeta$ is provided by boost-invariant one-dimensional expansion with the velocity in the $z$ direction, $v_z$, proportional to $z$ co-ordinate~\cite{Bjorken:1982qr}
\begin{eqnarray}
u^\mu_{\mathrm{BJ}}=\frac{x^\mu}{\tau} = \frac{t}{\tau}\left(1,0,0,\frac{z}{t}\right)\,,
~~\tau=\sqrt{t^2-z^2}\,.
\label{eq:be0}
\end{eqnarray}

After inserting this solution into the constitutive Equation~(\ref{eq:bp}), we arrive at the equation of motion~\cite{Hirano:2008hy}:
\begin{equation}
\frac{d\epsilon}{d\tau}=-\frac{\epsilon+p_s}{\tau}\left(1-\frac{4}{3\tau T}\frac{\eta}{s}-\frac{1}{\tau T}\frac{\zeta}{s}\right)\,.
\label{eq:be1}
\end{equation}

The last two terms on the right-hand side in Equation~(\ref{eq:be1}) describe a compression of the energy density due to viscous corrections. Two dimensionless coefficients in the viscous correction, $\eta/s$ and $\zeta/s$, where $s$ is the entropy density, reflect the intrinsic properties of the fluids. It is worth mentioning  that neglecting   $\eta$ and $\zeta$ in Equation~(\ref{eq:be1}), i.e.,~for the ideal fluid EoS with $p_s=\frac{1}{3}\epsilon$, one obtains the celebrated Bjorken solution of ideal hydrodynamics~\cite{Bjorken:1982qr} 
\begin{equation}
\frac{d\epsilon}{d\tau}=-\frac{\epsilon+p_s}{\tau} = -\frac{4}{3}\frac{\epsilon}{\tau} \qquad
\Rightarrow \qquad \epsilon=\tau^{-4/3} \,,
\label{eq:be2}
\end{equation}
frequently used to discuss the salient features of the ultra-relativistic~HICs. 

The one-dimensional character of the Bjorken flow (\ref{eq:be0}) makes it possible to replace the $z$ co-ordinate with the radial one $r$. Radial flow in the transverse direction, i.e.,~when $r_{\perp}=\sqrt{x^2+y^2}$, was studied in Ref.~\cite{Chojnacki:2004ec}. For~this case and the constant sound velocity $c_s$, analytic solutions of relativistic viscous hydrodynamics describing expanding fireballs were developed in Ref.~\cite{Csanad:2019lcl}. In~a three-dimensional case, a new class of exact fireball solutions of relativistic dissipative hydrodynamics for arbitrary shear and bulk viscosities, as~well as for other dissipative coefficients, was studied in Ref.~\cite{Csorgo:2020iug}. The~common property of these solutions is the presence of the relativistic Hubble~flow.

However, the~analogy between the solutions describing HICs and expansion of the early universe must not be pushed too far, since in the latter case form of the energy-momentum tensor $T_{\mu \nu}$ and particle four-current $j_{\mu}$ of the matter, cf.~\mbox{Equations~(\ref{Tmn}) and (\ref{jm})} is strongly constrained by the symmetries of the FLRW metric (\ref{eq:FLRW}). In~particular, due to the local momentum isotropy, the term $\left<\nabla^{\mu}u^\nu\right>$ appearing in the viscous shear-stress tensor $\pi^{\mu\nu}$, cf.~Equation~(\ref{eq:bp}), vanishes~\cite{Maartens:1996vi}. Consequently, the~term proportional to $\eta$ in Equation~(\ref{eq:be1}) disappears. The~shear viscosity $\eta$ also disappears in the theories with scalar perturbations of the metric tensor $g_{\mu\nu}$  \cite{Mukhanov:2005sc,Abe:2020sqb}. There, the fluctuations of energy density destroy the homogeneity but not the isotropy of the early universe FLRW metrics. Hence, in~the following discussion, we will consider mainly the bulk viscosity --- the property of expanding matter arising typically in mixtures. They can be either of different species, as~in a radiative fluid, or~of the same species but with different energies, as~in a Maxwell–Boltzmann gas. In~each of these instances, the bulk viscosity provides the internal ‘friction’ that sets in due to the different cooling rates in the expanding mixture~\cite{Maartens:1996vi}.

The relativistic Navier-Stokes description given by Equation~(\ref{eq:bp}) accounts only for terms that are \textit{linear} in velocity gradient. This leads, unfortunately, to~severe problems. In~particular, when  the thermodynamic force $\langle\nabla^{\mu}u^\nu\rangle$ or $\nabla_\mu u^\mu$ is suddenly switched off/on, the corresponding thermodynamic flux $\pi^{\mu\nu}$ or $\Pi$ which is a purely local function of the velocity gradient also instantaneously vanishes/appears~\cite{Muronga:2003ta}. The~linear proportionality between dissipative fluxes and forces causes an instantaneous (acausal) influence on the dissipative currents, leading to numerical instabilities~\cite{Hiscock:1983zz}. The solution of this problem requires the inclusion of terms that are second order in gradients~\cite{Gale:2013da}. The~resulting equations for the dissipative fluxes $\pi^{\mu\nu}$  and $\Pi$  then become the relaxation-type equations~\cite{Hirano:2008hy,Jaiswal:2016hex}.  The~latter encode the time delay between the appearance of thermodynamic gradients that drive the system out of local equilibrium and the associated build-up of dissipative flows in response to these gradients, thereby restoring causality~\cite{Jaiswal:2016hex}. Accounting for non-zero relaxation times at all stages of the evolution constrains departures from local equilibrium, thereby both stabilizing the theory and improving its quantitative~precision.  

Let us provide a few examples of this approach. The~$2^{\rm nd}$-order theory version of boost-invariant one-dimensional flow, cf.~Equation~(\ref{eq:be1}), can be found in Ref.~\cite{Hirano:2008hy}. Due to its length, we do not reproduce it here and refer the interested reader to the original publication. The second example can be found in Ref.~\cite{Medina:2019skp}, where  the relaxation time $\tau_{\pi}$ proportional to the shear viscosity parameter $\eta$  was used to study the evolution of the universe filled with QGP with nonzero shear viscosity. The~authors argue that in general relativity, the following modification of the shear-stress tensor
\begin{adjustwidth}{-\extralength}{0cm}
\begin{equation}
\pi^{\mu \nu} \rightarrow \pi^{\mu \nu}+
\tau_{\pi}\left[u^{\alpha}\pi^{\mu \nu}_{;\alpha}+\frac{4}{3}\pi^{\mu \nu}\nabla_{\alpha}u^{\alpha}  \right] = 2\eta\left<\nabla^{\mu}u^\nu\right> \,, \qquad 
\pi^{\mu \nu}_{;\alpha}\equiv \partial_{\alpha} \pi^{\mu \nu}+
\Gamma^{\mu}_{\alpha \beta}\pi^{\beta\nu}+
\Gamma^{\nu}_{\alpha \beta} \pi^{\mu\beta} \,,
\label{eq:med1}
\end{equation}
	\end{adjustwidth}
where $\pi^{\mu \nu}_{;\alpha}$ is a covariant derivative of $\pi^{\mu\nu}$ and $\Gamma^{\mu}_{\alpha \beta}$ are the Christoffel symbols, makes the resulting Navier-Stokes equations causal. Using the FLRW metric and taking into account that the compatibility with the isotropy and homogeneity of the universe demands $\pi^{\mu \nu}$ to be diagonal, the~solution of Equation~(\ref{eq:med1}) reads~\cite{Medina:2019skp}
\begin{equation}
\pi^{00}(t)=\pi^{00}(t_0)\left [\frac{a(t_0)}{a(t)}\right]^4 e^{-\frac{t-t_0}{\tau_{\pi}}} \,, \qquad
\pi^{ij}(t)=\pi^{ij}(t_0)\left [\frac{a(t_0)}{a(t)}\right]^6 e^{-\frac{t-t_0}{\tau_{\pi}}} \delta_{ij} \,.
\label{eq:med2}
\end{equation}

In the Friedmann equations, the effect of the traceless viscosity tensor shows up in the modification of the initial energy density $\epsilon(t_0)$ and in the behavior of the energy density at times $t\lesssim t_0 +\tau_{\pi}$, which at later times goes over to the standard expression~\cite{Medina:2019skp}
\begin{equation}
\epsilon(t)=\left[\epsilon(t_0)+\pi^{00}(t_0)\right]\left [ \frac{a(t_0)}{a(t)} \right ]^4
-\pi^{00}(t_0)\left [ \frac{a(t_0)}{a(t)} \right ]^4 e^{-\frac{t-t_0}{\tau_{\pi}}}\, .
\end{equation}

The third example is provided in Ref.~\cite{Bemfica:2020zjp} where causal general relativistic viscous fluid theory with the inclusion of all dissipative contributions (shear viscosity $\eta$, bulk viscosity $\zeta$, and~heat flow $W^{\mu}$) and the effects from nonzero baryon number are discussed. According to the authors, the applicability of this theory ranges from the modeling of viscous effects in neutron star mergers to low-energy~HICs.

\section{YM Equations of Motion in the Effective Action~Approach}
\label{App:C}

This section provides a short summary on the derivation of the YM equations of motion by means of variational methods with respect to the connections ${\mathcal A}_\mu^a$ when applied to the effective Lagrangian of \eq{YMeffectiveLagrangian}, as was performed in~Ref.~\cite{Addazi:2018fyo}.

When varying the effective action with respect to ${\mathcal A}_\mu^a$ and $\partial_\nu {\mathcal A}_\mu^a$, one arrives at the Euler-Lagrange equations of motion:
\beql{ELeom}
  \frac{\partial {\mathcal L}_\text{eff}}{\partial {\mathcal A}_\mu^a}
  -
  \nabla_\nu \frac{\partial {\mathcal L}_\text{eff}}{\partial (\partial_\nu {\mathcal A}_\mu^a)}
  =
  0,
  \qquad
  \nabla_\nu \frac{\partial {\mathcal L}_\text{eff}}{\partial (\partial_\nu {\mathcal A}_\mu^a)}
  =
  \frac{1}{\sqrt{-g}} \partial_\nu \bigg[
  \sqrt{-g}
  \frac{\partial {\mathcal L}_\text{eff}}{\partial (\partial_\nu {\mathcal A}_\mu^a)}
  \bigg].
\eeq

It is straightforward to compute the derivatives of the effective Lagrangian:
\begin{adjustwidth}{-\extralength}{0cm}
\begin{align}
  \label{eq:diffEffectiveLagrangian1}
  & \frac{\partial {\mathcal L}_\text{eff}}{\partial {\mathcal A}_\mu^a}
  =
  \frac{1}{4{\bar g}^2} \bigg[
    \frac{\partial {\mathcal J}}{\partial {\mathcal A}_\mu^a}
    -
    \frac{\mathcal J}{{\bar g}^2}
    \frac{\partial {\bar g}^2}{\partial {\mathcal A}_\mu^a}
  \bigg],
  \qquad
  \frac{\partial {\mathcal J}}{\partial {\mathcal A}_\mu^a}
  =
  \frac{4f^{abc}{\mathcal F}^{b\,\mu\nu}{\mathcal A}_\nu^c}{\sqrt{-g}},
  \qquad
  \frac{\mathcal J}{{\bar g}^2}
  \frac{\partial {\bar g}^2}{\partial {\mathcal A}_\mu^a}
  =
  \frac{\mathcal J}{{\bar g}^2}
  \frac{\partial {\bar g}^2}{\partial {\mathcal J}}
  \frac{\partial {\mathcal J}}{\partial {\mathcal A}_\mu^a},
  \\
  \label{eq:diffEffectiveLagrangian2}
  & \Rightarrow \quad
  \frac{\partial {\mathcal L}_\text{eff}}{\partial {\mathcal A}_\mu^a}
  =
  \frac{1}{{\bar g}^2}
  \frac{f^{abc} {\mathcal F}^{b\,\mu\nu}{\mathcal A}_\nu^c}{\sqrt{-g}}
  \Big[ 1 - \frac{\beta}{2} \Big],
  \\
  \label{eq:diffEffectiveLagrangian3}
  & \frac{\partial {\mathcal L}_\text{eff}}{\partial (\partial_\nu {\mathcal A}_\mu^a)}
  =
  \frac{4{\mathcal F}^{a\,\mu\nu}}{\sqrt{-g}},
  \\
  \label{eq:diffEffectiveLagrangian4}
  & \Rightarrow \quad
  \nabla_\nu
  \frac{\partial {\mathcal L}_\text{eff}}{\partial (\partial_\nu {\mathcal A}_\mu^a)}
  =
  \frac{1}{\sqrt{-g}} \partial_\nu\left(
    \sqrt{-g}
    \frac{1}{{\bar g}^2} \frac{{\mathcal F}^{a\,\mu\nu}}{\sqrt{-g}}
    \Big[ 1 - \frac{\beta}{2} \Big]
  \right),
\end{align}
	\end{adjustwidth}
where the RG equation for the exact $\beta$-function that conveniently recasts \eq{RGequation} as
\beql{RGequationRecast}
  \frac{\mathcal J}{{\bar g}^2} \frac{\partial {\bar g}^2}{\partial \mathcal J}
  \equiv
  \frac{\beta}{2}
  =
  \frac{\mathrm d \ln \lvert {\bar g}^2 \rvert}{\mathrm d \ln \lvert \mathcal J \rvert/\mu_0^4},
  \quad
  \beta = \beta({\bar g}^2),
\eeq
has been inserted. If~$\beta$ is known to all-loop order, the~running of the coupling and hence the solutions to the equations of motion may be found to all-loop order accuracy. In~the expression above, an~arbitrary dimensionful renormalisation parameter $\mu_0$ has been explicitly introduced as a reference scale. The~natural boundary condition is \mbox{$\bar g(\mathcal J) \to g_\text{YM}$} when \mbox{$\lvert \mathcal J \rvert \to \mu_0^4$}.

Inserting Equations~\eqref{eq:diffEffectiveLagrangian2} and \eqref{eq:diffEffectiveLagrangian4} into \eq{ELeom}, the~resulting equations of motion is
\beql{collectingEom}
  \frac{1}{{\bar g}^2}
  \frac{f^{abc}{\mathcal F}^{b\,\mu\nu}{\mathcal A}_\nu^c}{\sqrt{-g}}
  \Big[ 1 - \frac{\beta}{2} \Big]
  -
  \frac{1}{\sqrt{-g}} \partial_\nu
  \bigg(
    \sqrt{-g}\frac{1}{{\bar g}^2}
    \frac{{\mathcal F}^{a\,\mu\nu}}{\sqrt{-g}}
    \Big[ 1 - \frac{\beta}{2} \Big]
  \bigg)
  =
  0 \,.
\eeq

This can be rewritten on the operator form
\beql{operatorForm}
  \hat{\mathcal D}_\nu^{ab}
  \bigg[
    \frac{{\mathcal F}^{b\,\mu\nu}}{{\bar g}^2\sqrt{-g}}
    \bigg( 1 - \frac{\beta}{2} \bigg)
  \bigg]
  = 0 \,,
\eeq
  where the differential operator $\hat{\mathcal D}$ is given by
\beql{differentialOperator}
  {\hat{\mathcal D}}_\nu^{ab}
  \equiv
  \delta^{ab} \frac{\partial_\nu \sqrt{-g}}{\sqrt{-g}} - f^{abc}{\mathcal A}_\nu^{c}.
\eeq

The action of this differential operator on a function \mbox{$h(x)$} is defined as follows:
\beql{operatorAction}
  \big[ \hat{\mathcal D} h(x) \big]_\nu^{ab}
  \equiv
  \delta^{ab}\frac{\partial_\nu \big[ \sqrt{-g} h(x) \big]}{\sqrt{-g}}
  -
  f^{abc} {\mathcal A}_\nu^c h(x) \,.
\eeq

\section{One-Loop Effective YM~Lagrangian}
\label{sec:one-loop}

Let us briefly discuss the effective YM theory at the one-loop order. The~usefulness of studying the one-loop case is further motivated by a comparison of the one-loop and all-loop order expansion in Section~\ref{sec:SUNFRG}. The~standard one-loop \mbox{SU($N$)} $\beta$-function reads (see, e.g.,~Ref.~\cite{Deur:2016tte})
\beql{oneLoopBeta}
  \beta_1 \equiv -B_1 {\bar g}_1^2 \,,
  \quad
  B_1 = \frac{bN}{48\pi^2} \,,
  \quad
  b = 11 \,,
\eeq
and the corresponding solution of the RG equation (\eq{RGequationRecast}) is given by \beql{oneLoopCoupling}
  {\bar g}^2(\mathcal J)
  =
  \frac{{\bar g}_1^2(\mu_0^4)}{1 + \tfrac{B_1}{2} {\bar g}_1^2(\mu_0^4) \ln\big( \lvert \mathcal J \rvert / \mu_0^4 \big)} \,.
\eeq

Substituting this expression into the effective all-order Lagrangian in \eq{YMeffectiveLagrangian}, we obtain
\beql{oneLoopEffectiveLagrangian}
  {\mathcal L}_\text{eff}^{(1)}
  =
  \frac{\mathcal J}{4{\bar g}_1^2(\mu_0^4)}
  \bigg[ 1 + \frac{B_1}{2} {\bar g}_1^2(\mu_0^4) \ln\bigg( \frac{\lvert \mathcal J \rvert}{\mu_0^4} \bigg) \bigg] \,.
\eeq

Making trivial substitutions,
\beql{SavvidySubstitutions}
  \mathcal J \to -4 {\bar g}_1^2(\mu_0^4) \mathcal F \,,
  \quad
  \mu_0^4 \to 2e\mu^4 \,,
  \quad
  {\bar g}_1^2(\mu_0^4) \to g_\text{YM}^2 \,,
\eeq
one arrives at another form of the one-loop effective Lagrangian frequently used in the literature (e.g.,~Ref.~\cite{Savvidy:2019grj} and references therein),
\beql{SavvidyOneLoopLagrangian}
  {\mathcal L}_\text{eff}^{(1)}
  =
  -\mathcal F
  -
  \frac{bN}{96\pi^2} g_\text{YM}^2 \mathcal F
  \bigg[ \ln\bigg( \frac{2\lvert g_{YM}^2 \mathcal F \rvert}{\mu^4} \bigg) - 1 \bigg] \,.
\eeq

The compact form of the all-order effective Lagrangian used earlier in \eq{YMeffectiveLagrangian} straightforwardly produces the standard representation of the one-loop effective Lagrangian upon the redefinitions of \eq{SavvidySubstitutions}, which is reassuring. The~usual covariant renormalization condition on the effective Lagrangian~\cite{Savvidy:2019grj}
\beql{renormalisationCondition}
  \frac{\partial {\mathcal L}_\text{eff}}{\partial \mathcal F} \Big\lvert_{t=0} = -1 \,,
  \qquad
  t \equiv \frac{1}{2}\ln\bigg( \frac{2\lvert g_\text{YM}^2 \mathcal F \rvert}{\mu^4} \bigg) \,,
\eeq
is apparently satisfied for \eq{SavvidyOneLoopLagrangian}. Indeed,
\beql{renormalisationConditionValidation}
  \frac{\partial {\mathcal L}_\text{eff}^{(1)}}{\partial \mathcal F}
  =
  -1
  -
  \frac{bN}{96\pi^2} g_\text{YM}^2 \ln\bigg( \frac{2\lvert g_\text{YM}^2\mathcal F \rvert}{\mu^4} \bigg)
  \to
  -1
  \qquad
  \text{for}
  \qquad
  \ln\bigg( \frac{2\lvert g_\text{YM}^2\mathcal F \rvert}{\mu^4} \bigg) \to 0 \,.
\eeq

This condition has been originally employed in Refs.~\cite{Savvidy:1977as,Matinyan:1976mp} to derive the generic form of the one-loop effective Lagrangian in \eq{SavvidyOneLoopLagrangian} (see Ref.~\cite{Savvidy:2019grj} and references therein, for~a more elaborate review). In~a compact notation of \eq{YMeffectiveLagrangian}, the~latter condition reads
\beql{compactCondition}
  \frac{\partial {\mathcal L}_\text{eff}}{\partial \mathcal J} \Big\lvert_{t=0}
  =
  \frac{1}{4{\bar g}_1^2(\mu_0^4)} \,,
  \qquad
  t \equiv \frac{1}{2}\ln\bigg( \frac{e\lvert \mathcal J \rvert}{\mu_0^4} \bigg) \,.
\eeq

\begin{adjustwidth}{-\extralength}{0cm}
\printendnotes[custom]
\reftitle{References}

\end{adjustwidth}
\end{document}